%% file: main.tex
\documentclass[a4paper, twoside, titlepage]{llncs}
\input{packages}
\input{macros}
\usepackage{geometry}
\title{Monitoring Distributed Component-Based Systems}

\begin{document}
\input{header}
\input{abstract}
%
%%%%%%%%%%%%%%%%%%%%%%%%%%%%%%%%%
%%%%%%%%%%%%%%%%%%%%%%%%%%%%%%%%%
%
\input{introduction}
\input{preliminary}
\input{distcbs}
\input{globaltrace}
\input{complattice}

%\input{rvltl}
\input{implementation}
\input{evaluation}
\input{rw}

\input{conclusion}
%
%%%%%%%%%%%%%%%%%%%%%%%%%%%%%%%%%
%%%%%%%%%%%%%%%%%%%%%%%%%%%%%%%%%
%
\bibliographystyle{splncs}
%\bibliography{main}
\bibliography{MyLib}
\appendix
\input{appendix}
\end{document}

%% file: packages.tex
\usepackage{fullpage}
\usepackage[utf8]{inputenc}
\usepackage[T1]{fontenc}
\usepackage{stmaryrd}
\usepackage{amssymb}
\usepackage[pdftex]{graphicx}
\usepackage{verbatim}
\usepackage{amsmath}
\usepackage{times}
\usepackage{xspace}
\usepackage{tikz}
\usepackage{pgfplots}
\usetikzlibrary{arrows,automata,fit,arrows,decorations.pathreplacing,calc,positioning,decorations.text,petri,positioning, shapes,patterns,backgrounds}
\usepackage{algorithm}
\usepackage{algpseudocode}
\usepackage{xcolor}
\usepackage{todonotes}
\usepackage{multirow}
\usepackage{multicol}
\usepackage{tablefootnote}
\usepackage{mathtools}
\usepackage{gensymb}
\usepackage{url}
\usepackage{dsfont}

\usepackage[compatibility=false]{caption}
\usepackage{subcaption}
\usepackage{flexisym}
\usepackage{amsmath}
\usepackage{graphicx}
\usepackage{xcolor,colortbl}
\usepackage{mathpartir}

\usepackage{booktabs}  % publication-quality tables
\usepackage{eqparbox}  % sub-align table cells
\usepackage{booktabs}
\usepackage[]{hyperref}

\usepackage{upgreek}

%% file: macros.tex
\DeclareMathOperator{\init}{\mathit{init}}

\DeclareMathOperator{\make}{\textsc{Make}}
\DeclareMathOperator{\toolname}{\textsc{RVDist}}
\DeclareMathOperator{\joint}{joint}

\DeclareMathOperator{\meet}{meet}

\newcommand{\lmid}{\ \middle| \ }
\newcommand{\UPD}{\mathit{U\!P\!D}}
\newcommand{\PROG}{\mathit{P\!R\!O\!G}}
\DeclareMathOperator{\fromthequeue}{\mathit{from\textnormal{-}the\textnormal{-}queue}}
\DeclareMathOperator{\latticextend}{\mathit{lattice\textnormal{-}extend}}
\DeclareMathOperator{\eventused}{\mathit{event\textnormal{-}is\textnormal{-}used}}
\newcommand{\tooldist}[0]{\textsc{RVDist}}

\newcommand{\setofschedulers}[0]{\ensuremath{\mathbf{S}}}
\newcommand{\setofcomponents}[0]{\ensuremath{\mathbf{B}}}
\newcommand{\setofsharedcomponents}[0]{\ensuremath{\mathbf{B}_{\rm s}}}

\newcommand{\ie}{i.e.,\xspace}
\newcommand{\eg}{{e.g.,}\xspace}

\newcommand{\cf}{\textsl{cf.}\xspace}
\newcommand{\etc}{\textsl{etc.}\xspace}

\newcommand{\resp}{\textsl{resp.}\xspace}

\newcommand{\defref}[1]{Definition~\ref{#1}}
\newcommand{\exref}[1]{Example~\ref{#1}}
\newcommand{\propref}[1]{Property~\ref{#1}}
\newcommand{\proporef}[1]{Proposition~\ref{#1}}
\newcommand{\secref}[1]{Sec.~\ref{#1}}
\newcommand{\algoref}[1]{Algorithm~\ref{#1}}
\newcommand{\figref}[1]{Figure~\ref{#1}}

\newcommand{\ignore}[1]{}
\newcommand{\true}{\ensuremath{\mathtt{true}}}
\newcommand{\false}{\ensuremath{\mathtt{false}}}
\newcommand{\goesto}[1][]{\stackrel{#1}{\longrightarrow}}
\newcommand{\Actions}[0]{\ensuremath{\mathit{Act}}}
\newcommand{\IntActions}[0]{\ensuremath{\mathit{Int}}}
\newcommand{\GActions}[0]{\ensuremath{\mathit{GAct}}}
\newcommand{\setof}[1]{\ensuremath{\left\{ #1 \right\}}}
\newcommand{\trans}[2]{\stackrel{#2}{\ensuremath{\rightarrow}}_{#1}}
\newcommand{\longtrans}[2]{\ {\ensuremath{\xrightarrow{#2}}_{#1}}\ }

\DeclareMathOperator{\involved}{involved}
\DeclareMathOperator{\managed}{managed}
\DeclareMathOperator{\scope}{scope}
\DeclareMathOperator{\send}{\mathit{snd}}
\DeclareMathOperator{\rcv}{\mathit{rcv}}

\DeclareMathOperator{\Tr}{Tr}

\DeclareMathOperator{\upd}{upd}
\DeclareMathOperator{\last}{last}

\DeclareMathOperator{\length}{length}
\DeclareMathOperator{\pref}{pref}

\DeclareMathOperator{\equ}{equ}
\DeclareMathOperator{\Map}{map}

\DeclareMathOperator{\IP}{\mathit{IP}}
\DeclareMathOperator{\vc}{\mathit{vc}}
\DeclareMathOperator{\VC}{\mathit{VC}}

\DeclareMathOperator{\prog}{prog}
\DeclareMathOperator{\updfct}{upd}
\DeclareMathOperator{\LTL}{\mathit{LTL}}
\DeclareMathOperator{\drain}{\mathit{Drain}}
\DeclareMathOperator{\fil}{\mathit{Fill}}
\DeclareMathOperator{\tank}{\mathit{Tank}}
\DeclareMathOperator{\filling}{\mathit{filling}}
\DeclareMathOperator{\draining}{\mathit{draining}}

\DeclareMathOperator{\inc}{inc}

\DeclareMathOperator{\event}{event}

\DeclareMathOperator{\upda}{extend}
\DeclareMathOperator{\updb}{update}
\DeclareMathOperator{\rem}{remove}

\newcommand{\AP}[0]{\ensuremath{\mathit{AP}}}

\newcommand{\squishlist}{
 \begin{list}{$\bullet$}
  { \setlength{\itemsep}{0pt}
     \setlength{\parsep}{1pt}
     \setlength{\topsep}{1pt}
     \setlength{\partopsep}{0pt}
     \setlength{\leftmargin}{1.2em}
     \setlength{\labelwidth}{1.5em}
     \setlength{\labelsep}{0.4em} } }
\newcommand{\squishend}{
  \end{list}  }

\makeatletter
\newbox\xrat@below
\newbox\xrat@above
\newcommand{\xrightarrowtail}[1][]{%
  \setbox\xrat@below=\hbox{\ensuremath{\scriptstyle ~~~}}%
  \setbox\xrat@above=\hbox{\ensuremath{\scriptstyle #1}}%
  \pgfmathsetlengthmacro{\xrat@len}{max(\wd\xrat@below,\wd\xrat@above)+.6em}%
  \mathrel{\tikz [>->>,baseline=-.75ex]
                 \draw (0,-0.025) -- node[below=-2pt] {\box\xrat@below}
                                node[above=-2pt] {\box\xrat@above}
                       (\xrat@len,-0.025) ;}}
\makeatother

\newcommand{\model}[0]{\ensuremath{\mathbf{M}}}

\newcommand{\pdefref}[1]{Definition~\ref{#1}}
\newcommand{\psecref}[1]{Section~\ref{#1}}
\newcommand{\pfigref}[1]{Figure~\ref{#1}}
\newcommand{\pexref}[1]{Example~\ref{#1}}

\newcommand{\pproporef}[1]{Proposition~\ref{#1}}
\newcommand{\palgoref}[1]{Algorithm~\ref{#1}}

\newcommand{\ppdefref}[1]{Definition~\ref{#1}}
\newcommand{\ppsecref}[1]{Section~\ref{#1}}
\newcommand{\ppfigref}[1]{Figure~\ref{#1}}
\newcommand{\ppexref}[1]{Example~\ref{#1}}

\newcommand{\ppproporef}[1]{Proposition~\ref{#1}}
\newcommand{\ppalgoref}[1]{Algorithm~\ref{#1}}

\newcommand{\pgref}[1]{\ref{#1}}
\DeclareMathOperator{\upto}{,}
\DeclareMathOperator{\quant}{:}

\newcommand{\ra}[1]{\renewcommand{\arraystretch}{#1}}
\usepackage{adjustbox}

%% file: header.tex
\title{Monitoring Distributed Component-Based Systems}
\author{
Hosein Nazarpour\inst{1}, Yli\`es Falcone\inst{1}, Mohamad Jaber\inst{2}, Saddek Bensalem\inst{1}, Marius Bozga\inst{1}
}
%\author{}
%\institute{}
%
\institute{
Univ. Grenoble Alpes, Inria, CNRS, VERIMAG, LIG, Grenoble, France\\
\texttt{Firstname.Lastname@imag.fr} \and
American University of Beirut\\
\texttt{mj54@aub.edu.lb}
}
\titlerunning{Monitoring Distributed Component-Based Systems}
\authorrunning{H. Nazarpour, Y. Falcone, S. Bensalem, M. Bozga}
\maketitle

%% file: abstract.tex
\abstract{ 
This paper addresses the online monitoring of distributed component-based systems with multi-party interactions against user-provided properties expressed in linear-temporal logic and referring to global states.
We consider intrinsically independent components whose interactions are partitioned on distributed controllers.
In this context, the problem that arises is that a global state of the system is not available to the monitor.
Instead, we attach local controllers to schedulers to retrieve the concurrent local traces.
Local traces are sent to a global observer which reconstructs the set of global traces that are compatible with the local ones, in a concurrency-preserving fashion.
In this context, the reconstruction of the global traces is done on-the-fly using a lattice of partial states encoding the global traces compatible with the locally-observed traces.
We implemented our monitoring approach in a prototype tool called {\tooldist}.
{\tooldist} executes in parallel with the distributed model and takes as input the events generated from each scheduler and outputs the evaluated computation lattice.
Our experiments show that, thanks to the optimisation applied in the online monitoring algorithm, i) the size of the constructed computation lattice is insensitive to the the number of received events, (ii) the lattice size is kept reasonable and (iii) the overhead of the monitoring process is cheap.
}

%% file: introduction.tex
%
%%%%%%%%%%%%%%%%%%%%%%%%%%%%%%%%%%%%%%%%%
\section{Introduction}
\label{sec:intor}
%%%%%%%%%%%%%%%%%%%%%%%%%%%%%%%%%%%%%%%%%
%
%
%%%%%%%%%%%%%%%%%%%%%%%%%%%%%%%%%%
\paragraph{Runtime Verification}
%%%%%%%%%%%%%%%%%%%%%%%%%%%%%%%%%%
%
Runtime Verification (RV)~\cite{rvwww,HavelundG05,Leucker2008,FalconeHR13,SokolskyHL12,BauerLS10,DBLP:conf/rv/FalconeFM09} is a lightweight and effective technique to ensure the correctness of a system at runtime, that is whether or not the system respects or meets a desirable behavior.
It can be used in numerous application domains, and more particularly when integrating together unreliable software components. 
Runtime verification complements exhaustive verification methods such as model checking~\cite{clarke82,queille82}, and theorem proving~\cite{hoare69}, as well as incomplete solutions such as testing~\cite{boris90} and debugging~\cite{shapiro83}.
In RV, a run of the system under inspection is analyzed incrementally using a decision procedure: a \emph{monitor}. 
This monitor may be generated from a user-provided high level specification (\eg a temporal formula, an automaton). 
This monitor aims to detect violation or satisfaction w.r.t. the given specification. 
Generally, it is a state machine processing an execution sequence (step by step) of the monitored program, and producing a sequence of verdicts (truth-values taken from a truth-domain) indicating specification fulfillment or violation. 
For a monitor to be able to observe the runs of the system, the system should be instrumented in such a way that at runtime, the program sends relevant events that are consumed by the monitor. 
Usually, one of the main challenges when designing an RV framework is its performance. 
That is, adding a monitor in the system should not deteriorate executions of the initial system, time and memory wise.
%
%%%%%%%%%%%%%%%%%%%%%%%%%%%%%%%%%%
\paragraph{Component-Based Systems with Multi-Party Interactions (CBSs)}
%%%%%%%%%%%%%%%%%%%%%%%%%%%%%%%%%%
%
Component-based design consists in constructing complex systems from given requirements using a set of predefined components~\cite{sifakis2005framework}.
Components are abstract building blocks encapsulating behavior.
Each component is defined as an atomic entity with some actions and interfaces. 
Components communicate and interact with each other through their interfaces.
They can be composed in order to build composite components. 
Their composition should be rigorously defined so that it is possible to infer the behavior of composite components from the behavior of their constituents as well as global properties from the properties of individual components and the interactions between them.
Each multi-party interaction is a set of simultaneously-executed actions of the existing components~\cite{bliudze08}.

The execution of a CBS with multi-party interactions is carried on using schedulers (also known as processes or engines) managing the interactions.
In the distributed setting, the execution of interactions of a CBS is distributed among several independent schedulers.
In an implemented distributed CBS, schedulers and components are interconnected (\eg networked physical locations) and work together as a whole unit to meet some requirements.
The execution of a multi-party interaction is then achieved by sending/receiving messages between the scheduler in charge of the execution of the interaction and the components involved in the interaction~\cite{basu2008distributed}.
In this setting, each scheduler along with its associated components can be seen as a multi-threaded system, so that the computations of the components in the scope of the scheduler are done concurrently.
Moreover, the simultaneous execution of several interactions managed by several schedulers is possible.
Thus, the execution trace of a distributed system is a partial trace.
%
%\textbf{Although the execution trace of the system is defined, it is not observable.}
%
Each scheduler is aware of its execution trace, that is the \textit{locally observed partial-trace} consisting of a sequence of the partial states of components in the scope of the scheduler.
A set of the local partial-traces of the schedulers represents the execution trace of the system.
%
%
%
%
%
%
%Although the global execution is defined in this setting, it is not observable and each scheduler is aware of its local execution trace.
%%
%A set of local traces represents the global trace of the entire distributed system.
%
%
%
%%%%%%%%%%%%%%%%%%%%%%%%%%%%%%%%%%
\paragraph{Challenges of Monitoring Distributed CBSs}
%%%%%%%%%%%%%%%%%%%%%%%%%%%%%%%%%%
%
However, it is generally not possible to ensure or verify a desired behavior of such systems using static verification techniques such as model-checking or static analysis, either because of the state-space explosion problem or because the property can only be decided with information available at runtime (\eg from the user or the environment).
In this paper, we are interested in complementary verification techniques for CBSs such as runtime verification.
To this end, we propose techniques to runtime verify a component-based system against properties referring to the global state of the system.
This implies in particular that properties can not be projected and checked on individual components.
In the following we point out the problems that one encounters when monitoring CBSs at runtime.

The runtime monitoring of an asynchronous distributed system is a much more difficult task, because  in the distributed setting, (i) the execution of the system is more dynamic and parallel, in the sense that each scheduler executes its associated actions concurrently and we have a set of parallel executions, (ii) neither a global clock nor a shared memory is used, hence, schedulers can have different processing speeds and can suffer from clock drifts, and (iii) since the execution of interactions is based on sending/receiving messages and delays in the reception of messages in asynchronous communications are inevitable, the runtime monitor does not receive the events with the same order as they are actually occurred.
Therefore, events cannot be ordered based on time.
The absence of ordering between the execution of the interactions in different schedulers causes the main problem in the distributed setting that is (i) the global state of the system does not exist, and (ii) the actual partial trace of the system is not observable.
For monitoring such systems, we avoid synchronization to take global snapshots, which would go against the parallelism of the verified system.
The monitoring problem is even more complicated because no component of the system can be aware of the global trace and the monitor needs to reconstruct the global trace from the events emitted by schedulers at runtime, and then reason about their correctness. 
Our goal is to provide methods that can be used for the verification of such CBSs by applying instrumentation techniques to observe the global behavior of the systems while preserving their performance and initial behavior.
Consequently, the designed instrumentation technique should be defined formally and its correctness formally proved.
%
%%%%%%%%%%%%%%%%%%%%%%%%%%%%%%%%%%
\paragraph{Approach Overview}
%%%%%%%%%%%%%%%%%%%%%%%%%%%%%%%%%%
%
We define a \emph{monitoring hypothesis} based on the definition of an abstract semantic model of CBS.
The abstract semantic system is composed of a non-empty set of components $\setofcomponents$ and their joint actions which are managed by a non-empty set of scheduler $\setofschedulers$.
Each component $B \in \setofcomponents$ is endowed with a set of actions $\Actions_B$.
Joint actions of component, aka multi-party interactions, involve the execution of actions on several components.
An interaction is a non-empty subset of $\setof{\Actions_B \mid B \in \setofcomponents}$, such that at most one action of each component is involved in an interaction.
In addition, to model concurrent behavior, each atomic component $B \in \setofcomponents$ has internal actions which we model as a unique action $\beta$, such that each action of $B$ is followed by the internal action $\beta$.
The set of interactions in the system is distributed among a non-empty set of schedulers $\setofschedulers$.
Schedulers coordinate the execution of interactions and ensure that each multi-party interaction is jointly executed.
Our monitoring hypothesis is that the behavior of the monitored system complies to this model.
We argue that this model is abstract enough to encompass a variety of (component-based) systems, and serves the purpose of describing the knowledge needed on the verified system and later guides their instrumentation.
A property $\varphi$ specifies the desired runtime behavior of the system (referring to the global state of the system) which has to be evaluated while the system is running.
%
%Correctness of such an abstract model needs to be verified at runtime with respect to a property of components product state space $\varphi$.
%
In this context, each scheduler is only aware of its local partial-trace, that is a set of ordered \textit{local events} (\ie actions which change the state of the system).
Moreover, events from different schedulers are not totally ordered.
In order to evaluate the global behavior of the system consisting of several schedulers, it is necessary to find i) a set of possible ordering among the events of all schedulers, that is, the set of compatible partial traces that could possibly happen in the system, and ii) the set of  global traces corresponding to the compatible partial traces. 
Intuitively, our method consists in two steps, (i)  instrumentation of the abstract CBS to obtain system events and send them to the monitor, and (ii) reconstruction of the compatible global trace(s) of the system and evaluate them on-the-fly by the monitor.
The instrumentation is done as follows:
\begin{itemize}
	\item Each scheduler $S \in \setofschedulers$ is composed with a controller.
	 Each controller is in charge of detecting and sending the events that occurred in the corresponding scheduler.
	 \item A central monitor component (global observer) is added to the system. 
	 This new module receives the events which occurred in schedulers and are sent by their associated controllers.
	 The monitor works in parallel with the system and applies an online monitoring algorithm upon the reception of each event.
	 %
	%
%\textbf{	For an partial trace of a multi-threaded CBS, we define the notion of \textit{witness} trace, which is intuitively the unique trace of global states corresponding to the trace of the multi-threaded CBS as if this CBS was executed on a single thread (sequential setting) with the same sequence of the executed interactions.}
	%
	 \item In the distributed setting where (i) we have more than one schedulers, (ii) we have possibly some shared components (\ie components in the scope of more than one scheduler), and (iii) schedulers do not communicate together and only communicate with their own associated components, we compose each shared component with a controller.
	 The controller of a shared component only communicates with the controllers of the schedulers whenever the shared components and the schedulers communicate.
	Indeed, in our abstract model, what makes the events of different schedulers to be causally related is only the shared components which are involved in several multi-party interactions managed by different schedulers. 
	In other words, the executions of two actions managed by two schedulers and involving a shared component are definitely causally related, because each execution requires the termination of the other execution in order to release the shared component.
	To take into account these existing causalities among the events, in the distributed setting, we employ vector clocks to define the ordering of events.
	The controller of a shared component is used to resolve the ordering among the events involving the shared component.
	 Each event associated to the execution of a multi-party interaction is labeled by a vector clock.
	 Ordering of such events are defined based on their vector clocks.
	 %
	 %Moreover, in distributed setting events are partially ordered, so that the monitor receives the events of a unique scheduler in the same order as they happened in the scheduler, but the ordering of two events from two different schedulers are not defined.
	%
	The monitor receives the partially-ordered events representing the local partial-traces.
	%
%	 
%\textbf{	To find the global trace, the monitor must find the set of possible total order among the events, to obtain the set of compatible global traces that could possibly happen in the system.
%%
%%	We present such a set of compatible global traces in a form of general notion of \textit{computation lattice}.
%	To represent all the compatible global traces we use the general notion of \textit{computation lattice}.
%	%
%	A computation lattice has $|\setofschedulers|$ orthogonal axes, with one axis for each scheduler. 
%	%
%	A path in the lattice is a compatible global trace of the system.}
	 %
\end{itemize}
We propose an online monitoring method for distributed systems as follows:

In the distributed setting, the monitor is aware of the local partial-traces of the schedulers.
	The monitor computes all the compatible partial traces of the system with respect to the partial ordering of the received events.
	Each compatible partial trace could possibly happen in the system and would produce the same events.
	We introduce an online algorithm to reconstruct the corresponding global trace for each partial trace.
	To represent the set of reconstructed  compatible global traces we use the general notion of \textit{computation lattice}.
	A computation lattice has $|\setofschedulers|$ orthogonal axes, with one axis for each scheduler.
	The direction of each axis represents the system state evolution with respect to the execution of interactions managed by the associated scheduler.
	Each path in the lattice represents a compatible global trace of the system.
	We define a novel on-the-fly monitoring technique to evaluate any Linear Temporal Logic (LTL) properties over the computation lattice.
	%
%	to find the global trace, the monitor must find the set of possible total order among the events, to obtain the set of compatible global traces that could possibly happen in the system.
%
%	We present such a set of compatible global traces in a form of general notion of \textit{computation lattice}.
%	To represent all the compatible global traces we use the general notion of \textit{computation lattice}.
%	%
%	A computation lattice has $|\setofschedulers|$ orthogonal axes, with one axis for each scheduler. 
%	%
%	A path in the lattice is a compatible global trace of the system.}
	%
%	For distributed systems we propose an online algorithm constructing the computation lattice to evaluate any Linear Temporal Logic (LTL) properties that refers to the global ready-states of the system.
	%
%We define a novel on-the-fly progression method over the constructed computation lattice. 
	%
To this end, we define a new structure of the computation lattice in which each node $\eta$ of the lattice is augmented by a set of formulas representing the evaluation of all the possible global traces from the initial node of the lattice (\ie initial state of the system) up to node $\eta$.	
We show that the constructed lattice is correct in the sense that it encompasses all the compatible global traces (\pproporef{proposition:make_soundness}, and \pproporef{proposition:make_completeness}).
The given formula is monitored by progression over the constructed lattice, so that the frontier node of the lattice contains a set of formulas, each of which corresponding to the evaluation of a compatible global trace (Theorem~\ref{theorem:soundness}, p.~\pageref{theorem:soundness}, and Theorem~\ref{theorem:completeness}, p.~\pageref{theorem:completeness}). 
Furthermore, we introduce an optimization algorithm to keep the size of the constructed lattice small by removing the unnecessary nodes. 
We show that such an optimization on the one hand does not affect the evaluation of the system and on the other hand increases the performance of the monitoring process.

We present an implementation of our monitoring approach in a tool called $\toolname$.
$\toolname$ is a prototype tool written in the C++ programming language.
$\toolname$ takes as input an LTL formula and a sequence of events, then constructs and evaluate the computation lattice against the given LTL property.
%
%We show that the constructed lattice consists in the set of all compatible global traces of the system.
%
Moreover, we present the evaluation of our monitoring approach on several distributed systems carried out with $\toolname$.
Our experiments show that, thanks to the optimization applied in the online monitoring algorithm, (i) the size of the constructed computation lattice is insensitive to the the number of received events, (ii) the lattice size is kept reasonable and (iii) the overhead of the monitoring process is cheap.
\paragraph{Outline.}
The remainder of this paper is organized as follows.
Section~\ref{sec:prelim} introduces some preliminary concepts. 
Section~\ref{sec:distcbs} defines an original abstract model of distributed CBSs, suitable for monitoring purposes, and allowing to define a monitoring hypothesis for the runtime verification of distributed CBSs.
In \secref{sec:local-global-trace}, we present the instrumentation used to generate the events of each scheduler which are aimed to be used in the construction of the global trace of a distributed CBS.
In \secref{sec:lattice}, we construct the computation lattice by collecting the events from the different schedulers.
Runtime verification of distributed CBSs is presented in \secref{sec:ltl}.
Section~\ref{sec:implem} describes $\toolname$, a C++ implementation of the monitoring framework used to carry an evaluation of our approach described in \secref{sec:evaluation}.
Section~\ref{sec:rw} presents related work.
Section~\ref{sec:conc} concludes and presents future work. 
Proofs of the propositions are in Appendix~\ref{sec:proofs}.
%
%

%% file: preliminary.tex
%
%%%%%%%%%%%%%%%%%%%%%%%%%%%%%%%%%%%%%%%%%
\section{Preliminaries and Notations}
\label{sec:prelim}
%%%%%%%%%%%%%%%%%%%%%%%%%%%%%%%%%%%%%%%%%
%
\paragraph{Sequences.}
Considering a finite set of elements $E$, we define notations about sequences of elements of $E$. A sequence $s$ containing elements of $E$ is formally defined by a total function $s : I \rightarrow E$ where $I$ is either the integer interval $[0, n]$ for some $n \in \mathbb{N}$, or $\mathbb{N}$ itself (the set of natural numbers). 
Given a set of elements $E$, $e_1\cdot e_2\cdots e_n$ is a sequence or a list of length $n$ over $E$, where $\forall i\in [1,n]: e_i\in E$.
The empty sequence is noted $\epsilon$ or $[~]$, depending on the context.
The set of (finite) sequences over E is noted $E^*$.
$E^+$ is defined as $E^* \setminus \{\epsilon\}$.
The length of a sequence $s$ is noted $\length(s)$.
We define $s(i)$ as the $i^{\rm th}$ element of $s$ and $s(i\cdots j)$ as the factor of $s$ from the $i^{\rm th}$ to the $j^{\rm th}$ element.
$s(i\cdots j)=\epsilon$ if $i>j$.
We also note $\pref(s)$, the set of non-empty \textit{prefixes} of $s$, \ie $\pref(s)=\{s(1\cdots k) \mid 1 \leq k\leq \length(s)\}$.
Operator $\pref$ is naturally extended to sets of sequences.
We define function  $\last:E^+\rightarrow E$ as $\last(e)= s(\length(s))$.
For an infinite sequence  $s=e_1 \cdot e_2 \cdot e_3 \cdots $, we define $s(i\cdots)= e_i \cdot e_{i+1} \cdots $ as the suffix of sequence $s$ from index $i$ on.
\paragraph{Tuples.}
An $n$-tuple is an ordered list of $n$ elements, where $n$ is a strictly positive integer.  
By $t[i]$  we denote $i^{\rm th}$ element of tuple $t$.
\paragraph{Labeled transition systems.}
Labeled Transition Systems (LTSs) are used to define the semantics of CBSs.
An LTS is defined over an alphabet $\Sigma$ and is a 3-tuple $(\mathrm{State}, \mathrm{Lab}, \mathrm{Trans})$ where $\mathrm{State}$ is a non-empty set of states, $\mathrm{Lab}$ is a set of labels, and $\mathrm{Trans}\subseteq \mathrm{State}\times \mathrm{Lab} \times \mathrm{State}$ is the transition relation.
A transition $(q,a,q') \in \mathrm{Trans}$ means that the LTS can move from state $q$ to state $q'$ by consuming label $a$. 
We abbreviate $(q,a,q')\in\mathrm{Trans}$ by $q\longtrans{\mathrm{Trans}}{a} q'$ or by $q \longtrans{}{a} q'$ when clear from context. 
Moreover, relation $\mathrm{Trans}$ is extended to its reflexive and transitive closure in the usual way and we allow for regular expressions over $\mathrm{Lab}$ to label moves between states: if $\mathit{expr}$ is a regular expression over $\mathrm{Lab}$ (i.e., $\mathit{expr}$ denotes a subset of $\mathrm{Lab}^*$), $q \longtrans{}{\mathit{expr}} q'$ means that there exists one sequence of labels in $\mathrm{Lab}$ matching $\mathit{expr}$ such that the system can move from $q$ to $q'$.
\paragraph{Observational equivalence and bi-simulation.}
The \emph{observational equivalence} of two transition systems is based on the usual definition of weak bisimilarity~\cite{mil95}, where $\theta$-transitions are considered to be unobservable.
Given two transition systems $S_1 = (\mathrm{Sta}_1, \mathrm{Lab} \cup \{ \theta \},\rightarrow_2)$ and $S_2 = (\mathrm{Sta}_2 , \mathrm{Lab} \cup \{ \theta \}, \rightarrow_2)$, system $S_1$ \emph{weakly simulates} system $S_2$, if there exists a relation $R \subseteq \mathrm{Sta}_{1} \times \mathrm{Sta}_{2}$ that contains the 2-tuple made of the initial states of $S_1$ et $S_2$ and such that the two following conditions hold:
\begin{enumerate}
 \item $\forall ( q_1 , q_2 ) \in R, \forall a \in \mathrm{Lab} : q_1 \goesto[a]_1 q_1' \implies \exists q_2' \in \mathrm{Sta}_2 :\ \left((q_1' , q_2') \in R \wedge q_2 \longtrans{2}{\theta^*\cdot a \cdot \theta^*} q_2'\right)$, and
 \item  $\forall (q_1,q_2) \in R : \left(\exists q_1' \in \mathrm{Sta}_1 : q_1 \goesto[\theta]_1 q_1'\right) \implies \exists q_2' \in \mathrm{Sta}_2 : \left((q_1',q_2') \in R \wedge q_2 \goesto[\theta^*]_2 q_2' \right)$.
\end{enumerate}
%
%where $\goesto[]_{\mathrm{Trans}_1}$ and $\goesto[]_{\mathrm{Trans}_2}$ are noted $\goesto[]_{1}$ and $\goesto[]_{2}$, respectively.

Equation 1. states that if a state $q_1$ simulates a state $q_2$ and if it is possible to perform $a$ from $q_1$ to end in a state $q_1'$, then there exists a state $q_2'$ simulated by $q_1'$ such that it is possible to go from $q_2$ to $q_2'$ by performing some unobservable actions, the action $a$, and then some unobservable actions.
Equation 2. states that if a state $q_1$ simulates a state $q_2$ and it is possible to perform an unobservable action from $q_1$ to reach a state $q_1'$, then it is possible to reach a state $q_2'$ by a sequence of unobservable actions such that $q_1'$ simulates $q_2'$.
In that case, we say that relation $R$ is a weak simulation over $S_1$ and $S_2$ or equivalently that the states of $S_1$ are (weakly) similar to the states of $S_2$. Similarly, a weak bi-simulation over $S_1$ and $S_2$ is a relation $R$ such that $R$ and $R^{-1} = \{ (q_2,q_1) \in \mathrm{Sta}_2 \times \mathrm{Sta}_1 \mid (q_1,q_2)\in R\}$ are both weak simulations. In this latter case, we say that $S_1$ and $S_2$ are \emph{observationally equivalent} and we write $S_1 \sim S_2$ to express this formally.
\paragraph{Vector Clock.}
Lamport introduced logical clocks as a device to substitute for the global real time clock~\cite{Lamport78}.
Logical clocks are used to order events  based on their relative logical dependencies rather than on a ``time'' in the common sense. 
Mattern and Fidge's vector clocks~\cite{fidge87,mattern89} are a more powerful extension (\ie strongly consistent with the ordering of events) of Lamport's scalar logical clocks.
In a distributed system with a set of schedulers $ \setof{S_1,\ldots,S_m}$, $\VC=\setof{(c_1,\ldots,c_m) \mid j\in [1 \upto m] \wedge c_j \in \mathbb{N} }$ is the set of vector clocks, such that vector clock $\vc\in \VC$ is a tuple of $m$ scalar (initially zero) values $c_1,\ldots,c_m$ locally stored in each scheduler $S_j \in\setof{S_1,\ldots,S_m}$ where  $\forall k \in [1 \upto m] \quant \vc[k]=c_k$ holds the latest (scalar) clock value scheduler $S_j$ knows about scheduler $S_k\in\setof{S_1,\ldots,S_m}$.
Each event in the system is associated to a unique vector clock.
For two vector clocks $\vc_1$ and $\vc_2$, $\max(\vc_1,\vc_2)$ is a vector clock $\vc_3$ such that $\forall k\in [1 \upto m] \quant \vc_3[k]=\max (\vc_1[k], \vc_2[k])$. $\min (\vc_1,\vc_2)$ is defined in similar way.
Moreover two vector clocks can be compared together such that   
$\vc_1 < \vc_2 \Longleftrightarrow \forall k\in[1 \upto m] \quant \vc_1[k] \leq \vc_2[k] \wedge \exists z\in[1 \upto m] \quant \vc_1[z] < \vc_2[z]$.
\paragraph{Happened-before relation~\cite{Lamport78}.}
The relation $\rightarrowtail$ on the set of events of a system is the smallest relation satisfying the following
three conditions:
(1) If $a$ and $b$ are events in the same
scheduler, and $a$ comes before $b$, then $a \rightarrowtail b$.
(2) If $a$ is the sending of a message by one scheduler and $b$ is the reception of the same message by another scheduler, then $a \rightarrowtail b$.
(3) If $a \rightarrowtail b$ and $b \rightarrowtail c$ then $a \rightarrowtail c$.
Two distinct events $a$ and $b$ are said to be concurrent if $a \not\!\rightarrowtail b$ and $b \not\!\rightarrowtail a$.

Vector clocks are strongly consistent with happened-before relation. That is, for two events $a$ and $b$ with associated vector clocks $\vc_a$ and $\vc_b$ respectively, $\vc_a < \vc_b \Longleftrightarrow a \rightarrowtail b$. 
\paragraph{Computation lattice~\cite{mattern89}.}
The computation lattice of a distributed system is represented in the form of a directed graph with $m$ (\ie number of schedulers that are executed in distributed manner) orthogonal axes.
Each axis is dedicated to the state evolution of a specific scheduler.
A computation lattice expresses all the possible traces in a distributed system.
Each path in the lattice represents a global trace of the system that could possibly have happened.
A computation lattice $\mathcal{L}$ is a pair $(N,\rightarrowtail)$, where $N$ is the set of nodes (\ie global states) and $\rightarrowtail$ is the set of happened-before relations among the nodes. 
\paragraph{Linear Temporal Logic (LTL)~\cite{TemporalLogicOfProgram}.}
Linear temporal logic (LTL) is a formalism for specifying properties of systems.
An LTL formula is built over a set of atomic propositions $AP$.
LTL formulas are written with the following grammar:
\[\varphi::= \quad p \quad|\quad \neg \varphi \quad|\quad \varphi_1 \vee \varphi_2\quad| \quad \textbf{X} \varphi \quad|\quad \varphi_1 \textbf{U} \varphi_2 \]
where $p\in AP$ is an atomic proposition. 
Note that we use only the \textbf{X} and \textbf{U} modalities for defining the valid formulas in LTL. The other modalities such as \textbf{F} (eventually), \textbf{G} (globally), \textbf{R} (release), \etc in LTL can be defined using the \textbf{X} and \textbf{U} modalities.

Let $ \sigma=q_0\cdot q_1 \cdot q_2 \cdots$ be an infinite sequence of states and $\models $ denotes the satisfaction relation.
The semantics of LTL is defined inductively as follows:
\begin{itemize}
	\item{\makebox[2cm][l]{$\sigma \models p$} $ \Longleftrightarrow q_0 \models p \  (\text{\ie }\  p \in q_0)$, for any $p \in AP$}
	\item{\makebox[2cm][l]{$\sigma \models \neg \varphi$} $\Longleftrightarrow \sigma\not\models \varphi $}
	\item{\makebox[2cm][l]{$\sigma \models \varphi_1 \vee \varphi_2$} $ \Longleftrightarrow \sigma\models\varphi_1\vee\sigma\models\varphi_2$}
	\item{\makebox[2cm][l]{$\sigma \models \textbf{X} \varphi$} $ \Longleftrightarrow \sigma(1\cdots) \models \varphi$}
	\item{\makebox[2cm][l]{$\sigma \models \varphi_1 \textbf{U} \varphi_2$} $ \Longleftrightarrow \exists j\geqslant 0 : \sigma(j\cdots) \models \varphi_2 \wedge \sigma(i\cdots) \models \varphi_1 , 0 \leqslant i< j$}
\end{itemize}
An atomic proposition $p$ is satisfied by $\sigma$ when it is member of the first state of $\sigma$.
$\sigma$ satisfies formula $\neg \varphi$ when it does not satisfy $\varphi$.
Disjunction of $\varphi_1$ and $\varphi_2$ is satisfied when either $\varphi_1$ or $\varphi_2$ is satisfied by $\sigma$.
$\sigma$ satisfies formula $\textbf{X} \varphi$  when the sequence of states starting from the next state of $\sigma$, that is, $q_1$ satisfies $\varphi$.
$\varphi_1 \textbf{U} \varphi_2$ is satisfied when $\varphi_2$ is satisfied at some point and $\varphi_1$ is satisfied  until that point.
\paragraph{Pattern-matching.}
We shall use the mechanism of pattern-matching to concisely define some functions.
We recall an intuitive definition for the sake of completeness. Evaluating the expression:
\begin{center}
\begin{tabular}{l}
\texttt{match expression with}\\
\quad\quad $\mid$ \texttt{pattern\_1} $\rightarrow$ \texttt{expression\_1}\\
\quad \quad $\mid$ \texttt{pattern\_2} $\rightarrow$ \texttt{expression\_2}\\
\quad\quad\ldots\\
\quad \quad$\mid$ \texttt{pattern\_n} $\rightarrow$ \texttt{expression\_n}\\
\end{tabular}
\end{center}
consists in comparing successively \texttt{expression} with the patterns \texttt{pattern\_1}, \ldots, \texttt{pattern\_n} in order. When a pattern \texttt{pattern\_i} fits \texttt{expression}, then the associated \texttt{expression\_i} is returned.

%% file: distcbs.tex
%
%%%%%%%%%%%%%%%%%%%%%%%%%%%%%%%%%%%%%%%%%
\section{Distributed CBSs with Multi-Party Interactions}
\label{sec:distcbs}
%%%%%%%%%%%%%%%%%%%%%%%%%%%%%%%%%%%%%%%%%
%
In the following, we describe our assumptions on the considered distributed component-based systems with multi-party interactions.
To this end, we assume a general semantics to define the behavior of the distributed system under scrutiny in order to make our monitoring approach as general as possible. 
However, neither the exact model nor the behavior of the system are known.
How the behaviors of the components and the schedulers are obtained is irrelevant.
Inspiring from conformance-testing theory~\cite{Tretmans93}, we refer to this hypothesis as the \textbf{monitoring hypothesis}.

Consequently, our monitoring approach can be applied to (component-based) systems whose behavior can be modeled as described in the sequel.
The semantics of the following model is similar to and compatible with other models for describing distributed computations (see \secref{sec:rw} for a comparison with other models and possible translations between models).
The remainder of this section is organized as follows.
Subsection~\ref{subsec:semantics} defines an abstract distributed component-based model. 
Subsection~\ref{subsec:trace} defines the execution traces of the abstract model, later used for runtime verification.
%
%%%%%%%%%%%%%%%%%%%%%%%%%%%%%%%%%%%%%%%%%
\subsection{Semantics of a Distributed CBS with Multi-Party Interactions}
\label{subsec:semantics}
%%%%%%%%%%%%%%%%%%%%%%%%%%%%%%%%%%%%%%%%%
%
In the following, we present the architecture of our abstract semantic CBS which is used throughout this paper.
\paragraph{Architecture of the system.}
The system under scrutiny $\model$ is composed of \emph{components} in a non-empty set $\setofcomponents= \setof{ B_1, \ldots, B_{|\setofcomponents|}}$ and  \emph{schedulers} in a non-empty set $\setofschedulers= \setof{ S_1, \ldots, S_{|\setofschedulers|} }$.
Each component $B_i$ is endowed with a set of actions $\Actions_i$.
Joint actions of component, aka multi-party interactions, involve the execution of actions on several components.
An interaction is a non-empty subset of $\cup^{|\setofcomponents|}_{i=1} \Actions_i$ and we denote by $\IntActions$ the set of interactions in the system.
At most one action of each component is involved in an interaction: $\forall a \in \IntActions \quant |a \cap \Actions_i| \leq 1$.
In addition, to model concurrent behavior, each atomic component $B_i$ has internal actions which we model as a unique action $\beta_i$, such that each action of $B_i$ is followed by the internal action $\beta_i$.
Schedulers coordinate the execution of interactions and ensure that each multi-party interaction is jointly executed (see \pdefref{def:scheduler}).

Let us assume some auxiliary functions obtained from the architecture of the system.
\begin{itemize}
	\item[--] Function $\involved: \IntActions \rightarrow 2^{\setofcomponents} \setminus \setof{\emptyset} $ indicates the components involved in an interaction.
Moreover, we extend function $\involved$ to internal actions by setting $ \involved(\beta_i) = i $, for any $ \beta_i \in \setof{\beta_1,\ldots,\beta_{|\setofcomponents|}} $.
Interaction $ a \in \IntActions $ is a joint action if and only if $ |\involved(a)| \geq 2 $.
	\item[--] Function $ \managed : \IntActions \rightarrow \setofschedulers $ indicates the scheduler managing an interaction: for an interaction $ a \in \IntActions$, $\managed(a)=S_j$ if $a$ is managed by scheduler $ S_j $. 
	\item[--] Function $\scope : \setofschedulers \rightarrow 2^{\setofcomponents} \setminus \setof{ \emptyset }$ indicates the set of components in the scope of a scheduler such that $\scope(S_j) = {\displaystyle \bigcup_{a'\in \setof{a \in \IntActions \ \mid \ \managed(a) = S_j}} \involved(a')}$.
\end{itemize}
In the remainder, we describe the behavior of components, schedulers, and their composition.
\paragraph{Components.}
The behavior of an individual component is defined as follows.
\begin{definition}[Behavior of a component]
\label{def:component}
The behavior of a component $B$ is defined as an LTS $( Q_B, \Actions_B \cup \setof{ \beta_B }, \trans{B}{} )$ such that:
\begin{itemize}
	\item $Q_B = Q_B^{\rm r} \cup Q_B^{\rm b}$ is the set of states, where $Q_B^{\rm r}$ (resp. $Q_B^{\rm b}$) is the so-called set of \emph{ready} (resp. \emph{busy}) states,
	\item $\Actions_B$ is the set of actions, and $\beta_B$ is the internal action,
	\item $\trans{B}{} \subseteq \left( Q_B^{\rm r} \times \Actions_B \times Q_B^{\rm b} \right) \cup \left( Q_B^{\rm b} \times \setof{\beta_B} \times Q_B^{\rm r} \right)$ is the set of transitions.
\end{itemize}   
Moreover, $Q_B$ has a partition $\setof{Q_B^{\rm r} , Q_B^{\rm b}}$.
\end{definition}
Intuitively, the set of ready (resp. busy) states $Q_B^{\rm r}$ (resp. $Q_B^{\rm b}$) is the set of states such that the component is ready (resp. not ready) to perform an action.
Component $B$ (i) has actions in set $\Actions_B$ which are possibly shared with some of the other components, (ii) has an internal action $\beta_B$ such that $\beta_B \not \in \Actions_B$  which models internal computations of component $B$, and (iii) alternates moving from a ready state to a busy state and from a busy state to a ready state, that is component $B$ does not have busy to busy or ready to ready move (as defined in the transition relation above). 
\input{component_tank}
\begin{example}[Component]
\figref{fig:component} depicts a component $\tank$ whose behavior is defined by the LTS $(Q^{\rm r} \cup Q^{\rm b}, \Actions \cup \setof{ \beta }, \trans{}{} )$ such that: 
\begin{itemize}
	\item[--] $Q^{\rm r}= \setof{d,f}$ is the set of \emph{ready} states and $Q^{\rm b}=\setof{d^\bot,f^\bot}$ is the set of \emph{busy} states,
	\item[--] $\Actions= \setof{\drain,\fil}$ is the set of actions and $\beta$ is the internal action,
	\item[--] $\trans{}{}= \setof{(d,\fil,d^\bot),(d^\bot,\beta,f),(f,\drain,f^\bot),(f^\bot,\beta,d)}$ is the set of transitions.
\end{itemize}
On the border, each $\bullet$ represents an action and provides an interface for the component to synchronize with actions of other components in case of joint actions.
\end{example}
In the following, we assume that each component $B_i \in \setofcomponents$ is defined by the LTS $( Q_{B_i}, \Actions_{B_i} \cup \setof{ \beta_{B_i} }, \trans{B_i}{} )$ where $Q_{B_i}$ has a partition $\setof{Q_{B_i}^{\rm r} , Q_{B_i}^{\rm b}}$ of ready and busy states; as per \defref{def:component}.
\paragraph{Schedulers.}
The behavior of a scheduler is defined as follows.
\begin{definition}[Behavior of a scheduler]
\label{def:scheduler}
The behavior of a scheduler $S$ is defined as an LTS $( Q_{S}, \Actions_{S}, \trans{S}{})$ such that:
\begin{itemize}
	\item $Q_{S}$ is the set of states,
	\item $\Actions_{S} = \Actions^\gamma_{S} \cup \Actions^\beta_{S}$ is the set of actions, where $\Actions^\gamma_{S} = \setof{a \in \IntActions \mid \managed(a) = S}$ and\\ $\Actions^\beta_{S} = \setof{\beta_i \mid B_i \in \scope(S)}$,
	\item $\trans{S}{} \subseteq  Q_{S} \times \Actions_{S} \times Q_{S} $ is the set of transitions.
\end{itemize}  
\end{definition}
$\Actions^\gamma_{S} \subseteq \IntActions$ is the set of interactions managed by $S$, and $\Actions^\beta_{S}$ is the set of internal actions of the components involved in an action managed by $S$.

In the following, we assume that each scheduler $S_j \in \setofschedulers$ is defined by the LTS $( Q_{S_j}, \Actions_{S_j}, \trans{S_j}{})$ where $\Actions_{S_j}= \Actions^\gamma_{S} \cup \Actions^\beta_{S}$; as per \defref{def:scheduler}.
The coordination of interactions of the system \ie the interactions in $\IntActions$, is distributed among schedulers. 
Actions of schedulers consist of interactions of the system.
Nevertheless, each interaction of the system is associated to exactly one scheduler ($\forall a \in \IntActions, \exists! S \in \setofschedulers : a \in \Actions_{S}$).
Consequently, schedulers manage disjoint sets of interactions (\ie $\forall S_i, S_j \in \setofschedulers: S_i \neq S_j \implies \Actions^\gamma_{S_i} \cap \Actions^\gamma_{S_j} = \emptyset$).
Intuitively, when a scheduler executes an interaction, it triggers the execution of the associated actions on the involved components.
Moreover, when a component executes an internal action, it triggers the execution of the corresponding action on the associated schedulers and also sends the updated state of the component to the associated schedulers, that is, the component sends a message including its current state to the schedulers.
Note, we assume that, by construction, schedulers are always ready to receive such a state update.
\begin{remark}
Since components send their update states to the associated schedulers, we assume that the current state of a scheduler contains the last state of each component in its scope. 
\end{remark}
\input{scheduler}
\begin{example}[Scheduler of distributed CBS]
\label{xmp:system}
\figref{fig:scheduler} depicts a distributed component-based system consisting of three components each of which is an instance of the component in \figref{fig:component}.
The set of interactions is $\IntActions= \{ \setof{\drain_1},$ $\fil_{12}, \drain_{23}, \setof{\fil_3}\}$ where $\fil_{12}= \setof{\fil_1,\fil_2}$ and $\drain_{23}= \setof{\drain_2,\drain_3}$ are joint actions.
Two schedulers $S_1$ and $S_2$ coordinate the execution of interactions such that $\managed(\setof{\drain_1})= \managed(\fil_{12} )=S_1$ and $\managed(\setof{\fil_3})=\managed(\drain_{23})=S_2$.
For $j \in [1,2]$, scheduler $S_j$ is defined as $( Q_{S_j}, \Actions_{S_j} , \trans{{S_j}}{} )$ with:
\begin{itemize}
	\item $ Q_{S_j} = \setof{l_0, l_1, l_2 , l_3}$, 
	\item $ \Actions^\gamma_{S_1} = \setof{ \setof{\drain_1}, \fil_{12}}$, $\Actions^\beta_{S_1} = \setof{\beta_1, \beta_2}$,
	\item $\Actions^\gamma_{S_2} = \setof{\drain_{23}, \setof{\fil_3}}$, $\Actions^\beta_{S_2} = \setof{\beta_2, \beta_3}$,
	\item $\trans{{S_1}}{} = \{(l_0 , \beta_2 , l_0),(l_1 , \beta_2 , l_1),(l_1 , \beta_1 , l_0),(l_2 , \beta_2 , l_1) , (l_3 , \beta_2 , l_0),(l_2 , \beta_1 , l_3), (l_3 , \setof{\drain_1} , l_2)$,\\ $(l_0 , \setof{\drain_1} , l_1),(l_0 , \fil_{12} , l_2)\}$,
	\item $ \trans{{S_2}}{} = \{(l_0 , \beta_2 , l_0),(l_1 , \beta_2 , l_1),(l_1 , \beta_3 , l_0),(l_2 , \beta_2 , l_1),(l_3 , \beta_2 , l_0),(l_2 , \beta_3 , l_3)$, $(l_3 , \setof{\fil_3} , l_2),$\\ $(l_0 , \setof{\fil_3} , l_1),(l_0 , \drain_{23} , l_2)\}$.
\end{itemize}
\end{example}
\begin{definition}[Shared component]
The set of shared components is defined as
\[
\setofsharedcomponents = \setof{ B \in \setofcomponents \lmid | \setof{ S \in \setofschedulers \mid  B \in \scope(S)} | \geq 2  }.
\]
\end{definition}
A shared component $B \in \setofsharedcomponents$ is a component in the scope of more than one scheduler, and thus, the execution of the actions of $B$ are managed by more than one scheduler.
\begin{example}[Shared component]
In \figref{fig:scheduler}, component $\tank_2$ is a shared component because interaction $\fil_{12}$, which is a joint action of $\fil_1$ and $\fil_2$, is coordinated by scheduler $S_1$ and interaction $\drain_{23}$, which is a joint action of $\drain_2$ and $\drain_3$, is coordinated by scheduler $S_2$.
\end{example}
The global execution of the system can be described as the parallel execution of interactions managed by the schedulers.
\begin{definition}[Global behavior]
\label{def:global_behavior}
The global behavior of the system is the LTS $( Q , \GActions , \trans{}{}) $ where:
\begin{itemize}
	\item $ Q \subseteq  \bigotimes_{i=1}^{|\setofcomponents|} Q_i \times \bigotimes_{j=1}^{|\setofschedulers|} Q_{S_j}$ is the set of states consisting of the states of schedulers and components,
	\item $ \GActions \subseteq 2^{\Actions \ \cup\ \bigcup_{i=1}^{|\setofcomponents|} \setof{\beta_i}}  \setminus \setof{ \emptyset } $ is the set of possible global actions of the system consisting of either several interactions and/or several internal actions (several interactions can be executed concurrently by the system),
	\item  $\rightarrow \subseteq Q \times \GActions \times Q$ is the transition relation defined as the smallest set abiding to the following rule.   

A transition is a move from state $( q_1, \ldots , q_{|\setofcomponents|} , q_{s_1},$ $ \ldots , q_{ s_{|\setofschedulers|} } ) $ to state $ ( q'_1, \ldots , q'_{|\setofcomponents|} ,  q'_{s_1} , \ldots , q'_{s_{|\setofschedulers|}}) $ on global actions in set $ \alpha \cup \beta $, where $\alpha \subseteq \IntActions$ and $\beta \subseteq \bigcup_{i=1}^{|\setofcomponents|} \setof{ \beta_i } $, noted $ (  q_1, \ldots , q_{|\setofcomponents|} , q_{s_1} , \ldots , q_{s_{|\setofschedulers|}}) \longtrans{}{\alpha \cup \beta}$ $(  q'_1, \ldots , q'_{|\setofcomponents|}, $ $ q'_{s_1} , \ldots , q'_{s_{|\setofschedulers|}}) $, whenever the following conditions hold:
\begin{itemize}
	\item[$C_1$:]
	$\forall i \in [1,{|\setofcomponents|}]: | \left( \alpha \cap \Actions_i \right) \cup \left( \{ \beta_i \} \cap \beta \right) | \leq 1$,
	\item[$C_2$:]
	$ \forall a \in \alpha: \left( \exists S_j \in \setofschedulers: \managed(a) = S_j \right) \Rightarrow \left( q_{s_j} \trans{S_j}{a} q'_{s_j} \wedge \forall B_i \in \involved(a) : q_i \longtrans{B_i}{a \cap \Actions_i}  q'_i \right)$,
	\item[$C_3$:]
	$ \forall \beta_i \in \beta: q_i \longtrans{B_i}{\beta_i} q'_i  \wedge \forall S_j \in \setofschedulers: B_i \in \scope(S_j) : q_{s_j} \longtrans{S_j}{\beta_i} q'_{s_j}$,
	\item[$C_4$:]
	$ \forall B_i \in \setofcomponents \setminus \involved( \alpha \cup \beta ) : q_i = q'_i$,
	\item[$C_5$:]
	$ \forall S_j \in \setofschedulers \setminus \managed( \alpha ) : q_{s_j} = q'_{s_j}$. 
\end{itemize}
where functions $ \involved $ and $ \managed $ are extended to sets of interactions and internal actions in the usual way.
\end{itemize}
\end{definition}
The above rule allows the components of the system to execute independently according to the decisions of the schedulers.
It can intuitively be understood as follows:
\begin{itemize}
	\item
\textit{Condition $C_1$} states that a component can perform at most one execution step at a time.
The executed global actions ($ \alpha \cup \beta $) contains at most one interaction involving each component of the system.
	\item
\textit{Condition $C_2$} states that whenever an interaction $a$ managed by scheduler $S_j$ is executed, $S_j$ and all components involved in this multi-party interaction must be ready to execute it.
	\item
\textit{Condition $C_3$} states that internal actions are executed whenever the corresponding components are ready to execute them. Moreover, schedulers are aware of internal actions of components in their scope. Note that, the awareness of internal actions of a component results in transferring the updated state of the component to the schedulers. 
	\item
\textit{Conditions $C_4$ and $C_5$} state that the components and the schedulers not involved in an interaction remain in the same state.
\end{itemize}
An example illustrating the global behavior of the system depicted in \figref{fig:scheduler} is provided later and described in terms of execution traces (\cf \exref{exmp:global_trace}).
\begin{remark}
The operational description of a distributed CBS has is usually more detailed.
For instance, the execution of conflicting interactions in schedulers needs first  to be authorized by a conflict-resolution module which guarantees that two conflicting interactions are not executed at the same time. 
Moreover, schedulers follow the (possible) priority rules among the interactions, that is, in the case of two or more enabled interactions (interactions which are ready to be executed by schedulers), those with higher priority are allowed to be executed.
Since we only deal with the execution traces of a distributed system, we assume that the obtained traces are correct with respect to the conflicts and priorities.
Therefore, defining the other modules is out of the scope of this work.
\end{remark}
%
%\begin{definition}[Monitoring hypothesis]
%The behavior of the distributed CBS under scrutiny can be modeled as an LTS as per \defref{def:global_behavior}.
%\end{definition}
%
%%%%%%%%%%%%%%%%%%%%%%%%%%%%%%%%%%%%%%%%%%
\subsection{Traces of a Distributed CBS with Multi-Party Interactions}
\label{subsec:trace}
%%%%%%%%%%%%%%%%%%%%%%%%%%%%%%%%%%%%%%%%%%
%
\begin{definition}[Trace of a CBS]
\label{def:global-trace}
A trace of system $\model$ is a continuously-growing sequence $ ( q_1^0, \ldots , q_{|\setofcomponents|}^0 ) \cdot ( \alpha^0 \cup \beta^0 ) \cdot ( q_1^1, \ldots , q_{|\setofcomponents|}^1 )  \cdots ( q_1^k, \ldots , q_{|\setofcomponents|}^k ) \cdots $, such that $( q_1^0, \ldots , q_{|\setofcomponents|}^0 ) = \init$ is the initial state of the system where $ q_1^0, \ldots , q_{|\setofcomponents|}^0 $ are the initial states of  $ B_1, \ldots , B_{|\setofcomponents|} $ respectively and $\forall i \in [0 \upto k-1] \quant (q_1^i, \ldots , q_{|\setofcomponents|}^i ) \longtrans{}{ \alpha^i \cup \beta^i } (q_1^{i+1}, \ldots , q_{|\setofcomponents|}^{i+1} ) $, where $\trans{}{}$ is the transition relation of the global behavior of the system and the states of schedulers are discarded.
\end{definition}
The set of traces of system $\model$ is denoted by $\Tr(\model)$.
%
%Indeed, each trace $t\in \Tr(\model)$ is a partial trace, since trace $t$ is defined over the partial states of $\model$.
%
Since trace $t\in \Tr(\model)$ has partial states where at least one component is busy with its internal computation, trace $t$ is referred to as a \textit{partial trace}.
\begin{example}[Trace]
\label{exmp:global_trace}
Two possible partial traces of the system in \ppexref{xmp:system} (depicted in~\pfigref{fig:scheduler}) are:\footnote{To facilitate the description of the trace, we represent each busy state as $\bot$.}
\begin{itemize}
	\item $t_1=(d_1,d_2,d_3) \cdot \setof{\fil_{12}} \cdot (\bot, \bot, d_3) \cdot \setof{\beta_1} \cdot (f_1, \bot, d_3) \cdot \setof{\setof{\drain_1}, \setof{\fil_3}} \cdot (\bot, \bot , \bot ) \cdot \setof{\beta_2} \cdot (\bot, f_2 , \bot)$,
	\item $t_2=(d_1,d_2,d_3) \cdot \setof{\fil_{12} ,\setof{\fil_3}} \cdot (\bot, \bot, \bot) \cdot \setof{\beta_3} \cdot (\bot,\bot, f_3) \cdot \setof{\beta_2} \cdot (\bot, f_2 , f_3)\cdot \setof{ \setof{\drain_{23}} ,\beta_1} \cdot (f_1, \bot , \bot )$.
\end{itemize}
Traces $t_1$ and $t_2$ are obtained following the global behavior of the system (\pdefref{def:global_behavior}).
\begin{itemize}
\item
In trace $t_1$, the execution of interaction $\fil_{12}$ represents the simultaneous execution of (i) action $\fil_{12}$ in scheduler $S_1$, (ii) action $\fil_1$ in component $\tank_1$, and (iii) action $\fil_2$ in component $\tank_2$.
After interaction $\fil_{12}$, component $\tank_1$ and $\tank_2$ move to their busy state whereas the state of component $\tank_3$ remains unchanged.
Moreover, the execution of internal action $\beta_2$ in trace $t_1$ represents the simultaneous execution of (i) internal action $\beta_2$ in component $\tank_2$, (ii) action $\beta_2$ in scheduler $S_1$ and (iii) action $\beta_2$ in scheduler $S_2$.
After the internal action $\beta_2$, component $\tank_2$ goes to ready state $f_2$.
\item
In trace $t_2$, the execution of global action $\setof{\fil_{12},\setof{\fil_3}}$ represents the simultaneous execution of two interactions $\fil_{12}$ and  $\setof{\fil_3}$, that is the simultaneous executions of (i) action $\fil_{12}$ in scheduler $S_1$, (ii) action $\fil_3$ in scheduler $S_2$, (iii) action $\fil_1$ in component $\tank_1$, (iv) action $\fil_2$ in component $\tank_2$, and (v) action $\fil_3$ in component $\tank_3$. 
Trace $t_2$ ends up with the simultaneous execution of interaction $\drain_{23}$ and the internal action of component $\tank_1$.
\end{itemize}
\end{example}
\begin{remark}
The operational description of a CBS is usually more detailed.
For instance, the execution of conflicting interactions in schedulers needs first  to be authorized by a conflict-resolution module which guarantees that two conflicting interactions are not executed at the same time. 
Moreover, schedulers follow the (possible) priority rules among the interactions, that is, in the case of two or more enabled interactions (interactions which are ready to be executed by schedulers), those with higher priority are allowed to be executed.
Since we only deal with the execution traces of a distributed system, we assume that the obtained traces are correct with respect to the conflicts and priorities.
Therefore, defining the other modules is out of the scope of this work.
\end{remark}
\begin{definition}[Monitoring hypothesis]
The behavior of the CBS under scrutiny can be modeled as an LTS as per \ppdefref{def:global_behavior}.
\end{definition}

In the following we consider a partial trace $ t = ( q_1^0, \ldots , q_{|\setofcomponents|}^0 ) \cdot ( \alpha^0 \cup \beta^0 ) \cdot ( q_1^1, \ldots , q_{|\setofcomponents|}^1 )  \cdots $, as per \ppdefref{def:global-trace}.
Each scheduler $ S_j \in \setofschedulers $ for $j \in [1 \upto |\setofschedulers|]$, observes a local partial-trace $ s_j (t) $ which consists in the sequence of the states of the components in its scope and actions it manages.
%
%\begin{definition}[The observable local partial-trace of a scheduler obtained from a partial trace]
\begin{definition}[Locally observed partial-trace]
\label{def:local-traces}
The local partial-trace $s_j(t)$ observed by scheduler $ S_j $ is inductively defined on the partial trace $t$ as follows:
\begin{itemize}
	\item 
	$ s_j \left( \left( q_1^0, \ldots , q_{|\setofcomponents|}^0 \right) \right) = \left( q_1^0, \ldots , q_{|\setofcomponents|}^0 \right) $, and
	\item
	$
	s_j \left( t \cdot ( \alpha \cup \beta ) \cdot q \right)
	= \left\{
	\begin{array}{ll}
		t & \hbox{if } S_j \notin \managed( \alpha ) \wedge \left(\involved ( \beta )\ \cap\ \scope ( S_j ) = \emptyset \right) \\
		t \cdot \theta \cdot q' & \hbox{otherwise}
	\end{array}
	\right.
	$\\
	where
	\begin{itemize}
	\item $q = \left( q_1, \ldots , q_{|\setofcomponents|} \right)$,
	\item
	$ \theta =  \left( \alpha \ \cap\ \setof{ a \in \IntActions  \mid \managed(a) = S_j } \right)\ \cup\ \left( \beta \cap \setof{ \beta_i  \mid B_i \in \scope ( S_j ) } \right) $,
	\item
	$ q' = ( q_1', \ldots , q_{|\setofcomponents|}' ) $ with 
	
	$
	q'_i
	= \left\{
	\begin{array}{ll}
		\last(s_j(t))[i] & \hbox{if } B_i \in \overline{\involved ( \theta )} \cap \scope ( S_j ), \\
		q_i & \hbox{if } B_i \in \involved ( \theta ) \cap \scope ( S_j ), \\
		?  & \hbox{otherwise } (B_i \not\in \scope ( S_j )).
	\end{array}
	\right.
	$
	\end{itemize}
\end{itemize}
\end{definition}
We assume that the initial state of the system, that is $\init = \left( q_1^0, \ldots , q_{|\setofcomponents|}^0 \right)$, is observable by all schedulers. 
An interaction $a \in \IntActions$ is observable by scheduler $S_j$ if $S_j$ manages the interaction (\ie $S_j \in \managed(a))$.
Moreover, an internal action $\beta_i$, with $i \in [1 \upto |\setofcomponents|]$, is observable by scheduler $S_j$ if $B_i$ is in the scope of $S_j$ (\ie $B_i \in \scope(S_j)$). 
The state observed after an observable interaction or internal action consists of the states of components in the scope of $S_j$, that is a state $(q_1, \ldots, q_{|\setofcomponents|})$ where $q_i$ is the new state of component $B_i$ if $ B_i \in \scope(S_j) $ and $?$ otherwise.
\begin{example}[Locally observed partial-trace]
\label{exmp:local_trace}
The associated locally observed partial-trace of $t_1$ and $t_2$ of \ppexref{exmp:global_trace} are: 
\begin{itemize}
	\item $s_1(t_1)= (d_1,d_2,d_3) \cdot \setof{\fil_{12}} \cdot (\bot, \bot, ?) \cdot \setof{\beta_1} \cdot (f_1, \bot, ?) \cdot \setof{\setof{\drain_1}} \cdot (\bot, \bot, ?) \cdot \setof{\beta_2} \cdot (\bot, f_2 , ?)$,
	\item $s_2(t_1)= (d_1,d_2,d_3) \cdot \setof{\setof{\fil_3}} \cdot (?, d_2, \bot) \cdot \setof{\beta_2} \cdot (?, f_2 , \bot)$,
	\item $s_1(t_2)= (d_1,d_2,d_3) \cdot \setof{\fil_{12}} \cdot (\bot, \bot, ?) \cdot \setof{\beta_2} \cdot (\bot, f_2, ?) \cdot \setof{\beta_1} \cdot (f_1, f_2, ?)$,
	\item $s_2(t_2)= (d_1,d_2,d_3) \cdot \setof{\setof{\fil_3}} \cdot (?, d_2, \bot) \cdot \setof{\beta_3} \cdot (?, d_2, f_3) \cdot \setof{\beta_2} \cdot (?, f_2, f_3)\cdot \setof{\drain_{23}} \cdot (?, \bot, \bot)$.
\end{itemize}

For instance, the local partial-trace $s_1(t_2)$ shows that scheduler $S_1$ is aware of the execution of interaction $\fil_{12}$ but it is not aware of the occurrence of internal action $\beta_3$ because component $\tank_3$ is not in the scope of scheduler $S_1$ and consequently the state of component $\tank_3$ in the local partial-trace of scheduler $S_1$ is denoted by $?$ (except for the initial state).
Moreover, scheduler $S_1$ is aware of the occurrences of internal actions $\beta_2$ and $\beta_1$ but it is not aware of action $\drain_{23}$ because scheduler $S_1$ does not manage action $\drain_{23}$.
\label{xmp:dmodel}
\end{example}
%

%% file: component_tank.tex
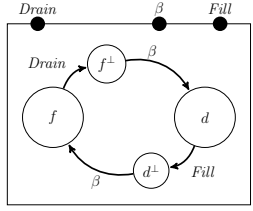
\begin{figure}[t]
    \centering
    \centering
    \Large
   \scalebox{.4} {   
\begin{tikzpicture}[->,>=stealth',shorten >=1pt,auto,node distance=5cm, semithick]
  \tikzstyle{every state}=[fill=none,draw=black,text=black]  
%%%%%%%%%%%%%%%%%%%%%%%%%%%%%%%%%%%
\filldraw[fill=none,draw=black](0,0.5)--(0,6.5)--(8,6.5)--(8,0.5)--(0,0.5)--cycle; 
\node at (1.5,3.4) [state] (A)        [minimum size=2cm]              {$f$};
\node            [state] (B)  [right of =A,minimum size=2cm]          {$d$};
\node [state] (A') [above right of =A,minimum size=0.5cm,node distance=2.5cm]    {$f^\perp$};
\node [state] (B') [below left of =B,minimum size=0.5cm,node distance=2.5cm]    {$d^\perp$};

\path [out=10,in=120,ultra thick](A') edge  [above left ] node[text centered] {$\beta$} (B);
\path [out=190,in=-60,ultra thick](B') edge  [below]  node[text centered] {$\beta$} (A);  
\path [out=-110,in=10,ultra thick] (B) edge  [below right]   node[text centered] { $\fil$} (B');
\path [out=70,in=190,ultra thick](A) edge  [above left]   node[text centered] { $\drain$} (A');  

\fill (1,6.5) circle (0.24cm);\node at (1,7) {$\drain$};
\fill (7,6.5) circle (0.24cm);\node at (7,7){$\fil$};
\fill (5,6.5) circle (0.24cm);\node at (5,7){$\beta$};  
%%%%%%%%%%%%%%%%%%%%%%%%%%%%%%%%%%%

\end{tikzpicture}}
\caption{Component $\tank$} 
\label{fig:component}   
    \end{figure} 

%% file: scheduler.tex
\pgfdeclarelayer{background}
\pgfdeclarelayer{foreground}
\pgfsetlayers{background,main,foreground}

    \tikzset{
        hatch distance/.store in=\hatchdistance,
        hatch distance=10pt,
        hatch thickness/.store in=\hatchthickness,
        hatch thickness=2pt
    }
\begin{figure}[t]    
\centering
  % \scalebox{0.77} {  
  \resizebox{\textwidth}{!}{ 
   %\hspace{-2.3cm}
\begin{tikzpicture}[->,>=stealth',shorten >=1pt,auto,node distance=5cm, semithick]
  \tikzstyle{every state}=[fill=none,draw=black,text=black]  
%%%%%%%%%%%%%%%%%%%%%%%%%%%%%%%%%%%
\draw[-] (-3,3.2)--(-3,3.9);\node  at (-3,4.2){\textbf{$\drain_1$}};
\draw[-] (-1.8,3.2)--(-1.8,3.9);\node  at (-1.8,4.2){\textbf{$\beta_1$}};
\draw[-] (0.6,3.2)--(0.6,3.9);\node  at (0.6,4.2){\textbf{$\beta_2$}};
\draw[-] (3.8,3.2)--(3.8,3.9);\node  at (3.8,4.2){\textbf{$\beta_3$}};
\draw[-] (-0.2,3.2)--(-0.2,3.9)--(-1.1,3.9)--(-1.1,3.2); \node at (-0.6,4.2)   {\textbf{$\fil_{12}$}};
\draw[-] (1.4,3.2)--(1.4,3.9)--(2.6,3.9)--(2.6,3.2); \node at (2,4.2)   {\textbf{$\drain_{23}$}};
\draw[-] [xshift=6cm](-1.5,3.9)--(-1.5,3.2);\node [xshift=6cm]  at (-1.5,4.2)   {\textbf{$\fil_3$}};

%%%%%%%%%%%%%%%%%%%%%%%%%%%%%%%%%%%
\filldraw[fill=none,draw=black](-0.5,2)--(-0.5,3.2)--(2,3.2)--(2,2)--(-0.5,2)--cycle;  
\node at (0.75,2.4)   {\textbf{Tank 2}};
\fill (1.4,3.2) circle (0.1cm);\node at (1.4,2.8)   {{\small $\drain_2$}};
\fill (-0.2,3.2) circle (0.1cm);\node at (-0.1,2.8)  {{\small $\fil_2$}};
\fill (0.6,3.2) circle (0.1cm);\node  at (0.6,2.8)   {{\small $\beta_2$}};
%%%%%%%%%%%%%%%%%%%%%%%%%%%%%%%%%%%
  \filldraw[fill=none,draw=black,xshift=-2.8cm](-0.5,2)--(-0.5,3.2)--(2,3.2)--(2,2)--(-0.5,2)--cycle;  
  \node [xshift=-2.8cm] at (0.75,2.4)   {\textbf{Tank 1}}; 
  \fill [xshift=-2.4cm](1.3,3.2) circle (0.1cm);\node [xshift=-2.5cm] at (1.3,2.8)   {{\small $\fil_1$}};
  \fill [xshift=-3.2cm](0.2,3.2) circle (0.1cm);\node [xshift=-2.9cm] at (0.2,2.8)  {{\small $\drain_1$}};
  \fill [xshift=-3.1cm](1.3,3.2) circle (0.1cm);\node [xshift=-3.1cm] at (1.3,2.8)   {{\small $\beta_1$}};
%%%%%%%%%%%%%%%%%%%%%%%%%%%%%%%%%%%
  \filldraw[fill=none,draw=black,xshift=2.8cm](-0.5,2)--(-0.5,3.2)--(2,3.2)--(2,2)--(-0.5,2)--cycle;  
  \node [xshift=2.8cm] at (0.75,2.4)   {\textbf{Tank 3}};
  \fill [xshift=2.8cm] (1.7,3.2) circle (0.1cm);\node [xshift=2.8cm] at (1.6,2.8)   {{\small $\fil_3$}};
  \fill [xshift=2.4cm] (0.2,3.2) circle (0.1cm);\node [xshift=2.8cm] at (0.1,2.8)  {{\small $\drain_3$}};
  \fill [xshift=2.5cm](1.3,3.2) circle (0.1cm);\node [xshift=2.5cm] at (1.3,2.8)   {{\small $\beta_3$}};
%%%%%%%%%%%%%%%%%%%%%%%%%%%%%%%%%%%
 \begin{pgfonlayer}{background}
   \draw[-,rounded corners=2em,line width=4em,blue!30,cap=round]
    (4.5,4.1)--(0.7,4.1);
   \draw[-,rounded corners=2em,line width=4em,yellow!50,cap=round]
    (-3.2,4.1)--(0.65,4.1);
    %\draw[-,rounded corners=2em,line width=4em,green!50,cap=round]
    %(0.6,4.1)--(0.65,4.1);
    %\draw[draw=none,preaction={fill=yellow!50}, pattern=horizontal lines, pattern color=blue!30] (0.6,4.1) circle (0.7cm);
    \draw[draw=none,right color=yellow!50,left color=blue!30] (0.6,4.1) circle (0.632cm);
\end{pgfonlayer}
%%%%%%%%%%%%%%%%%%%%%%%%%%%%%%%%%%%
  \filldraw[fill=yellow!50,draw=black,dashed](-7.5,5)--(-7.5,8)--(0.4,8)--(0.4,5)--(-7.5,5)--cycle;
\scalebox{0.8} {  
%  \node[state] (A) at (-7,9) {$l_1$};
%  \node[state] (B) [right of=A,node distance=5cm] {$l_2$}; 
%  \node[state] (C) [below of=B,node distance=2cm] {$l_3$};
%  \node[state] (D) [below of=A,node distance=2cm] {$l_4$};
%  
%  \path (A) edge  node {$\{\fil_{12}\}$} (B); 
%  \path (B) edge  node {$\{\beta_1\},\{\beta_1,\beta_2\}$} (C);
%  \path (C) edge  node {$\{\drain_1\},\{\drain_1,\beta_2\}$} (D);
%  \path (D) edge  node {$\{\beta_1\},\{\beta_1,\beta_2\}$} (A); 
%  \path (A) edge [loop left] node [left]{$\{\beta_2\}$} ();   
%  \path (B) edge [loop right] node [right]{$\{\beta_2\}$} ();
%  \path (C) edge [loop right] node [right]{$\{\beta_2\}$} ();   
%  \path (D) edge [loop left] node [left]{$\{\beta_2\}$} ();

  \node[state] (A) at (-7.68,8) {$l_1$};
  \node[state] (B) [right of=A,node distance=2.5cm] {$l_0$}; 
  \node[state] (C) [right of=B,node distance=2.5cm] {$l_2$};
  \node[state] (D) [right of=C,node distance=2.5cm] {$l_3$};
  
  \path (B) edge  node {$\fil_{12}$} (C);
  \path (C) edge  node {$\beta_1$} (D);
  \path (D) edge [bend left] node {$\drain_1$} (C);
  \path [out=145,in=35] (D) edge  node [above] {$\beta_2$} (B);
  \path [out=50,in=130](A) edge  node [above] {$\beta_1$} (B);
  \path (C) edge [bend left] node [below] {$\beta_2$} (A);
  \path (B) edge  node [above] {$\drain_1$} (A);
  \path [out=-150,in=150] (A) edge [loop] node [right]{$\beta_2$} ();
  \path [out=110,in=60] (B) edge [loop] node [above]{$\beta_2$} ();
  }
%%%%%%%%%%%%%%%%%%%%%%%%%%%%%%%%%%%
  \filldraw[fill=blue!30,draw=black,dashed,xshift=4.3cm](-3.5,5)--(-3.5,8)--(4.5,8)--(4.5,5)--(-3.5,5)--cycle;
\scalebox{0.8} {  
%  \node[state,xshift=5.4cm] (A) at (-2,9) {$l_1$};
%  \node[state] (B) [right of=A,node distance=5cm] {$l_2$}; 
%  \node[state] (C) [below of=B,node distance=2cm] {$l_3$};
%  \node[state] (D) [below of=A,node distance=2cm] {$l_4$};
%  
%  \path (A) edge  node {$\{\fil_3\},\{\fil_3,\beta_2\}$} (B); 
%  \path (B) edge  node {$\{\beta_3\},\{\beta_2,\beta_3\}$} (C);
%  \path (C) edge  node {$\{\drain_{23}\}$} (D);
%  \path (D) edge  node {$\{\beta_3\},\{\beta_2,\beta_3\}$} (A); 
%  \path (A) edge [loop left] node [left]{$\{\beta_2\}$} ();   
%  \path (B) edge [loop right] node [right]{$\{\beta_2\}$} ();
%  \path (C) edge [loop right] node [right]{$\{\beta_2\}$} ();   
%  \path (D) edge [loop left] node [left]{$\{\beta_2\}$} (); 

  \node[state,xshift=9.6cm] (A) at (-7.8,8) {$l_3$};
  \node[state] (B) [right of=A,node distance=2.5cm] {$l_2$}; 
  \node[state] (C) [right of=B,node distance=2.5cm] {$l_0$};
  \node[state] (D) [right of=C,node distance=2.5cm] {$l_1$};
  
  \path (C) edge  node [above]{$\drain_{23}$} (B);
  \path (B) edge  node [above] {$\beta_3$} (A);
  \path (A) edge [bend right] node [below]{$\fil_3$} (B);
  \path [out=35,in=145] (A) edge  node [above] {$\beta_2$} (C);
  \path [out=130,in=50](D) edge  node [above] {$\beta_3$} (C);
  \path (B) edge [bend right] node [below] {$\beta_2$} (D);
  \path (C) edge  node [above] {$\fil_3$} (D);
  \path [out=30,in=-30] (D) edge [loop] node [left]{$\beta_2$} ();
  \path [out=70,in=122] (C) edge [loop] node [above]{$\beta_2$} ();
   
  }
  \node  at (-3.5,8.2){Scheduler $S_1$};
  \node  at (4.7,8.2){Scheduler $S_2$};
%%%%%%%%%%%%%%%%%%%%%%%%%%%%%%%%%%%
\end{tikzpicture}}    
\caption{Abstract representation of a distributed CBS} 
\label{fig:scheduler}          
\end{figure}
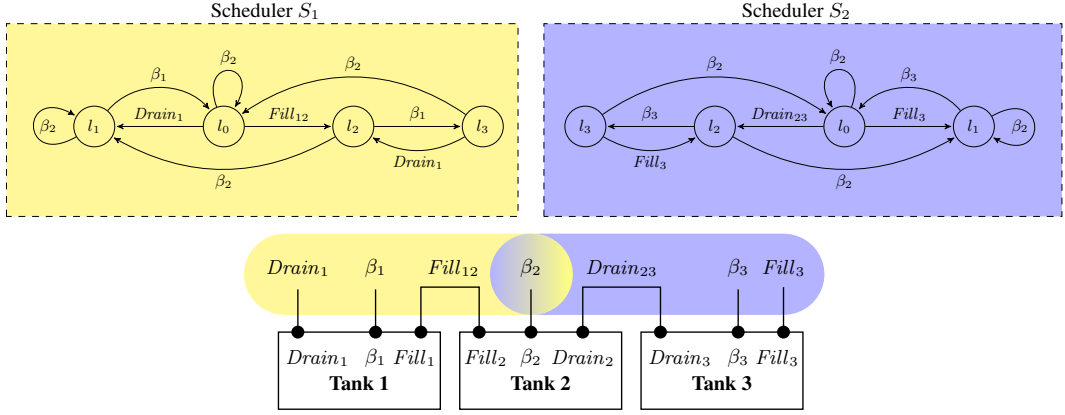

%% file: globaltrace.tex
%
%%%%%%%%%%%%%%%%%%%%%%%%%%%%%%%%%%%%%%%%%
\section{From Local Traces to Global Traces}
\label{sec:local-global-trace}
%%%%%%%%%%%%%%%%%%%%%%%%%%%%%%%%%%%%%%%%%
%
\input{observer}
We define a new component as a \textit{passive} observer which runs in parallel with the system and collects  local traces of schedulers and reconstruct the set of possible global traces compatible with the local traces (\figref{fig:passiveobserver}). 
The observer is always ready to receive information from schedulers. 
We use the term \textit{passive} for the observer since it does not force schedulers to send data and thus does not modify the execution of the monitored system.
We shall prove that such observer does not violate the semantics nor the behavior of the distributed system, that is, the observed system is observationally equivalent (see~\secref{sec:prelim}) to the initial system (\cf \propref{proposition:sim}).

For monitoring purposes, the observer should be able to order the execution of interactions from the received local traces appropriately. 
In our abstract model, since schedulers do not interact directly together by sending/receiving messages, the execution of an interaction by one scheduler seems to be concurrent with the execution of all interactions by other schedulers.   
Nevertheless, if scheduler $S_j$ manages interaction $a$ and scheduler $S_k$ manages interaction $b$ such that a shared component $B_i \in \setofsharedcomponents$  is involved in $a$ and $b$, \ie  $B_i \in \involved(a) \cap \involved(b)$, as a matter of fact, the execution of interactions $a$ and $b$ are causally related.
In other words, there exists only one possible ordering of $a$ and $b$ and they could not have been executed concurrently.
Ignoring the actual ordering of $a$ and $b$ would result in retrieving inconsistent global states (\ie states that does not belong to the original system).
We instrument the system by adding controllers to the schedulers and to the shared components.
The controllers of schedulers and the controllers of shared components interact whenever the scheduler and the shared components interact to transmit vector clocks and state update.
Each time a scheduler executes an interaction, the associated controller attaches a vector clock to this execution and notifies the observer.
Hence, the local trace of each scheduler is augmented by vector clocks and is then sent to the observer.

In the following, we define an instrumentation of abstract distributed systems to let schedulers send their local traces to an observer.
%
%%%%%%%%%%%%%%%%%%%%%%%%%%%%%%%%%%%%%%%%%
\subsection{Composing Schedulers and Shared Components with Controllers}
\label{sec:instrumentation}
%%%%%%%%%%%%%%%%%%%%%%%%%%%%%%%%%%%%%%%%%
%
We consider a distributed system consisting of a set of components $\setofcomponents = \setof { B_1 , \ldots , B_{|\setofcomponents|}} $ (as per \defref{def:component}) and a set of schedulers $\setofschedulers = \setof { S_1 , \ldots , S_{|\setofschedulers|} }$ where scheduler $S_j = ( Q_{S_j}, \Actions_{S_j}, \trans{s_j}{})$ manages the interactions in $ \Actions^\gamma_{S_j} $ and is notified by internal actions in $\Actions^\beta_{S_j}$, for $ S_j $ (as per \defref{def:scheduler}).
We attach to $S_j$ a local controller $\mathcal{C}^{\rm s}_j$ in charge of computing the vector clock and sending the local trace of $S_j$ to the observer.
Moreover, for each shared component $B_i \in \setofschedulers$, we attach a local controller $\mathcal{C}^{\rm b}_i$ to communicate with the controllers of the schedulers that have $B_i$ in their scope.

In the following, we define the controllers (instrumentation code) and the composition $\otimes$ as instrumentation process.
%
%%%%%%%%%%%%%%%%%%%%%%%%%%%%%%%%%%%%%%%%%
\subsubsection{Controllers of Schedulers}
\label{subsubsec:Controllers of Schedulers}
%%%%%%%%%%%%%%%%%%%%%%%%%%%%%%%%%%%%%%%%%
%
Controller $\mathcal{C}^{\rm s}_j$ is in charge of computing the correct vector clock of scheduler $S_j$ (\defref{def:controller_scheduler}).
It does so through the data exchange with the controllers of shared components, i.e., the controllers in the set
\[
\setof{\mathcal{C}^{\rm b}_i \lmid S_j \in \setofschedulers \wedge B_i \in \scope(S_j) },
\]
which are later defined in \defref{def:controller_shared}.
\begin{definition}[Controller of scheduler]
\label{def:controller_scheduler}
Controller $\mathcal{C}^{\rm s}_j$ is an LTS $( Q_{\mathcal{C}^{\rm s}_j}, \mathcal{R}_{\mathcal{C}^{\rm s}_j} , \trans{C^{\rm s}_j}{})$ such that:
\begin{itemize}
	\item $Q_{\mathcal{C}^{\rm s}_j}=2^{[1,{|\setofcomponents|}]} \times \VC$ is the set of states where $2^{[1,n]}$ is the set of subsets of component indexes and  $\VC$ is the set of vector clocks;	
	\item $\mathcal{R}_{\mathcal{C}^{\rm s}_j}=	 \setof{\left(\widehat{\beta_i},\emptyset\right) \lmid B_i \in \scope(S_j)} \bigcup \setof{\left(-,\setof{\rcv_i\lfloor\vc\rfloor}\right) \lmid B_i \in \scope(S_j) \cap {\setofsharedcomponents} \wedge \vc \in \VC } \\	
	 \bigcup \setof{\left(\widehat{a\lfloor\vc\rfloor}, \send \right) \lmid a\in \Actions^\gamma_{S_j} \wedge \vc \in \VC \wedge \send \subseteq \setof { \send_i\lfloor\vc\rfloor \lmid B_i \in \scope(S_j) \cap {\setofsharedcomponents} \wedge \vc \in \VC}} $\\ is the set of actions;
	\item $\trans{C^{\rm s}_j}{} \subseteq Q_{\mathcal{C}^{\rm s}_j} \times \mathcal{R}_{\mathcal{C}^{\rm s}_j} \times Q_{\mathcal{C}^{\rm s}_j}$ is the transition relation defined as:
\[
\begin{array}{l}
	 \setof{(\mathcal{I},\vc)\longtrans{C^{\rm s}_j}{\left(\widehat{a\lfloor\vc'\rfloor},\setof{\send_i\lfloor\vc'\rfloor \lmid i \in \involved(a) \wedge B_i \in {\setofsharedcomponents}}\right)}(\mathcal{I} \cup \involved(a),\vc') \lmid  a \in \Actions^\gamma_{S_j} \wedge 
    \vc' =\inc(\vc,j) }\\
	\bigcup \setof{(\mathcal{I},\vc) \longtrans{C^{\rm s}_j}{\left(\widehat{\beta_i},\emptyset\right)}(\mathcal{I} \setminus \setof{i},\vc) \lmid  \beta_i \in \Actions^\beta_{S_j}}\\
	\bigcup \setof{(\mathcal{I},\vc) \longtrans{C^{\rm s}_j}{\left(-,\setof{\rcv_i\lfloor\vc'\rfloor}\right)}(\mathcal{I},\max (\vc, \vc') \lmid \beta_i \in \Actions^\beta_{S_j} \wedge B_i \in {\setofsharedcomponents} }
\end{array}
\]
where $\inc(\vc,j)$ increments the $j^{\rm th}$ element of  vector clock $\vc$.
\end{itemize}
\end{definition}
When the controller $\mathcal{C}^{\rm s}_j$ is in state $(\mathcal{I},\vc)$, it means that (i) $\mathcal{I}$ is the set of busy components in the scope of scheduler $S_j$, (ii) the execution of their latest action has been managed by scheduler $S_j$, and (iii) $\vc$ is the current value of the vector clock of scheduler $S_j$.

An action in $\mathcal{R}_{\mathcal{C}^{\rm s}_j}$ is a pair $(x,y)$ where $x$ is associated to the actions  which send information from the controller to the observer and $y$ is associated to the actions in which the controller sends/receives information to/from the controllers of shared components, such that\\
$x \in  \setof {\widehat{a\lfloor\vc\rfloor} \lmid  a\in \Actions^\gamma_{S_j} \wedge \vc \in \VC} \cup \setof {\widehat{\beta_i} \lmid i \in \scope(j)} \cup \setof{-} $, and \\
$y \subseteq  \setof{ \send_i \lfloor \vc \rfloor, \rcv_i \lfloor \vc \rfloor \lmid  B_i \in \scope(S_j) \cap {\setofsharedcomponents} \wedge \vc \in \VC }$ can be intuitively understood as follows,
	\begin{itemize}
		\item action $\widehat{a\lfloor\vc\rfloor}$ consists in notifying the observer about the execution of interaction $a$ with vector clock $\vc$ attached.
		\item action $\widehat{\beta_i}$ consists in notifying the observer about the internal action of component $B_i$. The last state of component $B_i$ is also transmitted to the observer.
		\item action $-$ is used in the case when the controller does not interact with the observer,
		\item action $\send_i\lfloor\vc\rfloor$ consists in sending the value of the vector clock $\vc$ of the scheduler to the shared component $B_i$,
		\item action $\rcv_i\lfloor\vc\rfloor$ consists in receiving the value of the vector clock $\vc$ stored in the shared component $B_i$.
	\end{itemize} 
The set of transitions is obtained as the union of three sets which can be intuitively understood as follows:
	\begin{itemize}
		\item For each interaction $a\in \Actions^\gamma_{S_j}$ managed by scheduler $S_j$, we include a transition with action 
		\[
		\left(\widehat{a\lfloor\vc'\rfloor},\setof{\send_i\lfloor\vc'\rfloor \lmid  B_i \in \involved(a) \cap {\setofsharedcomponents}}\right),
		\]
		where $\widehat{a\lfloor\vc'\rfloor}$ is a notification to the observer about the execution of interaction $a$ along with the value of vector clock $\vc'$, and actions in set $\setof{\send_i\lfloor\vc'\rfloor \lmid  B_i \in \involved(a) \cap {\setofsharedcomponents}}$ send the value of the vector clock $\vc'$ to the shared components involved in interaction $a$.
Moreover, the set of indexes of the components involved in interaction $a$ (\ie in $\involved(a)$) is added to the set of busy components; and the current value of the vector clock is incremented.
		\item
		For each action associated to the notification of the internal action of component $B_i$ (that is, $\beta_i$), we include a transition labeled with action $( \widehat{\beta_i}, \emptyset )$ in the controller to send the updated state to the observer. 
	Moreover, this transition removes index $i$ from the set of busy components.
		\item
	For each action associated to the notification of internal action of a shared components $ B_i \in \setofsharedcomponents $, we include a transition labeled with action $\left( - , \setof{ \rcv_i \lfloor \vc' \rfloor } \right)$ in the controller to receive the value of the vector clock $\vc'$ stored in the shared component to update the vector clock of the scheduler by comparing the vector clock stored in the scheduler and the received vector clock from the shared component.
	\end{itemize}
Note that, to each shared component $B_i \in {\setofsharedcomponents}$, we also attach a local controller in order to exchange the vector clock among schedulers in the set $\setof{S_j \in \setofschedulers \lmid B_i \in \scope(S_j)}$; see \defref{def:controller_shared}. 

Below, we define how a scheduler is composed with its controller.
Intuitively, the controller of a scheduler ensures sending/receiving information among the scheduler, associated shared components and the observer. 
\begin{figure}[t]
\centering
\framebox{
	\begin{mathpar}
	\inferrule*[left=Cont-Sch1]
		{
		   a\in \IntActions  \and  q_s \longtrans{S_j}{a} q'_s \and  q_c \longtrans{C^{\rm s}_j}{\left( \widehat{a\lfloor\vc'\rfloor},\setof{\send_i\lfloor\vc'\rfloor \lmid  i \in \involved(a)\wedge B_i \in {\setofsharedcomponents}} \right)} q'_c
		}
		{
		   (q_s, q_c) \longtrans{sc_j}{\left(a, \left( \widehat{a\lfloor\vc'\rfloor},\setof{\send_i\lfloor\vc'\rfloor \lmid  i \in \involved(a)\wedge B_i \in {\setofsharedcomponents}} \right)\right)  } (q'_s, q'_c) 
		}
		
		\inferrule*[left=Cont-Sch2]
		{
		   i \in \mathcal{I} \and  q_s \longtrans{S_j}{\beta_i} q'_s  \and q_c \longtrans{C^{\rm s}_j}{\left(\widehat{\beta_i},\emptyset \right)} q'_c 
		}
		{
		   (q_s, q_c) \longtrans{sc_j}{\left(\beta_i,\left(\widehat{\beta_i},\emptyset \right)\right)} (q'_s, q'_c)
		}
		
		\inferrule*[left=Cont-Sch3]
		{
		   i \in \mathcal{I} \and  q_s \longtrans{S_j}{\beta_i} q'_s  \and q_c \longtrans{C^{\rm s}_j}{\left(\widehat{\beta_i},\emptyset \right)} q'_c 
		}
		{
		   (q_s, q_c) \longtrans{sc_j}{\left(\beta_i,\left(\widehat{\beta_i},\emptyset \right)\right)} (q'_s, q'_c)
		}
	\end{mathpar}
	}
	\caption{Semantics rules defining the composition controller / scheduler}
	\label{fig:semantics-controller-scheduler}
\end{figure}
\begin{definition}[Semantics of $S_j \otimes_{\rm s} \mathcal{C}^{\rm s}_j$]
\label{def:semantics-controller-scheduler}
The composition of scheduler $S_j$ and controller $\mathcal{C}^{\rm s}_j$, denoted by $S_j \otimes_{\rm s} \mathcal{C}^{\rm s}_j$, is the LTS $(Q_{S_j} \times Q_{\mathcal{C}^{\rm s}_j}, \Actions_{S_j} \times \mathcal{R}_{\mathcal{C}^{\rm s}_j}, \trans{sc_j}{} )$ where the transition relation $\trans{sc_j}{} \subseteq (Q_{S_j}  \times Q_{\mathcal{C}^{\rm s}_j}) \times (\Actions_{S_j} \times \mathcal{R}_{\mathcal{C}^{\rm s}_j}) \times (Q_{S_j} \times  Q_{\mathcal{C}^{\rm s}_j})$ is defined by the semantics rules in \figref{fig:semantics-controller-scheduler}.
\end{definition}
The semantics rules in \figref{fig:semantics-controller-scheduler} can be intuitively be understood as follows:
\begin{itemize}
	\item
	\textit{Rule \textsc{Cont-Sch1}.} When the scheduler executes an interaction $a\in \IntActions$, the controller (i) updates the vector clock by increasing its local clock, (ii) updates the set of busy components, (iii) notifies the observer  of the execution of $a$ along with the associated vector clock $\vc'$, and (iv) sends vector clock $\vc'$ to the shared components involved in $a$.
	\item
	\textit{Rule \textsc{Cont-Sch2}.} When the scheduler is notified of an internal action of component $B_i$ where $i \in \mathcal{I}$ (that is, the scheduler has managed the latest action of component $B_i$) through action $\beta_i$, the controller transfers the updated state of component $B_i$ to the observer through action $\widehat{\beta_i}$. 
	\item
	\textit{Rule \textsc{Cont-Sch3}.} When the scheduler is notified of an internal action of the shared component $B_i$ where $i \not \in \mathcal{I}$ (that is, the scheduler has not managed the latest action of component $B_i$), the controller receives the vector clock stored in component $B_i$ and updates the vector clock. 
\end{itemize}
\input{scheduler_controller_new}
\begin{example}[Controller of scheduler]
\figref{fig:scheduler_contoller} depicts the controller of scheduler $S_1$ (depicted in \figref{fig:scheduler}).
Actions $(\widehat{\beta_1},\emptyset)$ and $(\widehat{\beta_2},\emptyset)$ consist in sending the updated state to the observer. 
Actions $(\widehat{\drain_1},\emptyset)$ and $\widehat{(\fil_{12}},\{\send_2\})$ consist in notifying the observer about the occurrence of interactions managed by the scheduler. 
Moreover, $\send_2$ sends the vector clock to the shared component $\tank_2$. The controller receives the vector clock stored in the shared component $\tank_2$ through action $(-,\{\rcv_2\})$ and updates its vector clock.
For the sake of simplicity, variables attached to the  transition labels are not shown.
\end{example}
%
%
%%%%%%%%%%%%%%%%%%%%%%%%%%%%%%%%%%%%%%%%%
\subsubsection{Controllers of Shared Components}
\label{subsubsec:Controllers of Shared Components}
%%%%%%%%%%%%%%%%%%%%%%%%%%%%%%%%%%%%%%%%%
%
Below, we define the controllers attached to shared components.
Intuitively, the controller of a shared component ensures data exchange among the shared component and the corresponding schedulers. 
A scheduler sets it's current clock in a shared component's controller which can be used later by another scheduler.
\begin{definition}[Controller of shared component]
\label{def:controller_shared}
Local controller $\mathcal{C}^{\rm b}_i$ for a shared component $B_i \in {\setofsharedcomponents}$ with the behavior $( Q_i, \Actions_i \cup \setof{ \beta_i }, \trans{i}{} )$ is the LTS $( Q_{c^{\rm b}_i},\mathcal{R}_{c^{\rm b}_i}, \trans{C^{\rm b}_i}{})$, where
\begin{itemize}
	\item $Q_{c^{\rm b}_i} = \VC$ is the set of states,
	\item $\mathcal{R}_{c^{\rm b}_i}\subseteq\setof{\send_j\lfloor\vc\rfloor, \rcv_j\lfloor\vc\rfloor  \lmid S_j\in \setofschedulers \wedge B_i \in \scope(S_j) \wedge \vc \in \VC}$ is the set of actions, 
	\item $\trans{C^{\rm b}_i}{} \subseteq Q_{c^{\rm b}_i} \times \mathcal{R}_{c^{\rm b}_i} \times Q_{c^{\rm b}_i}$ is the transition relation defined as
	\[
	\begin{array}{l}
	\setof{\vc \longtrans{C^{\rm b}_i}{\setof{\rcv_j\lfloor\vc'\rfloor}} \max (\vc, \vc')
\mid a \in \IntActions \wedge a \cap \Actions_i \neq \emptyset  \wedge \managed(a)=S_j }
\\ \bigcup 
 \setof{\vc \longtrans{C^{\rm b}_i}{\setof{\send_j\lfloor\vc\rfloor }} \vc \lmid  S_j \in \setofschedulers \wedge B_i \in \scope(S_j) }.
	\end{array}
	\]
\end{itemize}
\end{definition}
The state of the controller $\mathcal{C}^{\rm b}_i$ is represented by its vector clock. 
Controller $\mathcal{C}^{\rm b}_i$ has two types of actions:
\begin{itemize}
	\item action $\rcv_j\lfloor\vc'\rfloor$ consists in receiving the vector clock $\vc'$ of scheduler $S_j$,
	\item action $\send_j\lfloor\vc\rfloor$ consists in sending the vector clock $\vc$ stored in the controller $\mathcal{C}^{\rm b}_i$ to scheduler $S_j$.
\end{itemize}

The two types of transitions can be understood as follow:
\begin{itemize}
	\item For each action of component $B_i$, which  is managed by scheduler $S_j$, we include a transition executing action $\rcv_j\lfloor\vc'\rfloor$ to receive the vector clock $\vc'$ of scheduler $S_j$ and to update the vector clock stored in controller $\mathcal{C}^{\rm b}_i$.
	\item We include a transition with a set of actions for all the schedulers that have component $B_i$ in their scope, that is $\setof{S_j \in \setofschedulers \lmid B_i \in \scope(S_j) }$, to send the stored vector clock of controller $\mathcal{C}^{\rm b}_i$ to the controllers of the corresponding schedulers, that is $\setof{\mathcal{C}^{\rm s}_j \lmid S_j \in \setofschedulers \wedge B_i \in \scope(S_j) }$. 
\end{itemize}
\begin{definition}[Semantics of $B_i \otimes_{\rm b} \mathcal{C}^{\rm b}_i$]
\label{def:semantics-controller-shared-component}
The composition of shared component $B_i$ and controller $\mathcal{C}^{\rm b}_i$, denoted by $B_i \otimes_{\rm b} \mathcal{C}^{\rm b}_i$, is the LTS $(Q_i  \times Q_{c^{\rm b}_i}, \left(\Actions_i \cup \setof{\beta_i} \right)\times \mathcal{R}_{c^{\rm b}_i}, \trans{bc_i}{} )$ where the transition relation $\trans{bc_i}{} \subseteq (Q_i  \times Q_{c^{\rm b}_i}) \times \left(\left(\Actions_i \cup \setof{\beta_i} \right)\times \mathcal{R}_{c^{\rm b}_i}\right)  \times (Q_i  \times Q_{c^{\rm b}_i})$ is defined by the semantics rules in \figref{fig:semantics-controller-shared-component}.

\begin{figure}[t]
	\centering
	\framebox{
	\begin{mathpar}
	\inferrule*[left=Cont-Sha1]
		{
		     a \in \IntActions  \and a \cap \Actions_i= \setof{a'}  \and \managed(a)=S_j \and q_b \longtrans{i}{a'} q'_b \and q_c \longtrans{C^{\rm b}_i}{\setof{\rcv_j\lfloor\vc'\rfloor}} q'_c
		}
		{
		   (q_b,q_c) \longtrans{bc_i}{\left(a',\setof{\rcv_j\lfloor\vc'\rfloor}\right)} (q'_b,q'_c)
		}
		
	\inferrule*[Left=Cont-Sha2]
		{
		   q_b \longtrans{i}{\beta_i} q'_b \and J=\setof{j\in[1,m] \lmid B_i \in \scope(S_j)}  \and q_c \longtrans{C^{\rm b}_i}{\setof{\send_j\lfloor\vc\rfloor \lmid  j \in J}}q'_c
		}
		{
		   (q_b,q_c) \longtrans{bc_i}{\left(\beta_i,{\setof{\send_j\lfloor\vc\rfloor \lmid j \in J}}\right)} (q'_b,q'_c)
		}

	\end{mathpar}
	}
	\caption{Semantics rules defining the composition controller / shared component}
	\label{fig:semantics-controller-shared-component}
\end{figure}
\end{definition}
The semantics rules in \figref{fig:semantics-controller-shared-component} can be intuitively understood as follows:
\begin{itemize}
	\item
	\textit{Rule \textsc{Cont-Sha1}.} applies when the scheduler notifies the shared component to execute an action part of an interaction.	
	  Controller $\mathcal{C}^{\rm b}_i$ receives the value of the vector clock of scheduler $S_j$ from the associated controller $\mathcal{C}^{\rm s}_j$ in order to update the value of the vector clock stored in controller $\mathcal{C}^{\rm b}_i$.
	\item
	\textit{Rule \textsc{Cont-Sha2}.} applies when the shared component $B_i$ finishes its computation by executing $\beta_i$, and controller $\mathcal{C}^{\rm b}_i$ notifies the controllers of the schedulers that have component $B_i$ in their scope, through actions $\send_j$, for $ j \in J$,  which sends the vector clock stored in controller $\mathcal{C}^{\rm b}_i$ to controllers $\mathcal{C}^{\rm s}_j$ with $ j \in J$, where $J$ is the set of indexes of schedulers which have the shared component $B_i$ in their scope.  
\end{itemize}
\input{controller_tank2}
\begin{example}[Controller of shared component]
\figref{fig:controller_tank2} depicts the controller of the shared component $\tank_2$ (depicted in \figref{fig:scheduler}).
Action $ \rcv_1$ (\resp $\rcv_2$) consists in the reception and storage of the vector clock from scheduler $S_1$ (\resp $S_2$ ) upon the execution of interaction $\fil_{12}$ (\resp $\drain_{23}$). 
Action $ \setof{\send_1,\send_2}$ sends the stored vector clock to the schedulers $S_1$ and $S_2$ when the component $\tank_2$ performs its internal action $\beta_2$.
\end{example}
%
%
%
%%%%%%%%%%%%%%%%%%%%%%%%%%%%%%%%%%%%%%%%%%%%%%%%%%%%%%%%%%%%%%%%%%%
\subsection{Correctness of Instrumentation}
%%%%%%%%%%%%%%%%%%%%%%%%%%%%%%%%%%%%%%%%%%%%%%%%%%%%%%%%%%%%%%%%%%%
%
Following the instrumentation defined in~\ppsecref{sec:instrumentation}, one can obtain a transformed system whose execution at runtime generates the associated events.
Furthermore, we shall prove that such an instrumentation does not modify the initial behavior of the original system.
\begin{definition}[Instrumented system]
\label{def:instrumented_system}
For a CBS consisting of a set of components $\setofcomponents = \setof { B_1 , \ldots , B_{|\setofcomponents|}} $ where $B_i= ( Q_i, \Actions_i \cup \setof{ \beta_i }, \trans{i}{} )$ (as per \pdefref{def:component}), a set of schedulers $\setofschedulers = \setof { S_1 , \ldots , S_{|\setofschedulers|} }$ where $S_j = ( Q_{S_j}, \Actions_{S_j}, \trans{s_j}{})$ (as per \pdefref{def:scheduler}), and with the global behavior $( Q , \GActions , \trans{}{}) $ (as per \pdefref{def:global_behavior}), the instrumented CBS is the set of components  $\{S_1 \otimes_{\rm s} \mathcal{C}^{\rm s}_1,\ldots,S_{|\setofschedulers|} \otimes_{\rm s} \mathcal{C}^{\rm s}_{|\setofschedulers|},$ $B'_1,\ldots,B'_{|\setofcomponents|}\}$ where $ B'_i = B_i \otimes_{\rm b} \mathcal{C}^{\rm b}_i$ if $B_i \in {\setofsharedcomponents}$ and $B'_i = B_i$ otherwise.
We define the behavior of the instrumented CBS as an LTS $( Q_c , \GActions_c , \trans{c}{}) $ where:
\begin{itemize}
	\item
	$Q_c \subseteq  \bigotimes_{i=1}^{{|\setofcomponents|}} Q'_i \times \bigotimes_{j=1}^{{|\setofschedulers|}} (Q_{S_j} \times Q_{\mathcal{C}^{\rm s}_j})$ is the set of states consisting of the states of schedulers and components with their controllers where $Q'_i = Q_i  \times Q_{c^{\rm b}_i} $ if $ B_i \in {\setofsharedcomponents} $ and $Q'_i = Q_i$ otherwise.
	\item
	$ \GActions_c = \GActions \times \{ \mathcal{R}_{\mathcal{C}^{\rm s}_j}, \mathcal{R}_{\mathcal{C}^{\rm b}_i} \mid S_j \in \setofschedulers \wedge B_i \in {\setofsharedcomponents} \} $ is the set of actions,
	\item $\trans{c}{}  \subseteq Q_c \times \GActions_c \times Q_c$ is the transition relation.
\end{itemize}
\end{definition}
Component $\mathcal{C}^{\rm s}_j$ is the controller of scheduler $S_j$ for $j \in [1 \upto |\setofschedulers|]$ as per  \ppdefref{def:controller_scheduler}, and component $\mathcal{C}^{\rm b}_i$ is the controller of shared component and $B_i$ as per  \ppdefref{def:controller_shared}.
An action of the instrumented system consists of two synchronous actions; an action of the original system and the action of the associated controllers to notify the observer about the occurrence of the action and/or the action of the controllers to exchange vector clocks.
\begin{proposition}
\label{proposition:sim}
$( Q , \GActions , \trans{}{}) \sim ( Q_c , \GActions_c , \trans{c}{}) $.
\end{proposition}
Proposition~\ref{proposition:sim} states that the LTS of the instrumented CBS (see \defref{def:instrumented_system}) is weakly bi-similar to the LTS of initial CBS.
Thus the composition of a set of controllers with schedulers and shared components defined in \ppsecref{sec:instrumentation} does not affect the semantics of the initial system.
\begin{proof}
The proof of Proposition~\ref{proposition:sim} is in Appendix~\ref{proof:sim} (p.~\pageref{proof:sim}).
\end{proof}
%
%%%%%%%%%%%%%%%%%%%%%%%%%%%%%%%%%%%%%%%%%%%%%%%%%%%%%%%%%%%%%%%%%%%
\subsection{Event Extraction from the Local Traces of the Instrumented System}
%%%%%%%%%%%%%%%%%%%%%%%%%%%%%%%%%%%%%%%%%%%%%%%%%%%%%%%%%%%%%%%%%%%
%
According to Definitions~\ref{def:controller_scheduler} and~\ref{def:controller_shared}, the first action in the semantics rules of a controlled scheduler or shared component corresponds to an interaction of the initial system.
Thus, the notion of trace is extended in the natural way by considering the additional semantics rules.
Elements of a trace are updated by including the new configurations and actions of controlled schedulers and shared components.
\begin{example}[Local traces of instrumented system]
\label{xmp:local-aug}
Consider \exref{xmp:dmodel}, the local traces of the instrumented system for two global traces $t_1$ and $t_2$ are:
\begin{itemize}
	\item $s_1(t_1)= (d_1,d_2,d_3) \cdot
	 \left(\fil_{12},\left(\widehat{\fil_{12}\lfloor (1,0) \rfloor} ,\send_2 \lfloor (1,0) \rfloor \right)\right)\cdot
	  (\bot, \bot, ?) \cdot
	   \left(\setof{\beta_1},\left( \widehat{\setof{\beta_1}},\emptyset\right)\right) \cdot
	    (f_1, \bot, ?) \cdot
	     \left(\setof{\drain_1},\widehat{\setof{\drain_1}\lfloor (2,0) \rfloor)} \right) \cdot
	      (\bot, \bot, ?) \cdot
	       \left(\setof{\beta_2},\left( \widehat{\setof{\beta_2}},\emptyset\right)\right) \cdot
	        (\bot, f_2 , ?)$,
	\item $s_2(t_1)= (d_1,d_2,d_3) \cdot
	 \left(\setof{\fil_3},\widehat{\setof{\fil_3}\lfloor (0,1) \rfloor } \right) \cdot
	  (?, d_2, \bot) \cdot
	   \left(\setof{\beta_2},\left( -,\rcv_1 \lfloor (1,0) \rfloor \right)\right) \cdot
	    (?, f_2 , \bot)$,
	\item $s_1(t_2)= (d_1,d_2,d_3) \cdot
	 \left(\fil_{12},\left(\widehat{\fil_{12}\lfloor (1,0) \rfloor } ,\send_2 \lfloor (1,0) \rfloor \right)\right) \cdot
	  (\bot, \bot, ?) \cdot
	   \left(\setof{\beta_2},\left( \widehat{\setof{\beta_2}},\emptyset\right)\right) \cdot
	    (\bot, f_2, ?) \cdot
	     \left(\setof{\beta_1},\left( \widehat{\setof{\beta_1}},\emptyset\right)\right) \cdot
	      (f_1, f_2, ?)$,
	\item $s_2(t_2)= (d_1,d_2,d_3) \cdot
	 \left(\setof{\fil_3},\widehat{\setof{\fil_3} \lfloor (0,1) \rfloor } \right) \cdot
	  (?, d_2, \bot) \cdot
	   \left(\setof{\beta_3},\left( \widehat{\setof{\beta_3}},\emptyset\right)\right) \cdot
	    (?, d_2, f_3) \cdot$\\ $
	      \left(\setof{\beta_2},\left( -,\rcv_1 \lfloor (1,0) \rfloor \right)\right) \cdot
	      (?, f_2, f_3)\cdot
	       \left(\drain_{23},\left(\widehat{\drain_{23} \lfloor (1,0) \rfloor} ,\send_2 \lfloor (1,2) \rfloor \right)\right) \cdot
	        (?, \bot, \bot)$.
\end{itemize}
In both traces, scheduler $S_2$ is notified of the state update of component $\tank_2$ (that is $\beta_2$), but scheduler $S_2$ does not sent it to the observer.
Indeed, following the semantics rules of composition of a scheduler and its controller (\defref{def:semantics-controller-scheduler}), a scheduler only sends the received state from a component only if the execution of the latest action on this component  has been managed by this scheduler.
\end{example}
\begin{definition}[Sequence of events]
\label{def:events}
Let $t$  be the global trace of the distributed system and $s_j(t)= q_0 \cdot \gamma_1 \cdot q_1 \cdots \gamma_{k-1} \cdot q_{k-1} \cdot \gamma_k \cdot q_k$, for $j \in [1,m]$, be the local trace of scheduler $S_j$ (as per \defref{def:local-traces}).
The sequence of events of $s_j(t)$ is inductively defined as follows:
\begin{itemize}
	\item 
	$ \event(q_0) = \epsilon$,
	\item
	$
	\event \left( s_j(t) \cdot \gamma \cdot q \right)
	= \left\{
	\begin{array}{ll}
		\event(s_j(t)) \cdot (a, \vc) & \hbox{if $\gamma$ is of the form $(*,(\widehat{a\lfloor\vc\rfloor},*))$ },  \\
		\event(s_j(t)) \cdot \beta_i & \hbox{if $\gamma$ is of the form $(*,(\widehat{\beta_i},*))$ },\\
		\event(s_j(t))  & \hbox{otherwise}.
	\end{array}
	\right.
	$
\end{itemize}
\end{definition}
Intuitively, any communication between the controller of scheduler $S_j$ and the observer is defined as an event of scheduler $S_j$.
According to the semantic rules of composition $S_j \otimes \mathcal{C}^{\rm s}_j$ (see \defref{def:semantics-controller-scheduler}), controller $\mathcal{C}^{\rm s}_j$ sends information to the observer (actions denoted by \textasciicircum \ over them) when scheduler $S_j$ (i) executes an interaction $a \in \Actions$, or (ii) is notified by the internal action of a component which the execution of its latest action has been managed by scheduler $S_j$.
\begin{example}[Sequence of events]
\label{xmp:seqevents}
The sequences of events of local traces in \exref{xmp:local-aug} are:
	\begin{itemize}
		\item $\event(s_1(t_1)) : (\fil_{12},(1,0)) \cdot \beta_1 \cdot(\drain_1,(2,0)) \cdot \beta_2$,
		\item $\event(s_2(t_1)) : (\fil_3,(0,1))$,
		\item $\event(s_1(t_2)) : (\fil_{12},(1,0)) \cdot \beta_2 \cdot \beta_1$,
		\item $\event(s_2(t_2)) : (\fil_3,(0,1)) \cdot \beta_3 \cdot (\drain_{23},(1,2))$.
	\end{itemize}
\end{example}
To interpret the state updates received by the observer, we use the notion of computation lattice, adapted to distributed CBSs with multi-party interactions in the next section.

%% file: observer.tex
\pgfdeclarelayer{background}
\pgfdeclarelayer{foreground}
\pgfsetlayers{background,main,foreground}
\begin{figure}[t]    
\centering
   \scalebox{1} {   
\begin{tikzpicture}[->,>=stealth',shorten >=1pt,auto,node distance=5cm, semithick]
  \tikzstyle{every state}=[fill=none,draw=black,text=black]
%%%%%%%%%%%%%%%%%%%%%%%%%%%%%%%%%%%
\draw[rounded corners=3em,line width=0.1em,red!60,cap=round,postaction={decorate,decoration={text along path,text align=center,text={event}}}]
  (0.3,6)--(2.5,7);  
\draw[rounded corners=3em,line width=0.1em,red!60,cap=round]
    (2.5,6)--(3,7);
\draw[rounded corners=3em,line width=0.1em,red!60,cap=round]
    (5.3,6)--(4,7);  
%%%%%%%%%%%%%%%%%%%%%%%%%%%%%%%%%%%
\filldraw[fill=none,draw=black](0,7)--(0,8)--(7,8)--(7,7)--(0,7)--cycle; 
\node at (3.5,7.5)   {Local traces $\longrightarrow$ Compatible Global traces};
%%%%%%%%%%%%%%%%%%%%%%%%%%%%%%%%%%%
\filldraw[fill=none,draw=black](0,5)--(0,6)--(2,6)--(2,5)--(0,5)--cycle;
\node at (1,5.5)   {$S_1$}; 
\filldraw[fill=yellow,draw=black](0,5)--(0,6)--(0.6,6)--(0.6,5)--cycle;
\node at (0.3,5.5)   {$C^s_1$};
\filldraw[fill=none,draw=black,xshift=2.2cm](0,5)--(0,6)--(2,6)--(2,5)--(0,5)--cycle; 
\node [xshift=2.2cm] at (1,5.5)   {$S_2$};
\filldraw[fill=yellow,draw=black,xshift=2.2cm](0,5)--(0,6)--(0.6,6)--(0.6,5)--cycle;
\node [xshift=2.2cm] at (0.3,5.5)   {$C^s_2$};
\node at (4.6,5.5)   {$ \cdots$};
\filldraw[fill=none,draw=black,xshift=4cm](1,5)--(1,6)--(3,6)--(3,5)--(1,5)--cycle; 
\node [xshift=4cm] at (2,5.5)   {$S_{\textit{\textbar}\setofschedulers\textit{\textbar}}$};
\filldraw[fill=yellow,draw=black,xshift=5cm](-0.1,5)--(-0.1,6)--(0.6,6)--(0.6,5)--cycle;
\node [xshift=5cm] at (0.26,5.5)   {$C^s_{\textit{\textbar}\setofschedulers\textit{\textbar}}$};
%%%%%%%%%%%%%%%%%%%%%%%%%%%%%%%%%%%
%\draw[-] (0.5,3.5)--(0.5,4)--(3.5,4)--(3.5,3.5); \fill (3.5,3.6) circle (0.1cm);\fill (0.5,3.6) circle (0.1cm);
%\draw[-] (1.5,4)--(1.5,3.5);\fill (1.5,3.6) circle (0.1cm);
%\draw[-] (2.5,4)--(2.5,3.5);\fill (2.5,3.6) circle (0.1cm);
%\node at (4,3.75)   {$ \cdots$};
%\draw[-] (4.5,3.5)--(4.5,4)--(6.5,4)--(6.5,3.5);\fill (6.5,3.6) circle (0.1cm);\fill (4.5,3.6) circle (0.1cm);
%\draw[-] (5.5,4)--(5.5,3.5);\fill (5.5,3.6) circle (0.1cm);
%%%%%%%%%%%%%%%%%%%%%%%%%%%%%%%%%%%
\filldraw[fill=none,draw=black,yshift=1.5cm](0,2)--(0,3)--(1,3)--(1,2)--(0,2)--cycle; 
\node [yshift=1.5cm] at (0.5,2.5)   {$B_1$};
\filldraw[fill=none,draw=black,xshift=1.5cm,yshift=1.5cm](0,2)--(0,3)--(1,3)--(1,2)--(0,2)--cycle; 
\node [xshift=1.5cm,yshift=1.5cm] at (0.5,2.5)   {$B_2$};
\node [xshift=2.5cm,yshift=1.5cm] at (0.5,2.5)   {$\cdots $};
\filldraw[fill=none,draw=black,xshift=3.5cm,yshift=1.5cm](0,2)--(0,3)--(1.5,3)--(1.5,2)--(0,2)--cycle; 
\node [xshift=3.5cm,yshift=1.5cm] at (1,2.5)   {$B_i$};
\node [xshift=3.5cm,red,yshift=1.5cm] at (1.05,2.15)   {{\footnotesize shared}};
\filldraw[fill=yellow,draw=black,xshift=3.5cm,yshift=1.5cm](0,2)--(0,3)--(0.6,3)--(0.6,2)--cycle;
\node [xshift=3.5cm,yshift=1.5cm] at (0.3,2.5)   {$C^b_i$};
\filldraw[fill=none,draw=black,xshift=6cm,yshift=1.5cm](0,2)--(0,3)--(1,3)--(1,2)--(0,2)--cycle; 
\node [xshift=5cm,yshift=1.5cm] at (0.5,2.5)   {$\cdots $};
\node [xshift=6cm,yshift=1.5cm] at (0.5,2.5)   {$B_{\textit{\textbar}\setofcomponents\textit{\textbar}}$};
%%%%%%%%%%%%%%%%%%%%%%%%%%%%%%%%%%%
 \begin{pgfonlayer}{background}
   \draw[-,line width=4em,blue!30,yshift=1.5cm]
    (-0.5,2.5)--(11,2.5); \node at (9,4)   {\textbf{Components}};
%   \draw[-,line width=3.6em,blue!30]
%    (-0.5,4)--(11,4); \node at (9,4)   {\textbf{Actions}};
   \draw[-,line width=4em,blue!10]
    (-0.5,5.5)--(11,5.5); \node at (9,5.5)   {\textbf{Schedulers}};
   \draw[-,line width=4em,green!50]
    (-0.5,7.5)--(11,7.5); \node at (9,7.5)   {\textbf{Observer}};
\end{pgfonlayer}
%%%%%%%%%%%%%%%%%%%%%%%%%%%%%%%%%%%
\end{tikzpicture}}    
\caption{Passive observer} 
\label{fig:passiveobserver}          
\end{figure}
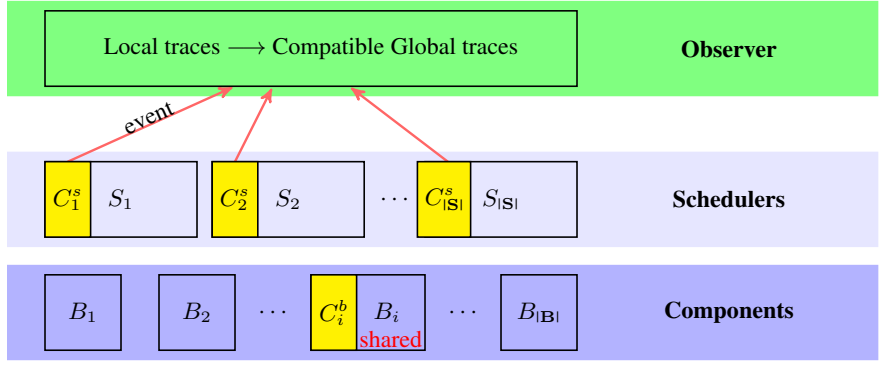

%% file: scheduler_controller_new.tex
\pgfdeclarelayer{background}
\pgfdeclarelayer{foreground}
\pgfsetlayers{background,main,foreground}
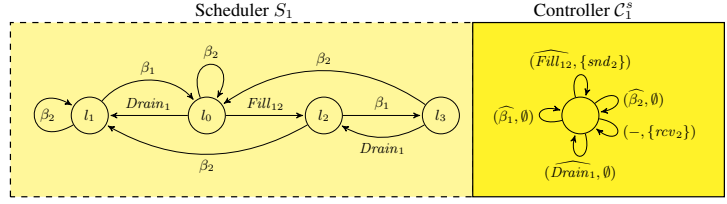
\begin{figure}[t]    
\centering
   \scalebox{0.77} {  
  %\resizebox{\textwidth}{!}{ 
   %\hspace{-2.3cm}
\begin{tikzpicture}[->,>=stealth',shorten >=1pt,auto,node distance=5cm, semithick]
  \tikzstyle{every state}=[fill=none,draw=black,text=black]  
%%%%%%%%%%%%%%%%%%%%%%%%%%%%%%%%%%%
  \filldraw[fill=yellow!50,draw=black,dashed](-7.5,5)--(-7.5,8)--(0.4,8)--(0.4,5)--(-7.5,5)--cycle;
  \filldraw[fill=yellow!90,draw=black](0.4,8)--(0.4,5)--(4.7,5)--(4.7,8)--cycle;
\scalebox{0.8} {  
  \node[state] (l) at (2.8,8) {};
  \path  (l) edge [loop below] node [below]{$(\widehat{\drain_1},\emptyset)$} ();
  \path  (l) edge [loop above] node [above]{$\widehat{(\fil_{12}},\{\send_2\})$} ();
  \path  [out=40,in=10] (l) edge [loop ] node [right]{$(\widehat{\beta_2},\emptyset)$} ();
  \path  [out=-10,in=-40] (l) edge [loop ] node [right]{$(-,\{\rcv_2\})$} ();
  \path  (l) edge [loop left] node [left]{$(\widehat{\beta_1}, \emptyset)$} ();

  \node[state] (A) at (-7.68,8) {$l_1$};
  \node[state] (B) [right of=A,node distance=2.5cm] {$l_0$}; 
  \node[state] (C) [right of=B,node distance=2.5cm] {$l_2$};
  \node[state] (D) [right of=C,node distance=2.5cm] {$l_3$};
  
  \path (B) edge  node {$\fil_{12}$} (C);
  \path (C) edge  node {$\beta_1$} (D);
  \path (D) edge [bend left] node {$\drain_1$} (C);
  \path [out=145,in=35] (D) edge  node [above] {$\beta_2$} (B);
  \path [out=50,in=130](A) edge  node [above] {$\beta_1$} (B);
  \path (C) edge [bend left] node [below] {$\beta_2$} (A);
  \path (B) edge  node [above] {$\drain_1$} (A);
  \path [out=-150,in=150] (A) edge [loop] node [right]{$\beta_2$} ();
  \path [out=110,in=60] (B) edge [loop] node [above]{$\beta_2$} ();
  }
%%%%%%%%%%%%%%%%%%%%%%%%%%%%%%%%%%%

  \node  at (-3.5,8.2){Scheduler $S_1$};
  \node  at (2.3,8.2){Controller $\mathcal{C}^s_1$};
%%%%%%%%%%%%%%%%%%%%%%%%%%%%%%%%%%%
\end{tikzpicture}}    
\caption{Controller attached to the scheduler} 
\label{fig:scheduler_contoller}          
\end{figure}

%% file: controller_tank2.tex
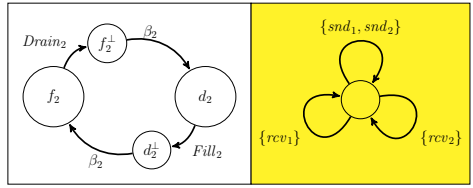
\begin{figure}[t]
    \centering
    \centering
    \Large
   \scalebox{.4} {   
\begin{tikzpicture}[->,>=stealth',shorten >=1pt,auto,node distance=5cm, semithick]
  \tikzstyle{every state}=[fill=none,draw=black,text=black]  
%%%%%%%%%%%%%%%%%%%%%%%%%%%%%%%%%%%
\filldraw[fill=none,draw=black](0,0.5)--(0,6.5)--(8,6.5)--(8,0.5)--(0,0.5)--cycle; 
\node at (1.5,3.4) [state] (A)        [minimum size=2cm]              {$f_2$};
\node            [state] (B)  [right of =A,minimum size=2cm]          {$d_2$};
\node [state] (A') [above right of =A,minimum size=0.5cm,node distance=2.5cm]    {$f_2^\perp$};
\node [state] (B') [below left of =B,minimum size=0.5cm,node distance=2.5cm]    {$d_2^\perp$};

\path [out=10,in=120,ultra thick](A') edge  [above left ] node[text centered] {$\beta_2$} (B);
\path [out=190,in=-60,ultra thick](B') edge  [below]  node[text centered] {$\beta_2$} (A);  
\path [out=-110,in=10,ultra thick] (B) edge  [below right]   node[text centered] { $\fil_2$} (B');
\path [out=70,in=190,ultra thick](A) edge  [above left]   node[text centered] { $\drain_2$} (A');  

%\fill (1,6.5) circle (0.24cm);\node at (1,7) {$\drain$};
%\fill (7,6.5) circle (0.24cm);\node at (7,7){$\fil$};
%\fill (5,6.5) circle (0.24cm);\node at (5,7){$\beta$};  
%%%%%%%%%%%%%%%%%%%%%%%%%%%%%%%%%%%
  \filldraw[fill=yellow!90,draw=black](8,0.5)--(8,6.5)--(15.2,6.5)--(15.2,0.5)--cycle;
  
  \node[state] (l) at (11.6,3.3) {};
  \path  [out=10,in=-60,ultra thick](l) edge [loop] node [below right]{$\{\rcv_2\}$} ();
  \path  [out=-120,in=-190,ultra thick](l) edge [loop] node [below left]{$\{\rcv_1\}$} ();
  \path  [out=125,in=55,ultra thick](l) edge [loop] node [above]{$\{\send_1, \send_2\}$} ();

%%%%%%%%%%%%%%%%%%%%%%%%%%%%%%%%%%%

\end{tikzpicture}}
\caption{Controller of shared component $\tank_2$} 
\label{fig:controller_tank2}   
    \end{figure} 

%% file: complattice.tex
%
%%%%%%%%%%%%%%%%%%%%%%%%%%%%%%%%%%%%%%%
\section{Computation Lattice of a Distributed CBS with Multi-party Interactions}
\label{sec:lattice}
%%%%%%%%%%%%%%%%%%%%%%%%%%%%%%%%%%%%%%%
%
In the previous section, we define how to instrument the system to have controllers generating events (\ie local partial-traces) sent to a central observer.
In this section, we define how the central observer constructs on-the-fly a computation lattice representing the possible global traces compatible with the local partial-traces received from the controllers of schedulers.

In the distributed setting several schedulers execute actions concurrently so that it is not possible to extract the actual partial trace of the system by only observing the local partial-traces.
One can find a set of compatible partial traces, each of which can be considered as the actual partial trace of the system with respect to the observed local partial-traces.
Intuitively, the projection of each compatible partial trace on a specific scheduler results the observed local partial-trace of the scheduler.
A naive monitoring solution is to construct the witness trace~\cite{NazarpourFBBC16} of each compatible partial-trace using the technique for the multi-threaded CBSs.
Such a solution would result a huge computation process overhead, because of the enormous number of compatible partial traces, specially for the distributed system with less number of shared component and more concurrent events.
Instead, we apply the global trace reconstruction on a computation lattice, that is a multi-dimensional execution trace, and introduce a novel technique for the monitoring of the computation lattice on-the-fly (see \psecref{sec:ltl}).
Such a computation lattice encompasses all the compatible global trace of the system.
%
%Such a set of traces is represented as a computation lattice of the distributed system, that is a multi-dimensional execution trace.
%
\paragraph{Computation Lattice}
The computation lattice is represented implicitly using vector clocks.
The construction of the lattice mainly performs the two following operations: (i) \textit{creations of new nodes} and (ii) \textit{updates} of existing nodes in the lattice. 
Action events lead to the creation of new nodes in the direction of the scheduler emitting the event while update events complete the information in the nodes of the lattice related to the state of the component related to the event.
The state of a nodes initially (after creation) is a partial state and by updating the lattice it becomes a global state. 
Since the received events are not totally ordered (because of potential communication delay), we construct the computation lattice based on the vector clocks attached to the received events.
Note that we assume the events received from a scheduler are totally ordered.

We first extend the notion of computation lattice.
\begin{definition}[Extended Computation lattice]
\label{def:computation_lattice}
An extended computation lattice $\mathcal{L}$ is a tuple $\left(N,\IntActions,\xrightarrowtail[] \right)$, where
\begin{itemize}
	\item $N \subseteq Q^l \times \VC$  is the set of nodes, with $Q^l = \bigotimes_{i=1}^{ |\setofcomponents| } \left( Q^{\rm r}_i \bigcup \setof{\bot_i^j \mid S_j \in \setofschedulers \wedge B_i \in \scope(S_j)} \right)$ and $\VC$ is the set of vector clocks,
	\item $\IntActions$ is the set of multi-party interactions as defined in \ppsecref{subsec:semantics},
	\item $\xrightarrowtail[] = \setof{(\eta,a,\eta')\in N \times \IntActions \times N \mid a \in \IntActions \wedge  \eta \rightarrowtail \eta' \wedge \eta.\mathit{state}\longtrans{}{a} \eta'.\mathit{state}}$,
\end{itemize} 
where  $\xrightarrowtail[]$ is the extended presentation of happened-before relation which is labeled by the set of multi-party interactions and $\eta.\mathit{state}$ referring to the state of node $\eta$.
\end{definition}
We simply refer to extended computation lattice as computation lattice.
Intuitively, a computation lattice consists of a set of partially connected nodes, where each node is a pair, consisting of a state of the system and a vector clock.
A system state consists in the states of all components.
The state of a component is either a ready state or a busy state (as per \pdefref{def:component}).
In this context we represent a busy state of component $B_i \in \setofcomponents $, by $\bot_i^j$ which shows that component $B_i$ is busy to finish the computation process of its latest action which has been managed by scheduler $ S_j \in \setofschedulers $.
A computation lattice $\mathcal{L}$ initially consists of an initial node $\init_{\mathcal{L}}= (\init,(0,\ldots,0))$, where $\init$ is the initial state of the system and $(0,\ldots,0)$ is a vector clock where all the clocks associated to the schedulers are zero.	
The set of nodes of computation lattice $\mathcal{L}$ is denoted by $\mathcal{L}.\mathit{nodes}$, and for a node $\eta = (q, \vc) \in \mathcal{L}.\mathit{nodes}$, $\eta.\mathit{state}$ denotes $q$ and $\eta.\mathit{clock}$ denotes $\vc$.
%
%By considering each node as an event associated with a vector clock, 
If (i) the event of node $\eta$ happened before the events of node $\eta'$, that is $\eta'.\mathit{clock} > \eta.\mathit{clock}$ and $\eta \rightarrowtail \eta'$, and (ii) the states of $\eta$ and $\eta'$ follow the global behavior of the system (as per \pdefref{def:global_behavior}) in the sense that the execution of an interaction $a \in \IntActions$ from the state of $\eta$ brings the system to the state of $\eta'$, that is $\eta.\mathit{state}\longtrans{}{a} \eta'.\mathit{state}$, then in the computation lattice it is denoted by $ \eta \xrightarrowtail[a] \eta'$ or by $\eta \xrightarrowtail[] \eta'$ when clear from context.

Two nodes $\eta$ and $\eta'$ of the computation lattice $\mathcal{L}$ are said to be concurrent if neither $\eta.\mathit{clock} > \eta'.\mathit{clock}$ nor $\eta'.\mathit{clock} > \eta.\mathit{clock}$.
For two concurrent nodes $\eta$ and $\eta'$ if there exists a node $\eta''$ such that $\eta'' \xrightarrowtail[] \eta$ and $\eta'' \xrightarrowtail[] \eta'$, then node $\eta''$ is said to be the \textit{meet} of $\eta$ and $\eta'$ denoted by $\meet(\eta,\eta',\mathcal{L})=\eta''$.	
\paragraph{} 
The rest of this section is structured as follows.
In \secref{subsec:intermediate-defs} some intermediate notions are defined in order to introduce our algorithm to construct the computation lattice in \ppsecref{subsec:algorithm}.
In \ppsecref{subsec:algorithm-correctness} we discuss the correctness of the algorithm.
%
%%%%%%%%%%%%%%%%%%%%%%%%%%%%%%%%%%%%%%%
\subsection{Intermediate Operations for the Construction of the Computation Lattice}
\label{subsec:intermediate-defs}
%%%%%%%%%%%%%%%%%%%%%%%%%%%%%%%%%%%%%%%
%
In the reminder, we consider a computation lattice $\mathcal{L}$ as per \ppdefref{def:computation_lattice}.
The reception of a new event either modifies $\mathcal{L}$ or is kept in a queue to be used later.
Action events extend $\mathcal{L}$ using operator $\upda$ (\pdefref{def:node-extend}), and update events update the existing nodes of $\mathcal{L}$ by adding the missing state information into them using operator $\updb$ (\pdefref{def:node_update}).
By extending the lattice with new nodes, one needs to further extend the lattice by computing joints of the created nodes (\pdefref{def:jointnode}) with the existing ones so as to complete the set of possible global traces.
\paragraph{Extension of the lattice.}
We define a function to extend a node of the lattice with an action event which takes as input a node of the lattice and an action event and outputs a new node.
\begin{definition}[Node extension]
\label{def:node-extend}
Function $\upda : (Q^l \times \VC) \times E_{\rm a} \rightarrow Q^l \times \VC$ is defined as follows.
For a node $\eta=((q_1,\ldots,q_{|\setofcomponents|}), \vc) \in Q^l \times \VC$ and an action event $e=(a,\vc')\in E_{\rm a}$, 

$
\upda(\eta,e)= 
\begin{dcases}	
((q'_1,\ldots,q'_{|\setofcomponents|}), \vc') & \text{if } \exists j \in [1 \upto |\setofschedulers|] \quant \\
&\quad (\vc'[j]=\vc[j]+1 \wedge \forall j'\in[1 \upto |\setofschedulers| ] \setminus \setof{j} \quant \vc'[j']=\vc[j']) \\
\text{ undefined } & \text{otherwise };
   \end{dcases}
$\\
with $\forall i\in[1 \upto |\setofcomponents| ] \quant q'_i=
\begin{dcases}
   q_i  & \text{ if } B_i \in \involved(a),  \\
   \bot_i^k & \text{ otherwise.}
   \end{dcases}
   $
   
   where $k=\managed(a).\mathit{index}$.
\end{definition}
Node $\eta$ said to be extendable by event $e$ if $\upda(\eta,e)$ is defined.
Intuitively, node $\eta=(q,\vc)$ represents a global state of the system and extensibility of $\eta$ by action event $e=(a,\vc')$ means that from the global state $q$, scheduler $S_j=\managed(a)$, could execute interaction $a$.
State $\bot_i^k$ indicates that component $B_i$ is busy and being involved in a global action which has been executed (managed) by scheduler $S_k$ for $k \in [1 \upto |\setofschedulers| ]$.

We say that computation lattice $\mathcal{L}$ is extendable by action event $e$ if there exists a node $\eta \in \mathcal{L}.\mathit{nodes}$ such that $\upda(\eta,e)$ is defined.
\begin{property}
\label{property:ext}
$\forall e \in E_{\rm a} \quant | \{ \eta \in \mathcal{L}.\mathit{nodes} \mid \exists \eta' \in Q^l \times \VC \quant  \eta' =  \upda(\eta,e) \} | \leq 1$.
\end{property} 
\propref{property:ext} states that for any update event $e$, there exists at most one node in the lattice for which function $\upda$ is defined (meaning that $\mathcal{L}$ can be extended by event $e$ from that node).
\begin{example}[Node extension]
\label{xmp:extend}
Considering the local partial-traces described in \ppexref{xmp:seqevents}, initially, the computation lattice consists of the initial node which has the initial state $\init$, with an associated vector clock $(0,0)$, \ie $\init_{\mathcal{L}}= ((d_1,d_2,d_3),(0,0))$.
As for the sequence of events in trace $t_1$, node $((d_1,d_2,d_3),(0,0))$ is extendable by event $(\fil_{12},(1,0))$ because, according to \defref{def:node-extend}, we have:
$
\upda(((d_1,d_2,d_3),(0,0)),(\fil_{12},(1,0)))= ((\bot_1^1,\bot_2^1,d_3),(1,0)).
$

Furthermore, to illustrate \propref{property:ext}, let us consider the extended lattice after event $(\fil_{12},(1,0))$ which consists of two nodes, initial node $((d_1,d_2,d_3),(0,0))$ and node $((\bot_1^1,\bot_2^1,d_3),(1,0))$.
When action event $(\fil_3,(0,1))$ is received, we have $\upda(\init_{\mathcal{L}},$ $(\fil_3,(0,1))=((d_1,d_2,\bot_3^2),(0,1))$ whereas $\upda(((\bot_1^1,\bot_2^1,d_3),(1,0)),$ $(\fil_3,(0,1)))$ is not defined which shows that \propref{property:ext} holds on the lattice.
\end{example}
We define a relation between two vector clocks to distinguish the concurrent execution of two interactions such that both could happen from a specific global state of the system.
\begin{definition}[Relation $\mathcal{J}_{\mathcal{L}}$]
\label{def:jrelation}
Relation $\mathcal{J}_{\mathcal{L}} \subseteq  \VC \times \VC$ is defined between two vector clocks as follows:
$\mathcal{J}_{\mathcal{L}}=\big\{(\vc,\vc') \in \VC \times \VC \mid \exists! k\in[1 \upto |\setofschedulers| ] \quant \vc[k]=\vc'[k]+1 \wedge \exists! l\in[1 \upto |\setofschedulers|] \quant \vc'[l]=\vc[l]+1 \wedge \forall j\in[1 \upto |\setofschedulers| ] \setminus \setof{k,l} \quant \vc[j]=\vc'[j]\big\}$.
\end{definition} 
For two vector clocks $\vc$ and $\vc'$ to be in relation $\mathcal{J}_{\mathcal{L}}$, $\vc$ and $vc'$ should agree on all but two clocks values related to two schedulers of indexes $k$ and $l$.
On one of these indexes, the value of one vector clock is equal to the value of the other vector clock plus 1, and the converse on the other index.
Intuitively, $(\eta.\mathit{clock}, \eta'.\mathit{clock}) \in \mathcal{J}_{\mathcal{L}}$ means that nodes $\eta$ and $\eta'$ are associated to two concurrent events (caused by the execution of two interactions managed by two different schedulers) that both could happen from a unique global state of the system which is the meet of $\eta$ and $\eta'$ (see \propref{property:meet}).
\exref{exmp:j_joint} illustrates relation $\mathcal{J}_{\mathcal{L}}$. 
\begin{property}
\label{property:meet}
$\forall \eta,\eta' \in \mathcal{L}.\mathit{nodes} \quant \left(\eta.\mathit{clock},\eta'.\mathit{clock}\right) \in \mathcal{J}_{\mathcal{L}} \implies \meet(\eta,\eta',\mathcal{L}) \in \mathcal{L}.\mathit{nodes}$.
\end{property} 
\propref{property:meet} states that for two nodes $\eta$ and $\eta'$ in lattice $\mathcal{L}$ such that $(\eta.\mathit{clock},\eta'.\mathit{clock}) \in \mathcal{J}_{\mathcal{L}}$, there exists a node in lattice $\mathcal{L}$ as the meet of $\eta$ and $\eta'$, that is $\meet(\eta,\eta',\mathcal{L}) \in \mathcal{L}.\mathit{nodes}$.

The joint node of $\eta$ and $\eta'$ is defined as follows.
\begin{definition}[Joint node]
\label{def:jointnode}
For two nodes $\eta,\eta' \in \mathcal{L}.\mathit{nodes}$ such that $(\eta.\mathit{clock},\eta'.\mathit{clock}) \in \mathcal{J}_{\mathcal{L}}$, the joint node of $\eta$ and $\eta'$, denoted by $\joint(\eta,\eta',\mathcal{L})=\eta''$, is defined as follows:
\begin{itemize}
	\item $\forall i\in [1 \upto |\setofcomponents|]\quant \eta''.\mathit{state}[i]=
	\begin{dcases}
   \eta.\mathit{state}[i]  & \text{ if } \eta.\mathit{state}[i] \neq  \eta_m.\mathit{state}[i], \\
   \eta'.\mathit{state}[i]  &\text{ otherwise;}
   \end{dcases}$
   \item $\eta''.\mathit{clock} = \max(\eta.\mathit{clock},\eta'.\mathit{clock})$;
\end{itemize}
where $\eta_m=\meet(\eta,\eta',\mathcal{L})$.
\end{definition}
According to \propref{property:meet}, for two nodes $\eta$ and $\eta'$ in relation $\mathcal{J}_{\mathcal{L}}$, their meet node exists in the lattice. 
The state of the joint node of $\eta$ and $\eta'$ is defined by comparing their states and the state of their meet.
Since two nodes in relation $\mathcal{J}_{\mathcal{L}}$ are concurrent, the state of component $B_i$ for $i \in [1 \upto |\setofcomponents|]$ in nodes $\eta$ and $\eta'$ is either equal to the state of component $B_i$ in their meet, or only one of the nodes $\eta$ and $\eta'$ has a different state than their meet (components can not be both involved in two concurrent executions).
The joint node of two nodes $\eta$ and $\eta'$ takes into account the latest changes of the state of the nodes $\eta$ and $\eta'$ compared to their meet.
Note that $\joint(\eta,\eta',\mathcal{L}) = \joint(\eta',\eta,\mathcal{L})$, because $\joint$ is defined for nodes whose clocks are in relation $\mathcal{J}_{\mathcal{L}}$.
\begin{example}[Relation $\mathcal{J}_{\mathcal{L}}$ and joint node]
\label{exmp:j_joint}
To continue \ppexref{xmp:extend}, the reception of action event $(\fil_3,(0,1))$ extends the lattice in the direction of scheduler $S_2$ because function $\upda$ is defined, that is:
\[
\upda(((d_1,d_2,d_3),(0,0)),(\fil_3,(0,1)))= ((d_1,d_2,\bot_3^2),(0,1)).
\]
After this extension, the lattice has three nodes which are $((d_1,d_2,d_3),(0,0))$, $((\bot_1^1,\bot_2^1,d_3),(1,0))$ and $((d_1,d_2,\bot_3^2),$ $(0,1))$.
According to \ppdefref{def:jrelation}, the vector clocks of the two nodes $((\bot_1^1,\bot_2^1,d_3),(1,0))$ and $((d_1,d_2,\bot_3^2),(0,1))$ are in relation $\mathcal{J}_{\mathcal{L}}$ (\ie $((1,0),(0,1)) \in \mathcal{J}_{\mathcal{L}}$).
Therefore, following \ppdefref{def:jointnode}, the joint node of the two above nodes is $((\bot_1^1,\bot_2^1,\bot_3^2),(1,1))$, and their meet is $((d_1,d_2,d_3),(0,0))$.
\end{example}
\paragraph{Update of the lattice.}
We define a function to update a node of the lattice which takes as input a node of the lattice and an update event and outputs the updated version of the input node.
\begin{definition}[Node update]
\label{def:node_update}
Function $\updb : (Q^l \times \VC) \times E_{\upbeta} \rightarrow Q^l \times \VC$ is defined as follows.
For a node $\eta=((q_1,\ldots,q_{|\setofcomponents|}), \vc)$ and an update event $e=(\beta_i,q'_i) \in E_{\upbeta}$ with $i \in [1 \upto |\setofcomponents| ]$ which is sent by scheduler $S_k$ with $k \in [1 \upto |\setofschedulers| ]$:

 $\updb(\eta,e)=((q_1,\ldots,q_{i-1},q''_i,q_{i+1},\ldots, q_{|\setofcomponents|}),\vc)$, 

with  $q''_i=
\begin{dcases}
   q_i'  &\text{ if } q_i=\bot_i^k,  \\
   q_i &\text{ otherwise.}
   \end{dcases}
   $ 
\end{definition}
An update event $(\beta_i,q'_i)$ contains an updated state of some component $B_i$.
By updating a node $\eta$ in the lattice with an update event which is sent from scheduler $S_k$, we update the partial state associated to $\eta$ by adding the state information of that component, if the state of component $B_i$ associated to node $\eta$ is $\bot_i^k$.
Intuitively means that a busy state which is caused by an execution of an action managed by scheduler $S_k$ can only be replace by a ready state sent by the same scheduler $S_k$.
Updating node $\eta$ does not modify the associated vector clock $\vc$.
\begin{example}[Node update]
\label{xmp:update}
To continue \exref{exmp:j_joint}, let us consider node $((\bot_1^1,\bot_2^1,d_3),(1,0))$ whose state is a partial state (because of the lack of the state information of $\tank_1$ and $\tank_2$), and update event $(\beta_1,f_1)$ sent by scheduler $S_1$. 
To obtain the updated node, we apply function $\updb$ over the node and the update event.
We have: $\updb(((\bot_1^1,\bot_2^1,d_3),(1,0)),(\beta_1,f_1))$ which results updated node $((f_1,\bot_2^1,d_3),(1,0))$.
Although the state of the updated node is a partial state, it has more state information then the state before update (\ie $(\bot_1^1,\bot_2^1,d_3)$).
Concerning the initial node of the lattice and update event $(\beta_1,f_1)$, $\updb(((d_1,d_2,d_3),(0,0)),(\beta_1,f_1))=((d_1,d_2,d_3),(0,0))$. 
\end{example}
\paragraph{Buffering events.}
The reception of an action event or an update event might not always lead to extending or updating the current computation lattice.
Due to communication delay, an event which has happened before another event might be received later by the observer.
It is necessary for the construction of the computation lattice to use events in a specific order.
Such events must be kept in a waiting queue to be used later.
For example, such a situation occurs when receiving action event $e$ such that function $\upda$ is not defined over $e$ and  none of the existing nodes of the lattice.
In this case event $e$  must be kept in the queue until obtaining another configuration of the lattice in which function $\upda$ is defined. 
Moreover, an update event $e'$ referring to an internal action of component $B_i$ is kept in the queue if there exists an action event $e''$ in the queue such that component $B_i$ is involved in $e''$, because we can not update the nodes of the lattice with an update event associated to an execution which is not yet taken into account in the lattice. 
\begin{definition}[Queue $\kappa$]
\label{def:queue}
A queue of events  is a finite sequence of events in $E$.
Moreover, for a non-empty queue $\kappa=e_1 \cdot e_2 \cdots e_r$, $\rem(\kappa,e)=\kappa(1\cdots z-1) \cdot \kappa(z+1\cdots r)$ with $e=e_z \in \{ e_1 , e_2, \ldots ,e_r \}$. 
\end{definition}
Queue $\kappa$ is initialized to an empty sequence. 
Function $\rem$ takes as input queue $\kappa$ and an event in the queue and outputs the version of $\kappa$ in which the given event is removed from the queue.  
\begin{example}[Event storage in the queue]
Let us consider trace $t_2$ in \ppexref{xmp:seqevents}, such that all the events of scheduler $S_2$ are received by the observer earlier than the events of scheduler $S_1$. 
After the reception of action event $(\fil_3,(0,1))$, since $\upda(((d_1,d_2,d_3),(0,0)),(\fil_3,(0,1)))$ is defined, the lattice is extended in the direction of scheduler $S_2$  and the new node $((d_1,d_2,\bot_3^2)(0,1))$ is created.
The reception of update event $(\beta_3,f_3)$ updates the newly created node $((d_1,d_2,\bot_3^2)(0,1))$ to $((d_1,d_2,f_3)(0,1))$.
After the reception of action event $(\drain_{23},(1,2))$, since there is no node in the lattice where function $\upda$ is defined over, event $(\drain_{23},(1,2))$ must be stored in the queue, therefore $\kappa= (\drain_{23},(1,2))$.
\end{example}
%
%%%%%%%%%%%%%%%%%%%%%%%%%%%%%%%%%%%%%%%
\subsection{Algorithm Constructing the Computation Lattice}
\label{subsec:algorithm}
%%%%%%%%%%%%%%%%%%%%%%%%%%%%%%%%%%%%%%%
%
In the following, we define an algorithm based on the above definitions to construct the computation lattice based on the events received by the global observer.
The algorithm consists of a main procedure (see \palgoref{alg:make}) and several sub-procedures using global variables lattice $\mathcal{L}$ (\pdefref{def:computation_lattice}) and queue $\kappa$ (\defref{def:queue}).

For an action event $e \in E_{\rm a}$ with $e=(a,\vc)$, $e.\mathit{action}$ denotes interaction $a$ and $e.\mathit{clock}$ denotes vector clock  $\vc$.
For an update event $e \in E_{\upbeta}$ with $e=(\beta_i,q_i)$, $e.\mathit{index}$ denotes index $i$.

Initially, after the reception of event $e$ from a controller of a scheduler, the observer calls the main procedure using $\make(e,\false)$.
In the following, we describe each procedure in detail.
\input{make}
\paragraph{\textsc{Make} (\algoref{alg:make}):}
Procedure \textsc{Make} takes two parameters as input: a boolean variable $\fromthequeue$ and an event $e$.
Parameters $e$ and $\fromthequeue$ vary based on the type of event $e$.
Boolean variable $\fromthequeue$ is true when the input event $e$ is picked up from the queue and false otherwise (\ie event $e$ is received from a controller of a scheduler). 
Procedure \textsc{Make} uses two sub-procedures, \textsc{ActionEvent} and \textsc{UpdateEvent}.
If the input event is an action event, sub-procedure \textsc{ActionEvent} is called, and if the input event is an update event, sub-procedure \textsc{UpdateEvent} is called.
Procedure \textsc{Make} updates the global variables.
\input{actionevent}
\paragraph{\textsc{ActionEvent} (\algoref{alg:actionevent}):}
Procedure \textsc{ActionEvent} is associated to the reception of action events and takes as input an action event $e$ and a boolean parameter $\fromthequeue$, which is false when event $e$ is received from a controller of a scheduler and true when event $e$ is picked up from the queue.
Procedure \textsc{ActionEvent} modifies global variables $\mathcal{L}$ and  $\kappa$.
Procedure \textsc{ActionEvent} has a local boolean variable $\latticextend$ which is true when an input action event could extend the lattice (\ie the current computation lattice is extendable by the input action event) and false otherwise.

By iterating over the existing nodes of lattice $\mathcal{L}$,  \textsc{ActionEvent} checks if there exists a node $\eta $ in $\mathcal{L}.\mathit{nodes}$ such that function $\upda$ is defined over event $e$ and node $\eta$ (\pdefref{def:node-extend}). 
If such a node $\eta$ is found, \textsc{ActionEvent} creates the new node $\upda(\eta,e)$, adds it to the set of the nodes of the lattice, invokes procedure \textsc{ModifyQueue}, and stops iteration.
Otherwise, \textsc{ActionEvent} invokes procedure \textsc{ModifyQueue} and terminates.

In the case of extending the lattice by a new node, it is necessary to create the (possible) joint nodes. 
To this end, in Line~\ref{line:make_joints} procedure \textsc{Joints} is called to evaluate the current lattice and create the joint nodes.  
For optimization purposes, after making the joint nodes procedure \textsc{RemoveExtraNodes} is called to eliminate unnecessary nodes to optimize the lattice size.

After making the joint nodes and (possibly) reducing the size of the lattice, if the input action event is not picked from the queue,  \textsc{ActionEvent} invokes procedure \textsc{CheckQueue} in Line~\ref{line:check_queue},  otherwise it terminates.
\input{updateevent}
\paragraph{\textsc{UpdateEvent} (\algoref{alg:updateevent}):}
Procedure \textsc{UpdateEvent} is associated to the reception of update events.
Recall that an update event $e$ contains the state update of some  component $B_i$ with $i\in[1 \upto |\setofcomponents|]$ ($e.\mathit{index}=i$).
Procedure \textsc{UpdateEvent} takes as input an update event $e$ and a boolean value associated to parameter $\fromthequeue$.
Procedure \textsc{UpdateEvent} modifies global variables $\mathcal{L}$ and $\kappa$.

First, \textsc{UpdateEvent} checks the events in the queue.
If there exists an action event $e'$ in the queue such that component $B_i$ is involved in $e'.\mathit{action}$, \textsc{UpdateEvent} adds update event $e$ to the queue using \textsc{ModifyQueue} and terminates.
Indeed, one can not update the nodes of the lattice with an update event associated to an execution which is not yet taken into account in the lattice. 

If no action event in the queue concerned component $B_i$, \textsc{UpdateEvent} updates all the nodes of the lattice (Lines~\ref{line:start_update_nodes}-\ref{line:end_update_nodes}) according to \ppdefref{def:node_update}.

Finally, the input update event is removed from the queue if it is picked from the queue, using \textsc{ModifyQueue}.
\input{modifyqueue}
\paragraph{\textsc{ModifyQueue} (\algoref{alg:modifyqueue}):}
Procedure \textsc{ModifyQueue} takes as input an event $e$ and two boolean variables $\fromthequeue$ and $\eventused$. 
Procedure \textsc{ModifyQueue} adds (\resp removes) event $e$ to (\resp from) queue $\kappa$ according to the following conditions.
If event $e$ is picked up from the queue (\ie  $\fromthequeue=\true$) and $e$ is used in the algorithm to extend or update the lattice (\ie  $\eventused=\true$), event $e$ is removed from the queue (Line~\ref{line:remove_queue}).
Moreover, if event $e$ is not picked up from the queue and it is not used in the algorithm, event $e$ is stored in the queue (Line~\ref{line:add_queue}).
\input{joints}
\paragraph{\textsc{Joints} (\algoref{alg:joints}):}
Procedure \textsc{Joints} extends lattice $\mathcal{L}$ in such a way that all the possible joints have been created.
First, procedure \textsc{Jcompute} is invoked to compute relation $\mathcal{J}_{\mathcal{L}}$ (\pdefref{def:jrelation}) among the existing nodes of the lattice and then creates the joint nodes and adds them to the set of the nodes of the lattice. 
Then, after the creation of the joint node of two nodes $\eta$ and $\eta'$, $(\eta.\mathit{clock},\eta'.\mathit{clock})$ is removed from relation $\mathcal{J}_{\mathcal{L}}$.
It is necessary to compute relation $\mathcal{J}_{\mathcal{L}}$ again after the creation of joint nodes, because new nodes can be in relation $\mathcal{J}_{\mathcal{L}}$.
This process terminates when $\mathcal{J}_{\mathcal{L}}$ is empty.
\input{jcompute}
\paragraph{\textsc{Jcompute} (\algoref{alg:jcompute}):}
Procedure \textsc{Jcompute} computes relation $\mathcal{J}_{\mathcal{L}}$ by pairwise iteration over all the nodes of the lattice and checks if the vector clocks of any two nodes satisfy the conditions in \ppdefref{def:jrelation}.
The pair of vector clocks satisfying the above conditions are added to relation $\mathcal{J}_{\mathcal{L}}$.
\input{checkqueue}
\paragraph{\textsc{CheckQueue} (\algoref{alg:checkqueue}):}
Procedure \textsc{CheckQueue} recalls the events stored in the queue $e \in \kappa$ and executes $\make(e, \true)$, to check whether the conditions for taking them into account to update the lattice hold.

Procedure \textsc{CheckQueue} checks the events in the queue until none of the events in the queue can be used either to extend or to update the lattice.
To this end, before checking queue $\kappa$, in Line~\ref{line:queue_copy} a copy of queue $\kappa$ is stored in $\kappa'$, and after iterating all the events in queue $\kappa$, the algorithm checks the equality of current queue and the copy of the queue before checking.
If the current queue $\kappa$ and copied queue $\kappa'$ have the same events, it means that none of the events in queue $\kappa$ has been used (thus removed), therefore the algorithm stops checking the queue again by breaking the loop in Line~\ref{line:break}.

Note, when the algorithm is iterating over the events in the queue, \ie when the value of variable $\fromthequeue$ is true, it is not necessary to iterate over the queue again (\palgoref{alg:actionevent}, Line~\ref{line:if_not_from_the_queue}).
Moreover, events in the queue are picked up in the same order as they have been stored in the queue (FIFO queue). 
\input{removeextranodes}
\paragraph{\textsc{RemoveExtraNodes} (\algoref{alg:removeextranodes}):}
For optimization reasons, after extending the lattice by an action event, procedure \textsc{RemoveExtraNodes} is called to eliminate some (possibly existing) nodes of the lattice.
A node in the lattice can be removed if the lattice no longer can be extended from that node.
Having two nodes of the lattice $\eta$ and $\eta'$ such that every clock in the vector clock of $\eta'$ is strictly greater than the respective clock of $\eta$, one can remove node $\eta$. 
This is due to the fact that the algorithm never receives an action event which could have extended the lattice from $\eta$ where the lattice has already took into account an occurrence of event which has greater clocks stamp than $\eta.\mathit{clock}$.
\begin{remark}
The reason to remove the extra nodes of the lattice can be explained as following. 
First, our online algorithm is used for runtime monitoring purposes, and second, each node $n$ represents the evaluation of system execution up to node $n$. 
Hence, the nodes which reflect the state of the system in the past are not valuable for the runtime monitor. 
\end{remark}
\input{lattice}
\begin{example}[Lattice construction]
\label{xmp:alg_make}
\figref{fig:lattice} depicts the computation lattice according to the received sequence of events concerning trace $t_2$ of \ppexref{xmp:seqevents}. 
Node $\left( \left( d_1,d_2,f_3 \right),\left( 0,1 \right) \right)$ is associated to event $(\fil_{12},(1,0))$ and node $\left( \left( f_1,f_2,d_3 \right),\left( 1,0 \right) \right)$ is associated to event $(\fil_3,(0,1))$.
Since these two events are concurrent, joint node $( ( f_1,f_2,f_3 ),$ $( 1,1 ) )$ is made.
Node $\left( \left( f_1,\bot_2^2,\bot_3^2 \right),\left( 1,2 \right) \right)$ is associated to event $(\drain_{23},(1,2))$.
Due to vector clock update technique, the node with vector clock of $(0,2)$ is not created.
\end{example}
%
%
%%%%%%%%%%%%%%%%%%%%%%%%%%%%%%%%%%%%%%%
\subsection{Insensibility to Communication Delay}
\label{subsec:algorithm-correctness}
%%%%%%%%%%%%%%%%%%%%%%%%%%%%%%%%%%%%%%%
%
%
Algorithm $\make$ can be defined over a sequence of events received by the observer $\zeta= e_1 \cdot e_2 \cdot e_3 \cdots e_z \in E^*$ in the sense that one can apply $\make$ sequentially from $e_1$ to $e_z$ initialized by taking event $e_1$, the initial lattice $\init_{\mathcal{L}}$ and an empty queue.
\begin{proposition}[Insensitivity to the reception order]
\label{proposition:make}
For any two sequences of events $\zeta ,\zeta' \in E^*$, we have\\ $ \left( \forall S_j \in \setofschedulers \quant \zeta \downarrow_{S_j} = \zeta'\downarrow_{S_j} \right) \implies  \make(\zeta) = \make(\zeta')$, where $\zeta\downarrow_{S_j}$ is the projection of $\zeta$ on scheduler $S_j$ which results the sequence of events  generated by $S_j$.
\end{proposition}
\proporef{proposition:make} states that different ordering of the events does not affect the output result of  Algorithm~$\make$.
Note, this proposition assumes that all events in $\zeta$ and $\zeta'$ can be distinguished.
For a sequence of events $\zeta \in E^*$, $\make(\zeta).\mathit{lattice}$ denotes the constructed computation lattice $\mathcal{L}$ by algorithm~$\make$.
%
%%%%%%%%%%%%%%%%%%%%%%%%%%%%%%%%%%%%%%%
\subsection{Correctness of Lattice Construction}
\label{subsec:construction-correctness}
%%%%%%%%%%%%%%%%%%%%%%%%%%%%%%%%%%%%%%%
%
Computation lattice $\mathcal{L}$ has an initial node $\init_{\mathcal{L}}$ which is the node with the smallest vector clock, and a \textit{frontier} node which is the node with the greatest vector clock.
A path of the constructed computation lattice $\mathcal{L}$ is a sequence of causally-related nodes of the lattice, starting from the initial node and ending up in the frontier node.
\begin{definition}[Set of the paths of a lattice]
\label{def:set-of-paths}
The set of the paths of a constructed computation lattice $\mathcal{L}$ is $\Pi(\mathcal{L}) = \bigg\{\eta_0 \cdot \alpha_1 \cdot \eta_1 \cdot \alpha_2 \cdot \eta_2 \cdots  \alpha_z \cdot \eta_z \mid  \eta_0 = \init_{\mathcal{L}} \wedge \forall r \in [1 \upto z] \quant \bigg( \eta_{r-1} \xrightarrowtail[\alpha_r] \eta_{r} \vee \big( \exists N \subseteq \mathcal{L}.\mathit{nodes} \quant \eta_{r-1}=\meet(N,\mathcal{L}) \wedge \eta_{r} =\joint(N,\mathcal{L}) \wedge \forall \eta \in N \quant \eta_{r-1} \xrightarrowtail[a_\eta] \eta \wedge \alpha_r = \bigcup_{\eta \in N} a_\eta \big) \bigg)  \bigg\}$, where the notions of $\meet$ and $\joint$ are naturally extended over a set of nodes.
\end{definition} 
A path is a sequence of nodes such that for each pair of adjacent
nodes either (i) the prior node is in $\xrightarrowtail[]$ relation with the next node or (ii) the prior and  the next node are the meet and the joint of a set of existing nodes respectively.
A path from a meet node to the associated joint node  represents an execution of a set of concurrent joint actions.
\begin{example}[Set of the paths of a lattice]
In the computation lattice $\mathcal{L}$ depicted in \ppfigref{fig:lattice}, there are three  distinct paths that begin from the initial node $\left( \left( d_1,d_2,d_3 \right),\left( 0,0 \right) \right)$ and end up to the frontier node $\left( \left( f_1,\bot^2_2,\bot^2_3 \right),\left( 1,2 \right) \right)$.
The set of paths is $\Pi(\mathcal{L})=\setof{\pi_1, \pi_2 ,\pi_3}$, where:
\begin{itemize}
	\item $\pi_1 = ((d_1,d_2,d_3),(0,0))\cdot \setof{\fil_{12}} \cdot  ((f_1,f_2,d_3),(1,0))\cdot \setof{\setof{\fil_3}} \cdot ((f_1,f_2,f_3),(1,1))\cdot \setof{\drain_{23}} \cdot  ((f_1,\bot^2_2,\bot^2_3),(1,2))$,
	\item $\pi_2 = ((d_1,d_2,d_3),(0,0))\cdot \setof{\setof{\fil_3}} \cdot ((d_1,d_2,f_3 ),(0,1))\cdot \setof{\fil_{12}} \cdot ((f_1,f_2,f_3),(1,1))\cdot \setof{\drain_{23}} \cdot ((f_1,\bot^2_2,\bot^2_3),(1,2))$,
	\item $\pi_3 = ((d_1,d_2,d_3),(0,0))\cdot \setof{\fil_{12} ,\setof{\fil_3}} \cdot((f_1,f_2,f_3),(1,1))\cdot \setof{\drain_{23}} \cdot((f_1,\bot^2_2,\bot^2_3),(1,2))$.
\end{itemize}
\end{example}
%
% consisting of a set of components $ \setofcomponents $ (as per \defref{def:component}) and a set of schedulers $\setofschedulers $ (as per \defref{def:scheduler})
%
Let us consider system $\model$ with the global behavior $( Q , \GActions , \trans{}{}) $ as per \ppdefref{def:global_behavior}.
At runtime, the execution of such a system produces a global trace 
$t= q^0 \cdot ( \alpha^1 \cup \beta^1 ) \cdot q^1 \cdot ( \alpha^2 \cup \beta^2) \cdots ( \alpha^{k} \cup \beta^{k} ) \cdot q^k$ as per \ppdefref{def:global-trace}.
Since the actual partial trace $t$ is not observable due to the occurrence of simultaneous and concurrent interactions and internal actions, partial trace $t$ can be represented as a set of compatible partial traces, which could have happened in the system at runtime.
\begin{definition}[Compatible partial traces of a partial trace]
\label{def:compatible-trace}
The set of all compatible partial traces of partial trace $t$ is $\mathcal{P}(t)=\{ t' \in Q \cdot (\GActions \cdot Q)^*\mid \forall j\in[1 \upto |\setofschedulers| ] \quant t'\downarrow_{S_j} = t \downarrow_{S_j} = s_j(t)\}$.
\end{definition} 
Partial trace $t'$ is compatible with the partial trace $t$ if the projection of both $t$ and $t'$ on scheduler $S_j$, for $j \in [1 \upto |\setofschedulers| ]$, results the local partial-trace of scheduler $S_j$.
In a partial trace, for each global action which consists of several concurrent interactions and internal actions of different schedulers, one can define different ordering of those concurrent interactions, each of which represents a possible execution of that global action.
Consequently, several compatible partial traces can be encoded from a partial trace of the distributed system $\model$.

Note that two compatible traces with only difference in the ordering of their internal actions are considered as a unique compatible trace. 
What matters in the compatible traces of a partial trace is the different ordering of interactions.
\begin{example}[The set of compatible partial traces]
\label{xmp:possible_global}
Let us consider the partial trace $t_1$ described in \ppexref{exmp:global_trace}, that is
$t_1=(d_1,d_2,d_3) \cdot \setof{\fil_{12}} \cdot (\bot, \bot, d_3) \cdot \setof{\beta_1} \cdot (f_1, \bot, d_3) \cdot \setof{\setof{\drain_1}, \setof{\fil_3}} \cdot (\bot, \bot , \bot ) \cdot \setof{\beta_2} \cdot (\bot, f_2 , \bot)$.
The projection of $t_1$ on each scheduler is represented as follow:
\begin{itemize}
 \item $t_1 \downarrow_{S_1}= (d_1,d_2,d_3) \cdot \setof{\fil_{12}} \cdot (\bot, \bot, ?) \cdot \setof{\beta_1} \cdot (f_1, \bot, ?) \cdot \setof{\setof{\drain_1}} \cdot (\bot, \bot , ? ) \cdot \setof{\beta_2} \cdot (\bot, f_2 , ?)$,
 \item $t_1 \downarrow_{S_2}= (d_1,d_2,d_3) \cdot \setof{\fil_3} \cdot (?, d_2,\bot)$. 
 \end{itemize}	
The set of compatible partial traces is
$\mathcal{P}(t_1)= \setof{t^1_1,t^2_1 ,t^3_1,t^4_1,t^5_1}$ where:
\begin{itemize}	
	\item $t^1_1 = (d_1,d_2,d_3) \cdot \setof{\fil_{12}} \cdot (\bot, \bot, d_3) \cdot \setof{\beta_1} \cdot (f_1, \bot, d_3) \cdot \setof{\setof{\fil_3}, \setof{\drain_1}} \cdot (\bot, \bot , \bot ) \cdot \setof{\beta_2} \cdot (\bot, f_2 , \bot)$,
	\item $t^2_1 =  (d_1,d_2,d_3) \cdot \setof{\fil_{12}} \cdot (\bot, \bot, d_3) \cdot \setof{\beta_1} \cdot (f_1, \bot, d_3) \cdot \setof{\setof{\drain_1}} \cdot (\bot, \bot, d_3) \cdot \setof{\setof{\fil_3}} \cdot (\bot, \bot , \bot ) \cdot \setof{\beta_2} \cdot (\bot, f_2 , \bot)$,
	\item $t^3_1 =  (d_1,d_2,d_3) \cdot \setof{\fil_{12}} \cdot (\bot, \bot, d_3) \cdot \setof{\beta_1} \cdot (f_1, \bot, d_3) \cdot \setof{\setof{\fil_3}} \cdot (f_1, \bot, \bot) \cdot \setof{\setof{\drain_1}} \cdot (\bot, \bot , \bot ) \cdot \setof{\beta_2} \cdot (\bot, f_2 , \bot)$,
	\item $t^4_1 =  (d_1,d_2,d_3) \cdot \setof{\fil_{12},\setof{\fil_3}} \cdot (\bot,\bot, \bot) \cdot  \setof{\beta_1} \cdot (f_1, \bot, \bot) \cdot \setof{\setof{\drain_1}} \cdot (\bot, \bot, \bot) \cdot \setof{\beta_2} \cdot (\bot, f_2 , \bot)$,
	\item $t^5_1 =  (d_1,d_2,d_3) \cdot \setof{\setof{\fil_3}} \cdot (d_1,d_2, \bot) \cdot \setof{\fil_{12}} \cdot (\bot, \bot, \bot)\cdot \setof{\beta_1} \cdot (f_1, \bot, \bot) \cdot \setof{\setof{\drain_1}} \cdot (\bot, \bot, \bot) \cdot \setof{\beta_2} \cdot (\bot, f_2 , \bot)$.
\end{itemize}
\end{example}

Since the desired property is defined over the global states, for monitoring purposes it is necessary to obtain the global trace of the system with a sequence of global states.
To this end, by inspiring the technique introduced in~\cite{NazarpourFBBC16} to reconstruct the witness trace, we define a function which takes as input a partial trace of the distributed system (\ie a sequence of partial states) and outputs an equivalent global trace in which all the internal actions ($\beta$) are removed from the trace and  instead the updated state after each internal action is used to complete the states of the global trace.
\begin{definition}[Function refine $\mathcal{R}_\beta$]
\label{def:refine}
Function $\mathcal{R}_\beta : Q \cdot (\GActions \cdot Q)^* \longrightarrow Q \cdot (\IntActions \cdot Q)^*$ is defined as:
\begin{itemize}
\item[$\bullet$] $\mathcal{R}_\beta(\init)=\init$,
\item[$\bullet$] $\mathcal{R}_\beta(\sigma\cdot (\alpha \cup \beta)\cdot q)=\begin{dcases}
	\mathcal{R}_\beta(\sigma)\cdot \alpha \cdot q  &\text{ if } \beta = \emptyset, \\
   \Map \ [ x \mapsto \upd(q,x)] \ (\mathcal{R}_\beta(\sigma))  &\text{ if } \alpha = \emptyset, \\   
   \Map \ [ x \mapsto \upd(q,x)] \ (\mathcal{R}_\beta(\sigma)\cdot \alpha \cdot q) &\text{ otherwise;}
   \end{dcases}$
\end{itemize}
with $\upd: Q \times (Q \cup 2^\IntActions)\longrightarrow Q \cup 2^\IntActions$ defined as:
\begin{itemize}
\item[--] $\upd((q_1,\ldots,q_{|\setofcomponents|} ),\alpha)=\alpha $,
\item[--] $\upd\bigg((q_1,\ldots,q_{|\setofcomponents|}),(q'_1,\ldots ,q'_{|\setofcomponents|})\bigg)=(q''_1,\ldots ,q''_{|\setofcomponents|})$,\\
where $ \forall k \in [1 \upto {|\setofcomponents|}] \quant q''_k =
	\begin{dcases}
   q_k  &\text{ if } (q_k\notin Q_k^{\rm b}) \wedge (q'_k\in Q_k^{\rm b}) \\
   q'_k &\text{ otherwise.}
   \end{dcases}$
\end{itemize}
\end{definition}
Function $\mathcal{R}_\beta$ uses the (information in the) state after internal actions in order to update the partial states using function $\upd$.

By applying function $\mathcal{R}_\beta$ over the set of compatible partial traces $\mathcal{P}(t)$, we obtain a new set of compatible global traces  which is (i) equivalent to $\mathcal{P}(t)$, (ii) internal actions are discarded in the presentation of each  global trace and (iii) contains maximal global states that can be built with the information contained in the partial states observed so far.

A refined global trace $\mathcal{R}_\beta(t)$ is said to be equal with a path $\eta_0 \cdot \alpha_1 \cdot \eta_1 \cdot \alpha_2 \cdot \eta_2 \cdots  \alpha_z \cdot \eta_z$ if $\mathcal{R}_\beta(t)=  (\eta_0.\mathit{state}) \cdot \alpha_1 \cdot (\eta_1.\mathit{state}) \cdot \alpha_2 \cdot (\eta_2.\mathit{state}) \cdots  \alpha_z \cdot (\eta_z.\mathit{state}) $.
%
%%%%%%%%%%%%%%%%%%%%%%%%%%%%%%%%
%
\begin{example}[Applying function $\mathcal{R}_\beta$]
\label{xmp:r_beta}
By applying function $\mathcal{R}_\beta$ over the set of compatible partial traces in \ppexref{xmp:possible_global} we have the refined traces (compatible global traces):
\begin{itemize}
	\item $\mathcal{R}_\beta(t^1_1)= (d_1,d_2,d_3) \cdot \setof{\fil_{12}} \cdot (f_1, f_2, d_3) \cdot  \setof{\setof{\drain_1}, \setof{\fil_3}} \cdot (\bot, f_2 , \bot )$,
	\item $\mathcal{R}_\beta(t^2_1)= (d_1,d_2,d_3) \cdot \setof{\fil_{12}} \cdot (f_1, f_2, d_3) \cdot \setof{\setof{\drain_1}} \cdot (\bot, f_2, d_3) \cdot \setof{\setof{\fil_3}} \cdot (\bot, f_2 , \bot  ) $,
	\item $\mathcal{R}_\beta(t^3_1)= (d_1,d_2,d_3) \cdot \setof{\fil_{12}} \cdot (f_1, f_2, d_3) \cdot  \setof{\setof{\fil_3}} \cdot (f_1, f_2, \bot) \cdot \setof{\setof{\drain_1}} \cdot (\bot, f_2 , \bot  )$,
	\item $\mathcal{R}_\beta(t^4_1)= (d_1,d_2,d_3) \cdot \setof{\fil_{12},\setof{\fil_3}} \cdot (f_1, f_2, \bot)  \cdot \setof{\setof{\drain_1}} \cdot (\bot, f_2 , \bot  )$,
	\item $\mathcal{R}_\beta(t^5_1)= (d_1,d_2,d_3) \cdot \setof{\setof{\fil_3}} \cdot (d_1, d_2, \bot) \cdot \setof{\fil_{12}} \cdot (f_1, f_2, \bot)  \cdot \setof{\setof{\drain_1}} \cdot (\bot, f_2 , \bot  )$.
\end{itemize}
\end{example}
In \pdefref{def:local-traces} we defined $\setof{s_1(t),\ldots,s_m(t)}$, the set of observable local partial-traces of the schedulers obtained from partial trace $t$.
According to \ppdefref{def:events}, from each local partial-trace we can obtain  the sequences of events generated by the controller of each scheduler, such that the set of all the sequences of the events is
$\{\event(s_1(t)),$ $\ldots,\event(s_m(t))\}$ with $\event(s_j(t)) \in E^*$ for $j \in [1 \upto |\setofschedulers| ]$.

In the following, we define the set of all possible sequences of events that could be received by the observer. 
\begin{definition}[Events order]
Considering partial trace $t$, the set of all possible sequences of events that could be received by the observer is $\Theta(t)=\{\zeta \in E^* \mid \forall j\in [1 \upto |\setofschedulers| ]\quant \zeta\downarrow_{S_j}=\event(s_j(t)) \}$.
\end{definition}
Events are received by the observer in any order just under a condition in which the ordering of the local events of a scheduler is preserved.
\begin{proposition}[Soundness]
\label{proposition:make_soundness}
$\forall \zeta \in \Theta(t) , \forall \pi \in \Pi\big(\make\left(\zeta\right).\mathit{lattice}\big) , \forall j\in[1 \upto |\setofschedulers| ]  \quant  \pi\downarrow_{S_j} = \mathcal{R}_\beta(s_j(t))$.
\end{proposition}
\proporef{proposition:make_soundness} states that the projection of all paths in the lattice on a scheduler $S_j$ for $j\in[1 \upto |\setofschedulers| ]$ results in the refined local partial-trace of scheduler $S_j$.

The following proposition states the correctness of the construction in the sense that applying Algorithm~$\make$ over a sequence of observed events (\ie $\zeta \in \Theta$) at runtime, results in a computation lattice which encodes a set of the sequences of global states, such that each sequence represents a global trace of the system.
\begin{proposition}[Completeness]
\label{proposition:make_completeness}
Given a partial trace $t$ as per \ppdefref{def:global-trace}, we have 

$\forall \zeta \in \Theta(t) , \forall t' \in\mathcal{P}(t) , \exists ! \pi \in \Pi\bigg(\make\left(\zeta\right).\mathit{lattice}\bigg) \quant \pi = \mathcal{R}_\beta(t')$.

$\pi$ said to be the associated path of the compatible partial-trace $t'$.
\end{proposition}
Applying algorithm $\make$ over any of the sequence of events, constructs a computation lattice whose set of paths consists on all the compatible global traces. 
\input{proposition_example}
\begin{example}[Existence of the set of compatible global traces in  the constructed lattice]
Let us consider partial trace $t_1$ presented in \ppexref{exmp:global_trace} and the set of all associated event of  $t_1$ that is presented in \ppexref{xmp:seqevents}.
Events are received by the observer in order to make the lattice.
Figure~\ref{fig:latticet1}, illustrates the associated constructed computation lattice using algorithm $\make$ consists of 5 paths $\pi_1$ to $\pi_5$.
The set of compatible global traces (presented in \pexref{xmp:r_beta}) can be extracted from the reconstructed lattice, where $\pi_k = \mathcal{R}_\beta(t_1^k)$ for $k\in[1 \upto 5]$.
Paths $\pi_1$ to $\pi_5$ are associated paths of the compatible partial-traces $t_1^1$ to $t_1^5$ respectively.
\end{example}
%
%%%%%%%%%%%%%%%%%%%%%%
\section{LTL Runtime Verification by Progression on the Reconstructed Computation Lattice}
\label{sec:ltl}
%%%%%%%%%%%%%%%%%%%%%%
%

In this section, we address the problem of monitoring an LTL formula specifying the desired global behavior of the system.

In the usual case, evaluating whether an LTL formula holds requires the monitoring procedure to have access to the global state of the system.
In the previous section we introduced how to construct the computation lattice using the partially-ordered events.
Although by stabilizing the system to have the ready state of all components we could obtain a complete computation lattice using algorithm $\make$ (see~\psecref{subsec:algorithm}), that is the state of each node is a global state, instead, we propose an on-the-fly verification of an LTL property during the construction of computation lattice.

There are many approaches to monitor LTL formulas based on various finite-trace semantics (see~\cite{BauerLS10}).
One way of looking at the monitoring problem for some LTL formula $\varphi$ is described in~\cite{bauer12} based on formula rewriting, which is also known as formula progression, or just \textit{progression}.
Progression splits a formula into (i) a formula expressing what needs to be satisfied by the observed events so far and (ii) a new formula (referred to a \textit{future goal}), which has to be satisfied by the trace in the future.
We apply progression over a set of finite partial-traces, where each trace consists in a sequence of (possibly) partial states, encoded from the constructed computation lattice. 
An important advantage of this technique is that it often detects when a formula is violated or validated before the end of the execution trace, that is, when the constructed lattice is not complete, so it is suitable for online monitoring.

To monitor the execution of a distributed CBS with respect to an LTL property $\varphi$, we introduce a more informative computation lattice by attaching to the each node of lattice $\mathcal{L}$ a set of formula.
Given a computation lattice $\mathcal{L} = \left( N,\xrightarrowtail[] \right)$ (as per \pdefref{def:computation_lattice}), we define 
an augmented computation lattice $\mathcal{L}^\varphi$ as follow.
\begin{definition}[Computation lattice augmentation]
\label{def:lattice-agumentation}
$\mathcal{L}^\varphi$ is a pair  $\left( N^\varphi ,\xrightarrowtail[] \right)$, where $N^\varphi \subseteq Q^l \times \VC \times 2^{\LTL}$ is the set of nodes augmented by $2^{\LTL}$, that is the set of LTL formulas.
The initial node is $\init^{\varphi}_{\mathcal{L}}= (\init,(0,\ldots,0),\{\varphi\})$ with $\varphi \in \LTL$  the global desired property. 
\end{definition}
In the newly defined computation lattice, a set of LTL formulas is attached to each node. 
The set of formulas attached to a node represents the different evaluation of the property $\varphi$ with respect to different possible paths form the initial node to the node.
The state and the vector clock associated to each node and the happened-before relation are defined similar to the initial definition of computation lattice (see~\pdefref{def:computation_lattice}).

The construction of the augmented computation lattice requires some modifications to algorithm $\make$:
\begin{itemize}
	\item Lattice $\mathcal{L}^\varphi$ initially has node $\init^{\varphi}_{\mathcal{L}}= (\init,(0,\ldots,0),\{\varphi\})$.
	\item The creation of a new node $\eta$ in the lattice with $\eta.\mathit{state}=q$ and $\eta.\mathit{clock}=\vc$, calculates the set of formulas $\Sigma$ associated to $\eta$ using the progression function (see \defref{def:progression}).
	The augmented node is $\eta= ( q, \vc, \Sigma )$, where $\Sigma = \{ \prog(\LTL',q) \mid   \LTL' \in \eta'.\Sigma \wedge ( \eta' \xrightarrowtail[] \eta \vee \exists N \subseteq \mathcal{L}^\varphi.\mathit{nodes} \quant \eta'=\meet(N,\mathcal{L}) \wedge \eta =\joint(N,\mathcal{L}) )\}$.
	We denote the set of formulas of node $\eta \in \mathcal{L}^\varphi.\mathit{nodes}$ by $\eta.\Sigma$.
	%
%	The set of formulas represents the different evaluation of the property $\varphi$ with respect to different possible paths form the initial node to node $\eta$.
	%
	\item Updating node $\eta=( q, \vc, \Sigma )$ by   update event $e = (\beta_i,q_i)\in E_\upbeta, i\in [1 \upto |\setofcomponents|]$ which is sent by scheduler $S_j, j\in[1\upto |\setofschedulers|]$ updates all associated formulas $\Sigma$ to $\Sigma'$ using the update function (see \pdefref{def:update}), where
$\Sigma' = \setof{ \updfct_\varphi (\LTL,q_i,j) \mid \LTL \in \Sigma }$.
\end{itemize} 
\begin{definition}[Progression function]
\label{def:progression}
$\prog: \LTL \times Q^l \longrightarrow \LTL $ is defined using a pattern-matching with $p\in AP_{i\in[1\upto |\setofcomponents|]}$ and $q=(q_1,\dots,q_{|\setofcomponents|})\in Q^l$.
\[
\begin{array}{rcl}
 \prog(\varphi, q )  & = & \mathtt{match} (\varphi) \;\mathtt{with}\\
        & \mid &  p\in AP_{i\in[1\upto |\setofcomponents|]} \rightarrow \begin{dcases}   
T &\text{ if } q_i \in Q^{\rm r}_i \wedge p\in q_i \\
F &\text{ if } q_i \in Q^{\rm r}_i \wedge p\not\in q_i\\
\textbf{X}^k_\beta p  &\text{ otherwise }(q_i = \bot^k_i, k \in[1\upto |\setofschedulers|])\\
   \end{dcases} \\
        & \mid &  \textbf{X}^k_\beta p \rightarrow 
        \textbf{X}^k_\beta p\\
        & \mid &  \varphi_1 \vee \varphi_2 \rightarrow \prog(\varphi_1, q)\vee \prog(\varphi_2,q)\\
        & \mid &  \varphi_1 \textbf{U} \varphi_2 \rightarrow \prog(\varphi_2, q)\vee \prog(\varphi_1,q) \wedge \varphi_1 \textbf{U} \varphi_2\\        
		& \mid &  \textbf{G} \varphi \rightarrow \prog(\varphi, q)\wedge \textbf{G}\varphi\\		
		& \mid & \textbf{F} \varphi \rightarrow \prog(\varphi, q)\vee \textbf{F}\varphi\\		
		& \mid &  \textbf{X} \varphi \rightarrow \varphi\\		
		& \mid &  \neg \varphi \rightarrow \neg \prog(\varphi, q)\\	
        & \mid &  T \rightarrow T
\end{array}
\]
We define a new modality $\textbf{X}_\beta$ such that $\textbf{X}^k_\beta p$ for $p\in AP_{i\in[1\upto |\setofcomponents|]}$ and $k \in[1\upto |\setofschedulers|]$ means that atomic proposition $p$ has to hold at next ready state of component $B_i$ which is sent by scheduler $S_k$.
For a sequence of partial states obtained at runtime $ \sigma=q_0\cdot q_1 \cdot q_2 \cdots$ such that  $\sigma_j=q_j $, we have  $\sigma_j \models \textbf{X}^k_\beta p \Leftrightarrow \sigma_z \models p$ where  $z=\min\left(\setof{r>j \lmid \left( \sigma_{r-1}\downarrow_{S_k} \right) \goesto[\beta_i]_{S_k} \left( \sigma_r\downarrow_{S_k} \right) } \right)$.
\end{definition}
The truth value of the progression of an atomic proposition $p\in AP_i$ for $i \in [1\upto |\setofcomponents|]$ with a partial state  $q=(q_1,\dots,q_{|\setofcomponents|})$ is evaluated by true (\resp false) if the state of component $B_i$ (that is $q_i$) is a ready state and satisfies (\resp does not satisfy) the atomic proposition $p$.
If the state of component $B_i$ is not a ready state, the evaluation of the atomic proposition $p$ is postponed to the next ready state of component $B_i$.
\begin{definition}[Formula update function]
\label{def:update}
$\updfct_\varphi: \LTL \times \setof{Q^{\rm r}_i}^{|\setofcomponents|}_{i=1}  \times [1\upto |\setofschedulers|] \rightarrow \LTL$ is defined using a pattern-matching with $q_i \in Q^{\rm r}_i$ for $i \in [1\upto |\setofcomponents|]$.
\[
\begin{array}{rcl}
 \updfct_\varphi(\varphi, q_i, j)  & = & \mathtt{match} (\varphi) \;\mathtt{with}\\
        & \mid &  \textbf{X}^k_\beta p \rightarrow \begin{dcases}
T & \text{ if }  p \in \AP_i \cap q_i \wedge k=j\\
F & \text{ if }  p\in \AP_i \cap \overline{q_i} \wedge k=j\\
\textbf{X}^k_\beta p &\text{ otherwise } (p\not\in AP_i \vee k\neq j)
   \end{dcases} \\
        & \mid &  \varphi_1 \vee \varphi_2 \rightarrow \updfct_\varphi(\varphi_1,q_i)\vee \updfct_\varphi(\varphi_2,q_i)\\
        & \mid &  \varphi_1 \wedge\varphi_2 \rightarrow \updfct_\varphi(\varphi_1,q_i)\wedge \updfct_\varphi(\varphi_2,q_i)\\        
        & \mid & \varphi_1 \textbf{U} \varphi_2 \rightarrow \updfct_\varphi(\varphi_1, q_i) \textbf{U} \updfct_\varphi(\varphi_2,q_i)\\                
		& \mid &  \textbf{G} \varphi \rightarrow \textbf{G} \updfct_\varphi(\varphi, q_i)\\				
		& \mid & \textbf{F} \varphi \rightarrow \textbf{F}\updfct_\varphi(\varphi, q_i)\\				
		& \mid &  \textbf{X} \varphi \rightarrow \textbf{X}\updfct_\varphi(\varphi, q_i)\\				
		& \mid &  \neg \varphi \rightarrow  \neg \updfct_\varphi(\varphi ,q_i)\\	
        & \mid &  T \rightarrow T\\
        & \mid &  p\in AP_{ i \in[1\upto |\setofcomponents|] } \rightarrow p
\end{array}
\]
\end{definition}
Update function updates a progressed LTL formula with respect to a ready state of a component.
Intuitively, a formula consists in an atomic proposition  whose truth or falsity depends on the next ready state of component $B_i$ sent by scheduler $S_k$, that is  $\textbf{X}^k_\beta p$ where $p\in \AP_i$, can be evaluated using update function by taking the first ready state of component $B_i$ received from scheduler $S_k$ after the formula rewrote to $\textbf{X}^k_\beta p$.
\input{table}
\begin{example}[Formula progression and formula update over an augmented computation lattice] 
Let consider the system presented in \ppexref{xmp:dmodel} with partial trace $t_2$ and the associated sequence of events presented in \ppexref{xmp:seqevents} and desired property $\varphi = \textbf{G} (d_3 \vee f_1)$.
$\mathcal{L}^\varphi$ initially has node $\init^{\varphi}_{\mathcal{L}}= ((d_1,d_2,d_3),(0,0),\{\varphi\})$.
By observing the action event $(\fil_{12},(1,0))$, algorithm $\make$ creates new node $\eta_1 = ((\bot, \bot, d_3),(1,0), \setof{\varphi })$, because 
$\prog(\textbf{G} (d_3 \vee f_1),(\bot, \bot,d_3))=\prog((d_3 \vee f_1),(\bot, \bot,d_3)) \wedge \textbf{G} (d_3 \vee f_1) = T \wedge \textbf{G} (d_3 \vee f_1) = \textbf{G} (d_3 \vee f_1) = \varphi$.

By observing the action event $(\fil_3,(0,1))$, new node $\eta_2 = ((d_1, d_2, \bot),(0,1), \setof{\textbf{X}_\beta d_3 \wedge \varphi })$ is created, because
$\prog(\textbf{G} (d_3 \vee f_1),(d_1, d_2,\bot))=\prog((d_3 \vee f_1),(d_1, d_2,\bot)) \wedge \textbf{G} (d_3 \vee f_1) = \textbf{X}_\beta d_3 \wedge \textbf{G} (d_3 \vee f_1)= \textbf{X}_\beta d_3 \wedge \varphi $.
Formula $\textbf{X}_\beta d_3$ means that the evaluation of partial state $(d_1, d_2,\bot)$ with respect to formula $(d_3 \vee f_1)$ is on hold until the next ready state of component $\tank_3$.

Consequently joint node 
$\eta_3 = ((\bot, \bot, \bot),(1,1), \{ (\textbf{X}_\beta d_3) \wedge (\textbf{X}_\beta d_3 \vee \textbf{X}_\beta f_1)  \wedge \varphi, (\textbf{X}_\beta d_3 \vee \textbf{X}_\beta f_1)\wedge \varphi, (\textbf{X}_\beta d_3 \vee \textbf{X}_\beta f_1)\wedge \varphi\})$ is made.
The update event $(\beta_3,f_3)$ updates both state and the set of formulas associated to each node as follows.
Although $\init^{\varphi}_{\mathcal{L}}$ and $\eta_1$ remain intact, but node $\eta_2$ is updated to $((d_1, d_2, f_3),(0,1), \setof{ F })$ because
$\updfct_\varphi(\textbf{X}_\beta d_3 \wedge \varphi,f_3)=F$. 
Moreover, node $\eta_3$ is updated to $\eta_3 = ((\bot, \bot, f_3),(1,1), \setof{ F , ( \textbf{X}_\beta f_1)\wedge \varphi, ( \textbf{X}_\beta f_1)\wedge \varphi})$.

The update event $(\beta_2,f_2)$  updates nodes $\eta_1$ and $\eta_3$ such that $\eta_1 = ((\bot, f_2, d_3), (1,0), \{\varphi \})$ and 
$\eta_3 = ((\bot, f_2, f_3),$ $(1,1), \setof{ F , ( \textbf{X}_\beta f_1)\wedge \varphi , ( \textbf{X}_\beta f_1)\wedge \varphi})$.

By observing the action event $(\drain_{23},(1,2))$, the new node $\eta_4 = ((\bot, \bot, \bot),(1,2), \{F ,(\textbf{X}_\beta d_3 \vee \textbf{X}_\beta f_1) \wedge ( \textbf{X}_\beta f_1)\wedge \varphi,(\textbf{X}_\beta d_3 \vee \textbf{X}_\beta f_1) \wedge ( \textbf{X}_\beta f_1)\wedge \varphi\})$ is created.

The update event $(\beta_1,f_1)$ updates nodes $\eta_1$, $\eta_3$ and $\eta_4$ such that $\eta_1 = ((f_1, f_2, d_3),(1,0), \setof{\varphi })$,  $\eta_3 = ((f_1, f_2, f_3),$ $(1,1), \setof{ F ,  \varphi,  \varphi})$ and $\eta_4 = ((f_1, \bot, \bot),(1,2), \setof{F ,  \varphi,  \varphi})$.

Table~\ref{table:examp} shows the step-by-step reconstructing and monitoring of the associated computation lattice.
The highlighted nodes are the removed nodes using  \ppalgoref{alg:removeextranodes}, but for the sake of better understanding we show them.
\end{example}
%   
%
%%%%%%%%%%%%%%%%%%%%%%%%%%%%%%%%%%%%%%%%%%%%%%%%%%%%%%%%%
\subsection{Correctness of Formula Progression on the Lattice}
%%%%%%%%%%%%%%%%%%%%%%%%%%%%%%%%%%%%%%%%%%%%%%%%%%%%%%%%%
%
In \ppsecref{sec:lattice}, we introduced how from an unobservable partial trace $t$ of a distributed CBS one can construct a set of paths representing the set of compatible global-traces of $t$ in form of a lattice.
Furthermore, in \ppsecref{sec:ltl} we adapted formula progression over the constructed lattice with respect to a given LTL formula $\varphi$.
What we obtained is a directed lattice $\mathcal{L}^\varphi$ starting from the initial node and ending up with frontier node $\eta^f$.
The set of formulas attached to the frontier node, that is $\eta^f.\Sigma$, represents the progression of the initial formula over the set of path of the lattice.   
\begin{definition}[Progression on a partial trace]
\label{def:bigprog}
Function $\PROG :  \LTL \times Q \cdot (\GActions \cdot Q)^*  \rightarrow \LTL $  is defined as:
\begin{itemize}
	\item
	$\PROG(\varphi, \init) = \varphi$,
	\item
	$\PROG(\varphi, \sigma)= \varphi'$
	\item
	$\PROG\left(\varphi, \sigma \cdot \left(\alpha \cup \beta\right)\cdot q\right)= \prog \left( \UPD \left(\PROG \left( \varphi , \sigma \right), \mathcal{Q} \right), q \right)$ where
	\begin{itemize}
		\item $\mathcal{Q} = \setof{q[i] \mid \beta_i \in \beta}$ is the set of updated states,
		\item function $\UPD: \LTL \times Q^R   \rightarrow \LTL$  is defined as:  
		\begin{itemize}
			\item $ \UPD \left( \varphi, \setof{\epsilon} \right)= \varphi$,
			\item $ \UPD \left( \varphi, Q^r \cup \setof{q_i} \right) = \updfct_\varphi \left(\UPD \left( \varphi, Q^r \right),q_i \right)$.
		\end{itemize}
	with $Q^R \subseteq \setof{ q \in Q^{\rm r}_i \lmid i \in [1\upto |\setofcomponents|]}$ the set of subsets of ready states of the components.
	\end{itemize}	
\end{itemize}
\end{definition}
Function $\PROG$ uses functions $\prog$ (\pdefref{def:progression}), $\updfct_\varphi$ (\pdefref{def:update}) and function $\UPD$.
Since after each global action in the partial trace we reach a partial state(see \pdefref{def:global-trace}), function $\updfct_\varphi$ does not need to check among multiple partial states to find whose formula must be updated. 
That is why $\PROG$ uses the simplified version of function $\updfct_\varphi$ by eliminating the scheduler index input. 
Moreover functions $\prog$ modified in such way to take as input a partial state in $Q$ instead of a state in $Q^l$ because as we above mentioned, the index of schedulers does not play a role in the progression of an LTL formula on a partial trace. 

Given a partial state $t$ as per \ppdefref{def:global-trace} and an LTL property $\varphi$, by $\eta^f.\Sigma$ we denote the set of LTL formulas of the frontier node of the constructed computation lattice $\mathcal{L}^\varphi$, we have the two following proposition and theorems:
\input{overview}
\begin{proposition}
\label{proposition:prog}
Given an LTL formula $\varphi$ and  a partial trace $t$, there exists a partial trace $t'$ such that $\PROG(\varphi, t') =\mathit{progression} (\varphi, \mathcal{R}_\beta(t'))$ with:

\[
t' = \begin{dcases}
    t  &\text{ if } \last(t)[i] \in Q_i^{\rm r} \text{ for all } i \in [1\upto |\setofcomponents|], \\
   t \cdot \beta \cdot q  &\text{ otherwise.}
   \end{dcases}
\]
%
%
%
%
%\[
%\PROG(\varphi, t) = \begin{dcases}
%    \mathit{progression} (\varphi, \mathcal{R}_\beta(t))  &\text{ if } \last(t)[i] \in Q_i^{\rm r} \text{ for all } i \in [1\upto |\setofcomponents|], \\
%   \mathit{progression} (\varphi, \mathcal{R}_\beta(t'))  &\text{ otherwise.}
%   \end{dcases}
%\]
%
%
%For a global trace $t$ there exists a global trace $t'$ such that $\PROG(\varphi, t') =\mathit{progression} (\varphi, \mathcal{R}_\beta(t'))$ where 
%$
%t' = \begin{dcases}
%    t  &\text{ if } \last(t)[i] \in Q_i^{\rm r} \text{ for all } i \in [1\upto |\setofcomponents|], \\
%   t \cdot \beta \cdot q  &\text{ otherwise.}
%   \end{dcases}
%$
%
%
%
Where $ \beta \subseteq \bigcup_{i \in [1\upto |\setofcomponents|]} \setof{ \beta_i }$, $\forall i \in [1\upto |\setofcomponents|], q[i] \in Q_i^{\rm r}$ and $\mathit{progression}$ is the standard progression function on a global trace as described in~\cite{bauer12}.
\end{proposition}
Proposition~\ref{proposition:prog} states that progression of an LTL formula on a partial trace of a distributed system (as per~\pdefref{def:global-trace}) using $\PROG$ results similar to the standard progression of the LTL formula on the corresponding refined global trace using $\mathit{progression}$ if we allow the system to be stabilized by the execution of $\beta$ actions of busy components.
Intuitively, our progression method over on a trace of a distributed system follows the standard progression technique on a trace of a sequential system where the global state of the system is always defined.
\begin{theorem}[Soundness]
\label{theorem:soundness}
For a partial trace $t$ and LTL formula $\varphi$, we have   
\[
\forall \varphi' \in \eta^f.\Sigma , \exists t' \in \mathcal{P}(t) \quant \PROG(\varphi, t')= \varphi'.
\]
\end{theorem}
Theorem~\ref{theorem:soundness} states that each formula of the frontier node is derived from the progression of formula $\varphi$ on a compatible partial-trace of $t$.
\begin{theorem}[Completeness]
\label{theorem:completeness}
For a partial trace $t$ and an LTL formula $\varphi$, we have:
\[
\eta^f.\Sigma = \setof{\PROG(\varphi, t') \lmid t' \in \mathcal{P}(t) }.
\]
\end{theorem}
Theorem~\ref{theorem:completeness} states that the set of formulas in the frontier node is equal to the set of progression of $\varphi$ on all the compatible partial-traces of $t$.

\ppfigref{fig:over-view-ltl} depicts our monitoring approach for a distributed CBS.

%% file: make.tex
\algnewcommand\algorithmicinput{\textbf{Global variables:}}
\algnewcommand\INPUT{\item[\algorithmicinput]}
\begin{algorithm}[t]
\caption{$\make$}
\label{alg:make}
\begin{algorithmic}[1]
%\INPUT{$\mathcal{L},\kappa, V$}
\INPUT{$\mathcal{L}$ initialized to $\init_{\mathcal{L}}$,

$\qquad \qquad \ \kappa$ initialized to $\epsilon$,

$\qquad \qquad \ V$ initialized to $(0,\ldots,0)$.}
%\INPUT{$\kappa$ initialized to an empty sequence}
%\INPUT{$V$ initialized to an empty sequence}
\Procedure{Make}{$e, \fromthequeue$}
\If {$e \in E_{\rm a}$} \label{line:start_action_event}
\Comment{if $e$ is an action event.}
	\State $\textsc{ActionEvent}	(e,\fromthequeue)$
\ElsIf {$e \in E_{\upbeta}$} \label{line:start_update_event}
\Comment{if $e$ is an update event.}
	\State $\textsc{UpdateEvent}	(e,\fromthequeue)$
\EndIf \label{line:end_update_event}
\EndProcedure
\end{algorithmic}
\end{algorithm}

%% file: actionevent.tex
\begin{algorithm}
\caption{\textsc{ActionEvent}}
\label{alg:actionevent}
\begin{algorithmic}[1]
\Procedure{ActionEvent}{$e, \fromthequeue$}
\State{$\latticextend \gets \false$}
\For {all $\eta \in \mathcal{L}.\mathit{nodes}$}	
		\If {$\exists \eta' \in Q^l \times \VC : \eta' = \upda(\eta,e)$}
			\State { $\mathcal{L}.nodes \gets \mathcal{L}.\mathit{nodes} \cup  \setof{\eta'}$}
			\Comment{extend the lattice with the new node.}
			\State $\textsc{ModifyQueue}(e,\fromthequeue,\true)$
			\Comment{event $e$ is removed from the queue if it was picked up from the queue.}
			\State{$\latticextend \gets \true$}
			\State {\textbf{break}}
			\Comment{stop iteration  when the lattice is extended (Property~\ref{property:ext}).}
		\EndIf
	\EndFor
	\If {$\neg \latticextend$}
%		\State {$\kappa \gets \kappa \cdot e$}
%		\Comment{event $e$ is added to the queue if it can't extend the lattice.}
		\State $\textsc{ModifyQueue}(e,\fromthequeue,\false)$
		\Comment{event $e$ is added to the queue if it was not picked up from the queue.}
		\State {\textbf{return}}
	\EndIf
	\State $\textsc{Joints(~)}$ \label{line:make_joints}
	\Comment{extend the lattice with joint nodes.}
	\State $\textsc{RemoveExtraNodes}()$
	\Comment{lattice size reduction.}
%	\State $\textsc{UpdateV}\textsc{(}e\textsc{)}$ \label{line:global_variables_update}
%	\Comment{update global variables based on action event $e$.}
	\If {$\neg \fromthequeue$}
	\label{line:if_not_from_the_queue}
		\State $\textsc{CheckQueue(~)}	$ \label{line:check_queue}
		\Comment{recall the events stored in the queue.}
%	\Else 
%		\State {\textbf{return}}
%		\label{line:return_if_from_queue}
%		\Comment{return if event $e$ can not extend the lattice.}	
	\EndIf
\EndProcedure
\end{algorithmic}
\end{algorithm}

%% file: updateevent.tex
\begin{algorithm}[t]
\caption{\textsc{UpdateEvent}}
\label{alg:updateevent}
\begin{algorithmic}[1]
\Procedure{UpdateEvent}{$e, \fromthequeue$}
	\For {all $e' \in \kappa  $}	
		\If {$e' \in E_{\rm a} \wedge e.\mathit{index} \in \involved(e'.\mathit{action})$}
		\Comment{check if there exists an action event in the queue concerning component $B_{e.\mathit{index}}$.}
		\State $\textsc{ModifyQueue}	(e,\fromthequeue,\false)$
		\Comment{event $e$ is added to the queue if it was not picked up from the queue.}
%			\State {$\kappa \gets \kappa \cdot e$}
%			\Comment{event is added to the queue if the update condition does not hold.}
			\State {\textbf{return}}			
		\EndIf
	\EndFor
%	\If {$B_{e.index} \in B_{\rm s}$} 
		%\State {$e.\mathit{clock} \gets \vc_{e.\mathit{sender}}^{e.\mathit{index}}$}
	%\Comment{if the event is associated to a shared component.}
	%\State $\textsc{SharedComponentEvent}	(e)$
	%\label{line:shared_component_event}
	%\Else 
	\For {all $\eta \in \mathcal{L}.\mathit{nodes}$}	\label{line:start_update_nodes}
		\State {$\eta \gets \updb(\eta,e)$}
		\Comment {update nodes according to Definition~\ref{def:node_update}.}
	\EndFor		\label{line:end_update_nodes}
	%\EndIf
	\State $\textsc{ModifyQueue}	(e,\fromthequeue,\true)$
\EndProcedure
\end{algorithmic}
\end{algorithm}

%% file: modifyqueue.tex
\begin{algorithm}[t]
\caption{\textsc{ModifyQueue}}
\label{alg:modifyqueue}
\begin{algorithmic}[1]
\Procedure{ModifyQueue}{$e,\fromthequeue,\eventused$}
			\If {$\fromthequeue \wedge \eventused$}
				\State {$\kappa \gets \rem(\kappa,e)$}
				\label{line:remove_queue}
				\Comment{event $e$ is removed from the queue if it is picked from queue and used.}
			\ElsIf {$\neg \fromthequeue \wedge \neg \eventused$}
				\State {$\kappa \gets \kappa \cdot e$}
				\label{line:add_queue}
				\Comment{event $e$ is added to the queue if it is not picked from queue and could not be used.}
			\EndIf
\EndProcedure
\end{algorithmic}
\end{algorithm}

%% file: joints.tex
\begin{algorithm}[t]
\caption{\textsc{Joints}}
\label{alg:joints}
\begin{algorithmic}[1]
\Procedure{Joints}{}
    \State $\mathcal{J}_{\mathcal{L}}\gets\textsc{Jcompute}$       
	\Comment{compute the pairs of the vector clocks of the nodes which are in $\mathcal{J}_{\mathcal{L}}$.}
	\While {$\mathcal{J}_{\mathcal{L}} \neq \emptyset$ }
		\For {all $\eta,\eta' \in\mathcal{L}.\mathit{nodes} $ such that $ (\eta.\mathit{clock},\eta'.\mathit{clock}) \in \mathcal{J}_{\mathcal{L}}$}	
		\State {$\mathcal{L}.\mathit{nodes} \gets \mathcal{L}.\mathit{nodes} \cup \setof{\joint(\eta,\eta',\mathcal{L})}$}
		\Comment{extend the lattice with the new joint node.}
		\State {$\mathcal{J}_{\mathcal{L}} \gets \mathcal{J}_{\mathcal{L}} \setminus \{(\eta.\mathit{clock},\eta'.\mathit{clock})\}$}
		\EndFor
   		\State $\mathcal{J}_{\mathcal{L}}\gets\textsc{Jcompute}$       
	\EndWhile	
	%\State \textbf{return} $\mathcal{L}$
\EndProcedure
\end{algorithmic}
\end{algorithm}

%% file: jcompute.tex
\begin{algorithm}
\caption{\textsc{Jcompute}}
\label{alg:jcompute}
\begin{algorithmic}[1]
\Procedure{Jcompute}{}
		\For {all $\eta,\eta',\eta'' \in\mathcal{L}.\mathit{nodes}$}	
			\If {$\eta''\rightarrowtail \eta \wedge \eta''\rightarrowtail \eta'$} 	
			\Comment{if $\eta$ and $\eta'$ are associated to two concurrent events.}		
				\State {$\mathcal{J}_{\mathcal{L}}=\mathcal{J}_{\mathcal{L}} \cup \setof{(\eta.\mathit{clock},\eta'.\mathit{clock})}$}
				\Comment{$\eta.\mathit{clock}$ and $\eta'.\mathit{clock}$ are added to relation $\mathcal{J}_{\mathcal{L}}$.}
			\EndIf
		\EndFor 
		\State {\textbf{return} $\mathcal{J}_{\mathcal{L}}$}  
\EndProcedure
\end{algorithmic}
\end{algorithm}

%% file: checkqueue.tex
\begin{algorithm}[t]
\caption{\textsc{CheckQueue}}
\label{alg:checkqueue}
\begin{algorithmic}[1]
\Procedure{CheckQueue}{}
%	\State $\textsc{ModifyQueue}	(e,\kappa,\fromthequeue,\false)$
	\While {$\true $ } \label{line:while_update}
		\State {$\kappa' \gets \kappa$}
		\label{line:queue_copy}
		\For {all $z\in [1,\length(\kappa)]$}	
			\State {$\make(\kappa(z),\true)$}
			\Comment{recall the events of the queue.}
		\EndFor 
		%\State {$\fromthequeue \gets \false$}
		\If {$\kappa=\kappa'$} \label{line:start_check_equality}
			\State {\textbf{break}}	\label{line:break}		
			\Comment{break if none of the events in the queue is used.}
		\EndIf \label{line:end_check_equality}
	\EndWhile 
\EndProcedure
\end{algorithmic}
\end{algorithm}

%% file: removeextranodes.tex
\begin{algorithm}
\caption{\textsc{RemoveExtraNodes}}
\label{alg:removeextranodes}
\begin{algorithmic}[1]
\Procedure{RemoveExtraNodes}{}
	\For {all $\eta \in \mathcal{L}.\mathit{nodes}$} 
		\If {$\forall j \in [1,m], \exists \eta'\in \mathcal{L}.\mathit{nodes}: \eta'.\mathit{clock}[j] > \eta.\mathit{clock}[j]$}
		\Comment{if there exists a node with a strictly greater clocks in the vector clock.}
			\State {$\rem(\mathcal{L}.\mathit{nodes}, \eta)$}
			\Comment{the node with the smaller vector clock is removed.}
		\EndIf
	\EndFor
\EndProcedure
\end{algorithmic}
\end{algorithm}

%% file: lattice.tex
%\begin{figure}[t]
%	\centering
%    \scalebox{.6} {
%\begin{tikzpicture}
%  \node (init) at (0,-2) {$((d_1,d_2,d_3),(0,0))$};
%  \node (f12) at (2,0) {$((f_1,f_2,d_3),(1,0))$};
%  \node (f3) at (-2,0) {$((d_1,d_2,f_3),(0,1))$};
%  \node (a) at (0,2) {$((f_1,f_2,f_3),(1,1))$};
%  \node (b) at (-2,4) {$((f_1,\bot_2^2,\bot_3^2),(1,2))$};
%  \node (c) at (-4,2) [fill=red]{$(?,(0,2))$};
%  
%%  \draw[->] (init) -- node [below right] {$f_{12}$} (f12); 
%%  \draw[->] (init) -- node [below left] {$f_3$} (f3); 
%%  \draw[->] (a) -- node [above right] {$d_{23}$} (b);
%%  \draw[dashed,->] (f12) -- node [above right] {$f_3$} (a);
%%  \draw[dashed,->] (f3) -- node [above left] {$f_{12}$}(a); 
%  
%  \draw[->] (init) --  (f12); 
%  \draw[->] (init) -- (f3); 
%  \draw[->] (a) --  (b);
%  \draw[dashed,->] (f12) --  (a);
%  \draw[dashed,->] (f3) -- (a);
%   
%  \draw[->,red] (f3) -- (c);
%  \draw[->,red] (c) -- (b);
%  
% % \draw[preaction={draw=white, -,line width=6pt}] (f) -- (d) -- (b);
%\end{tikzpicture}
%}
%\caption{Computation lattice associated to trace $t_2$ in Example~\ref{xmp:dmodel}}
%\label{fig:lattice}
%\end{figure}
%
%
%
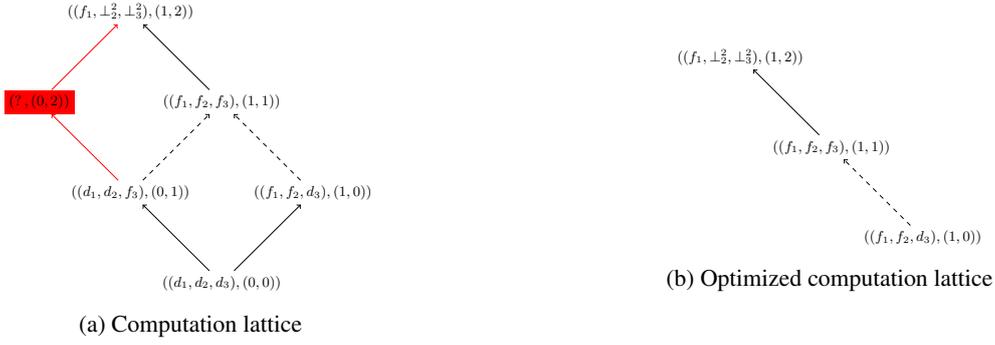
\begin{figure}[t]
    \centering
     	\begin{subfigure}{0.49\textwidth}
     	\centering
			    \scalebox{.6} {
\begin{tikzpicture}
  \node (init) at (0,-2) {$((d_1,d_2,d_3),(0,0))$};
  \node (f12) at (2,0) {$((f_1,f_2,d_3),(1,0))$};
  \node (f3) at (-2,0) {$((d_1,d_2,f_3),(0,1))$};
  \node (a) at (0,2) {$((f_1,f_2,f_3),(1,1))$};
  \node (b) at (-2,4) {$((f_1,\bot_2^2,\bot_3^2),(1,2))$};
  \node (c) at (-4,2) [fill=red]{$(?,(0,2))$};

  \draw[->] (init) --  (f12); 
  \draw[->] (init) -- (f3); 
  \draw[->] (a) --  (b);
  \draw[dashed,->] (f12) --  (a);
  \draw[dashed,->] (f3) -- (a);
   
  \draw[->,red] (f3) -- (c);
  \draw[->,red] (c) -- (b);
\end{tikzpicture}
}
\caption{Computation lattice}
\label{fig:lattice}
		\end{subfigure}
     	\begin{subfigure}{0.49\textwidth}
     	\centering
			    \scalebox{.6} {
\begin{tikzpicture}
  \node (f12) at (2,0) {$((f_1,f_2,d_3),(1,0))$};
  \node (a) at (0,2) {$((f_1,f_2,f_3),(1,1))$};
  \node (b) at (-2,4) {$((f_1,\bot_2^2,\bot_3^2),(1,2))$};
  \draw[->] (a) --  (b);
  \draw[dashed,->] (f12) --  (a);
\end{tikzpicture}
}
\caption{Optimized computation lattice}
\label{fig:optlattice}
		\end{subfigure}
		\caption{Computation lattice associated to trace $t_2$ in Example~\ref{xmp:dmodel}}
\label{fig:lattice2}
		\end{figure}

%% file: proposition_example.tex
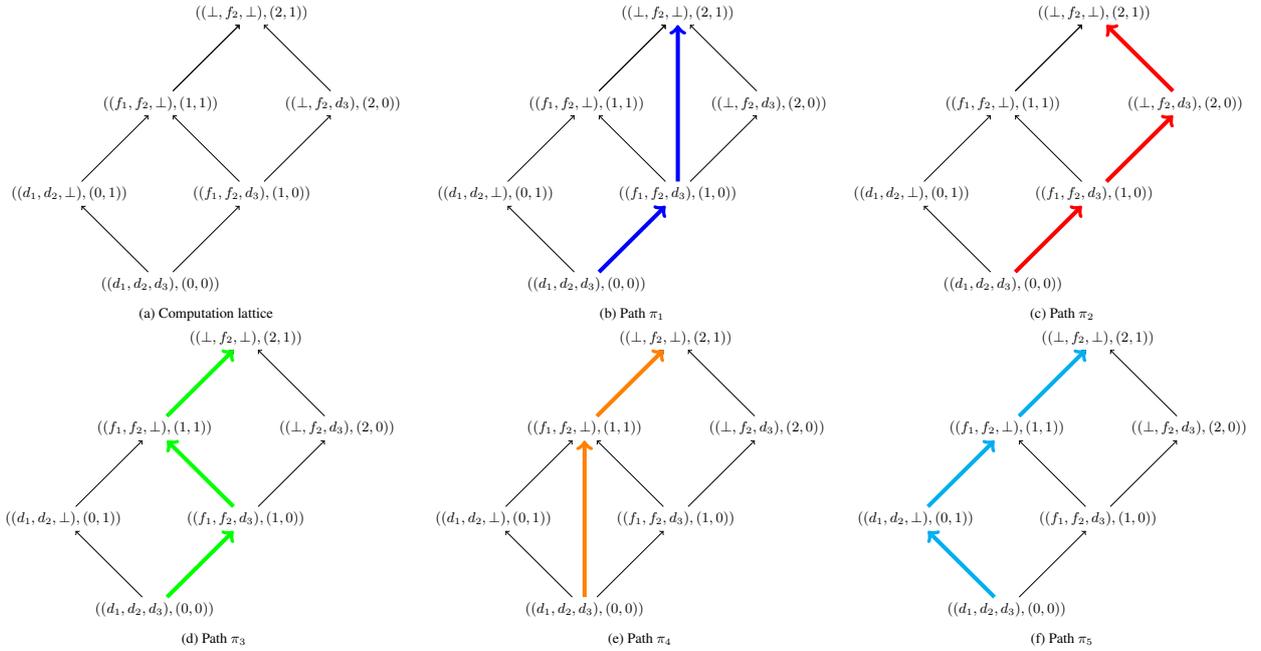
\begin{figure}[t]
	\centering
    \scalebox{0.6} {
    \centering
     	\begin{subfigure}{0.54\textwidth}
     	\centering
			\begin{tikzpicture}
  \node (init) at (0,-2) {$((d_1,d_2,d_3),(0,0))$};
  \node (f12) at (2,0) {$((f_1,f_2,d_3),(1,0))$};
  \node (f3) at (-2,0) {$((d_1,d_2,\bot),(0,1))$};
  \node (d1) at (4,2) {$((\bot,f_2,d_3),(2,0))$};
  \node (c) at (2,4) {$((\bot,f_2,\bot),(2,1))$};
  \node (a) at (0,2) {$((f_1,f_2,\bot),(1,1))$};  
  \draw[->] (init) --  (f12); 
  \draw[->] (init) -- (f3); 
  \draw[->] (f12) -- (d1);
  \draw[->] (a) --  (c);
  \draw[->] (f12) --  (a);
  \draw[->] (f3) -- (a);
  \draw[->] (d1) --  (c);
  \draw[->] (a) -- (c);
\end{tikzpicture}
\caption{Computation lattice}
		\end{subfigure}
     	\begin{subfigure}{0.55\textwidth}
     	\centering
			\begin{tikzpicture}
  \node (init) at (0,-2) {$((d_1,d_2,d_3),(0,0))$};
  \node (f12) at (2,0) {$((f_1,f_2,d_3),(1,0))$};
  \node (f3) at (-2,0) {$((d_1,d_2,\bot),(0,1))$};
  \node (d1) at (4,2) {$((\bot,f_2,d_3),(2,0))$};
  \node (c) at (2,4) {$((\bot,f_2,\bot),(2,1))$};
  \node (a) at (0,2) {$((f_1,f_2,\bot),(1,1))$};
  \draw[->,blue,line width=0.9mm] (init) --  (f12);
  \draw[->,blue,line width=0.9mm] (f12) -- (c); 
  \draw[->] (init) -- (f3); 
  \draw[->] (f12) -- (d1);
  \draw[->] (a) --  (c);
  \draw[->] (f12) --  (a);
  \draw[->] (f3) -- (a);
  \draw[->] (d1) --  (c);
  \draw[->] (a) -- (c);

\end{tikzpicture}
\caption{Path $\pi_1$}
		\end{subfigure}
		\begin{subfigure}{0.55\textwidth}
			\begin{tikzpicture}
  \node (init) at (0,-2) {$((d_1,d_2,d_3),(0,0))$};
  \node (f12) at (2,0) {$((f_1,f_2,d_3),(1,0))$};
  \node (f3) at (-2,0) {$((d_1,d_2,\bot),(0,1))$};
  \node (d1) at (4,2) {$((\bot,f_2,d_3),(2,0))$};
  \node (c) at (2,4) {$((\bot,f_2,\bot),(2,1))$};
  \node (a) at (0,2) {$((f_1,f_2,\bot),(1,1))$};

  \draw[->,red,line width=0.9mm] (init) --  (f12); 
  \draw[->] (init) -- (f3); 
  \draw[->,red,line width=0.9mm] (f12) -- (d1);
  \draw[->] (a) --  (c);

  \draw[->] (f12) --  (a);
  \draw[->] (f3) -- (a);
  \draw[->,red,line width=0.9mm] (d1) --  (c);
  \draw[->] (a) -- (c);

\end{tikzpicture}
\caption{Path $\pi_2$}
		\end{subfigure}		
		}\\
	\scalebox{0.6} {
        \centering
		\begin{subfigure}{0.55\textwidth}
			\begin{tikzpicture}
  \node (init) at (0,-2) {$((d_1,d_2,d_3),(0,0))$};
  \node (f12) at (2,0) {$((f_1,f_2,d_3),(1,0))$};
  \node (f3) at (-2,0) {$((d_1,d_2,\bot),(0,1))$};
  \node (d1) at (4,2) {$((\bot,f_2,d_3),(2,0))$};
  \node (c) at (2,4) {$((\bot,f_2,\bot),(2,1))$};
  \node (a) at (0,2) {$((f_1,f_2,\bot),(1,1))$};
  \draw[->,green,line width=0.9mm] (init) --  (f12);  
  \draw[->] (init) -- (f3); 
  \draw[->] (f12) -- (d1);
  \draw[->] (a) --  (c);
  \draw[->,green,line width=0.9mm] (f12) --  (a);
  \draw[->] (f3) -- (a);
  \draw[->] (d1) --  (c);
  \draw[->,green,line width=0.9mm] (a) -- (c);

\end{tikzpicture}
\caption{Path $\pi_3$}
		\end{subfigure}
		\begin{subfigure}{0.54\textwidth}
			\begin{tikzpicture}
  \node (init) at (0,-2) {$((d_1,d_2,d_3),(0,0))$};
  \node (f12) at (2,0) {$((f_1,f_2,d_3),(1,0))$};
  \node (f3) at (-2,0) {$((d_1,d_2,\bot),(0,1))$};
  \node (d1) at (4,2) {$((\bot,f_2,d_3),(2,0))$};
  \node (c) at (2,4) {$((\bot,f_2,\bot),(2,1))$};
  \node (a) at (0,2) {$((f_1,f_2,\bot),(1,1))$};
  \draw[->] (init) --  (f12); 
  \draw[->] (init) -- (f3); 
  \draw[->] (f12) -- (d1);
  \draw[->] (a) --  (c);
  \draw[->] (f12) --  (a);
  \draw[->] (f3) -- (a);
  \draw[->] (d1) --  (c);
  \draw[->,orange,line width=0.9mm] (init) --  (a);
  \draw[->,orange,line width=0.9mm] (a) -- (c);

\end{tikzpicture}
\caption{Path $\pi_4$}
		\end{subfigure}	
		\begin{subfigure}{0.54\textwidth}
			\begin{tikzpicture}
  \node (init) at (0,-2) {$((d_1,d_2,d_3),(0,0))$};
  \node (f12) at (2,0) {$((f_1,f_2,d_3),(1,0))$};
  \node (f3) at (-2,0) {$((d_1,d_2,\bot),(0,1))$};
  \node (d1) at (4,2) {$((\bot,f_2,d_3),(2,0))$};
  \node (c) at (2,4) {$((\bot,f_2,\bot),(2,1))$};
  \node (a) at (0,2) {$((f_1,f_2,\bot),(1,1))$};
  \draw[->] (init) --  (f12); 
  \draw[->,cyan,line width=0.9mm] (init) -- (f3); 
  \draw[->] (f12) -- (d1);
  \draw[->] (a) --  (c);
  \draw[->] (f12) --  (a);
  \draw[->,cyan,line width=0.9mm] (f3) -- (a);
  \draw[->] (d1) --  (c);
  \draw[->,cyan,line width=0.9mm] (a) -- (c);

\end{tikzpicture}
\caption{Path $\pi_5$}
		\end{subfigure}
		}
\caption{Computation lattice, all the associated paths and compatible traces associated to trace $t_1$ in Example~\ref{xmp:dmodel}}
\label{fig:latticet1}
\end{figure}

%% file: table.tex
\begin{table}[]
\centering
\ra{1.3}
\caption{On-the-fly construction and verification of computation lattice}
\label{table:examp}
\resizebox{\textwidth}{!}{
\begin{tabular}{@{}llr@{}}
\hline
step & event & \multicolumn{1}{c}{constructed lattice} \\ \hline
0    &   $\epsilon$    &      
     \scalebox{.6} {
\adjustbox{valign=c}{\begin{tikzpicture}
  \node (init) at (0,0) {$((d_1,d_2,d_3),(0,0), \setof {\varphi})$};
\end{tikzpicture}}}
      \\ \hline
1    &      \begin{tabular}[c]{@{}l@{}} - receiving event $\fil_{12}(1,0)$\\ - extending by making a new node\\ - the formula projection of new node \end{tabular}  & 
     \scalebox{.5} {
\adjustbox{valign=c}{\begin{tikzpicture}
%\begin{tikzpicture}
  \node (init) at (0,-2) {$((d_1,d_2,d_3),(0,0), \setof {\varphi})$};
  \node (f12) at (4,0) {$((\bot,\bot,d_3),(1,0),\setof{\varphi})$};
 
  \draw[->] (init) --  (f12); 
%    \node[
%      draw=red,
%      minimum width=\textwidth,
%      fit=(current bounding box.north west) (current bounding box.south east),
%    ]at (current bounding box.center){};
\end{tikzpicture}}}
     \\ \hline
2    &   \begin{tabular}[c]{@{}l@{}} - receiving event $\fil_3(0,1)$ \\ - extending by making a new node\\ - extending by making the joint node \\- the formula projection of new nodes \\- three formulas attached to the frontier\\~~represent the evaluation of three paths\\- the node with vector clock $(0,0)$ can be removed \end{tabular}     &  
     \scalebox{.5} {
\adjustbox{valign=c}{\begin{tikzpicture}
  \node [fill=red!20](init) at (0,-2) {$((d_1,d_2,d_3),(0,0), \setof {\varphi})$};
  \node (f12) at (4,0) {$((\bot,\bot,d_3),(1,0),\setof{\varphi})$};
  \node (f3) at (-4,0) {$((d_1,d_2,\bot),(0,1),\setof{\textbf{X}_\beta d_3 \wedge \varphi })$};
    \node (a) at (0,2) {$((\bot,\bot,\bot),(1,1),\setof{ (\textbf{X}_\beta d_3) \wedge (\textbf{X}_\beta d_3 \vee \textbf{X}_\beta f_1)  \wedge \varphi, (\textbf{X}_\beta d_3 \vee \textbf{X}_\beta f_1)\wedge \varphi, (\textbf{X}_\beta d_3 \vee \textbf{X}_\beta f_1)\wedge \varphi})$};

  \draw[->] (init) --  (f12); 
  \draw[->] (init) -- (f3); 
  \draw[dashed,->] (f12) --  (a);
  \draw[dashed,->] (f3) -- (a);
\end{tikzpicture}}}
     \\ \hline
\rule{0pt}{38pt} 3    &    \begin{tabular}[c]{@{}l@{}} - receiving event $(\beta_3,f_3)$ \\ - updating states of the existing nodes\\ - updating the formulas of existing nodes  \end{tabular}  &    
     \scalebox{.5} {
\adjustbox{valign=c}{\begin{tikzpicture}
  \node [fill=red!20] (init) at (0,-2) {$((d_1,d_2,d_3),(0,0), \setof {\varphi})$};
  \node (f12) at (4,0) {$((\bot,\bot,d_3),(1,0),\setof{\varphi})$};
  \node (f3) at (-4,0) {$((d_1,d_2,f_3),(0,1),\setof{F})$};
    \node (a) at (0,2) {$((\bot,\bot,f_3),(1,1),\setof{F, ( \textbf{X}_\beta f_1)\wedge \varphi, ( \textbf{X}_\beta f_1)\wedge \varphi})$};

  \draw[->] (init) --  (f12); 
  \draw[->] (init) -- (f3); 
  \draw[dashed,->] (f12) --  (a);
  \draw[dashed,->] (f3) -- (a);
\end{tikzpicture}}}
     \\ \hline
\rule{0pt}{38pt} 4    &   \begin{tabular}[c]{@{}l@{}} - receiving event $(\beta_2,f_2)$  \\ - updating states of the existing nodes\\ - updating the formulas of existing nodes  \end{tabular}   &   
     \scalebox{.5} {
\adjustbox{valign=c}{\begin{tikzpicture}
  \node [fill=red!20] (init) at (0,-2) {$((d_1,d_2,d_3),(0,0), \setof {\varphi})$};
  \node (f12) at (4,0) {$((\bot,f_2,d_3),(1,0),\setof{\varphi})$};
  \node (f3) at (-4,0) {$((d_1,d_2,f_3),(0,1),\setof{F})$};
  \node (a) at (0,2) {$((\bot,f_2,f_3),(1,1),\setof{F, ( \textbf{X}_\beta f_1)\wedge \varphi, ( \textbf{X}_\beta f_1)\wedge \varphi})$};

  \draw[->] (init) --  (f12); 
  \draw[->] (init) -- (f3); 
  \draw[dashed,->] (f12) --  (a);
  \draw[dashed,->] (f3) -- (a);
\end{tikzpicture}}
}      \\ \hline
\rule{0pt}{53pt} 5    &    \begin{tabular}[c]{@{}l@{}} - receiving event $\drain_{23}(1,2)$\\ - extending by making a new node\\ - the formula projection of new node\\- the node with vector clock $(0,1)$ can be removed \end{tabular}   & 
     \scalebox{.5} {
\adjustbox{valign=c}{\begin{tikzpicture}
  \node [fill=red!20] (init) at (0,-2) {$((d_1,d_2,d_3),(0,0), \setof {\varphi})$};
  \node (f12) at (4,0) {$((\bot,f_2,d_3),(1,0),\setof{\varphi})$};
  \node [fill=red!20] (f3) at (-4,0) {$((d_1,d_2,f_3),(0,1),\setof{F})$};
  \node (a) at (0,2) {$((\bot,f_2,f_3),(1,1),\setof{F, ( \textbf{X}_\beta f_1)\wedge \varphi, ( \textbf{X}_\beta f_1)\wedge \varphi})$};
  \node (b) at (-4,4) {$((\bot,\bot,\bot),(1,2),\setof{F , (\textbf{X}_\beta d_3 \vee \textbf{X}_\beta f_1) \wedge ( \textbf{X}_\beta f_1)\wedge \varphi, (\textbf{X}_\beta d_3 \vee \textbf{X}_\beta f_1) \wedge ( \textbf{X}_\beta f_1)\wedge \varphi})$};

  \draw[->] (init) --  (f12); 
  \draw[->] (init) -- (f3); 
  \draw[->] (a) --  (b);
  \draw[dashed,->] (f12) --  (a);
  \draw[dashed,->] (f3) -- (a);
\end{tikzpicture}}
}      \\ \hline
\rule{0pt}{53pt} 6    &    \begin{tabular}[c]{@{}l@{}} - receiving event $(\beta_1,f_1)$   \\ - updating states of the existing nodes\\ - updating the formulas of existing nodes  \end{tabular}  &   
     \scalebox{.5} {
\adjustbox{valign=c}{\begin{tikzpicture}
  \node [fill=red!20] (init) at (0,-2) {$((d_1,d_2,d_3),(0,0), \setof {\varphi})$};
  \node (f12) at (4,0) {$((f_1,f_2,d_3),(1,0),\setof{\varphi})$};
  \node [fill=red!20] (f3) at (-4,0) {$((d_1,d_2,f_3),(0,1),\setof{F})$};
  \node (a) at (0,2) {$((f_1,f_2,f_3),(1,1),\setof{F,\varphi,\varphi})$};
  \node (b) at (-4,4) {$((f_1,\bot,\bot),(1,2),\setof{F,\varphi,\varphi})$};

  \draw[->] (init) --  (f12); 
  \draw[->] (init) -- (f3); 
  \draw[->] (a) --  (b);
  \draw[dashed,->] (f12) --  (a);
  \draw[dashed,->] (f3) -- (a);
\end{tikzpicture}}
}      \\ \hline
\end{tabular}
}
\end{table}

%% file: overview.tex
\pgfdeclarelayer{background}
\pgfdeclarelayer{foreground}
\pgfsetlayers{background,main,foreground}

\begin{figure}[t]
    \centering
\resizebox{\textwidth}{!}{             
\tikzstyle{block} = [rectangle, draw, fill=gray!10, 
    text centered]   
\tikzstyle{block2} = [rectangle, draw,fill=yellow,  
    text width=5em, text centered, minimum height=2.5em]   
\tikzstyle{circle1} = [rectangle, draw,fill=yellow,  
    text width=5em, text centered, minimum height=2.5em]    
\tikzstyle{block3} = [rectangle, draw,fill=cyan!40,  
    text width=5em, text centered, minimum height=7em]     
\tikzstyle{line} = [draw, -latex']  
\def\myarm{1cm}
\def\myangle{0}
\tikzset{
  arm/.default=1cm,
  arm/.code={\def\myarm{#1}},
  angle/.default=0,
  angle/.code={\def\myangle{#1}} 
}   
\tikzset{
    myncbar/.style = {to path={
        let
            \p1=($(\tikztotarget)+(\myangle:\myarm)$)
        in
            -- ++(\myangle:\myarm) coordinate (tmp)
            -- ($(\tikztotarget)!(tmp)!(\p1)$)
            -- (\tikztotarget)\tikztonodes
    }}
} 
\begin{tikzpicture}[node distance = 1.5cm, auto]   
 \node[xshift=1.5cm,cloud,dashed, cloud puffs=9.7, cloud ignores aspect, minimum width=6cm, minimum height=4cm, align=center, draw,fill=green!40] (Tg) at (0cm, -0.5cm) {};
 \node [xshift=1.5cm,yshift=-1.2cm]  {Unobservable Global Trace $t$};
 \node [block,text width=10em,node distance = 1cm, xshift=1.5cm, yshift=-0.69cm]  (Sg) {Local Partial-Traces};
 \node [block2, text width=10em,xshift=1.5cm,]  (BIPg) {Distributed CBS};
 \node [xshift=14.8cm, yshift=-1.4cm, text width=10em, align=center]  {Sequence of\\ Partially-Ordered\\ Events};
 \node [block2,text width=8em, xshift=18cm, yshift=-0.5cm,fill=gray!40] (Tw) {Online Algortihm}; 
 \node [xshift=18cm, yshift=0.3cm]  {\algoref{alg:make}};

 \node [xshift=19.5cm, yshift=-2.8cm,rotate=-90]  {Lattice $\mathcal{L}$};
 \node [below of =Tw,yshift=-0.8cm,] (S){\frame{\includegraphics[width=17mm,angle =-10] {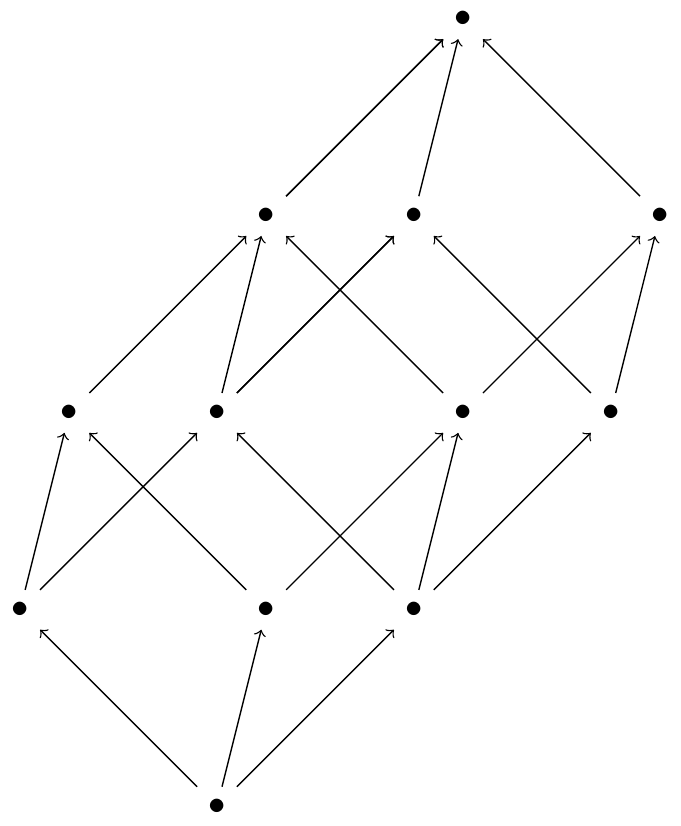}}};
 \node [block3,node distance = 1cm,text width=12em, draw=black!30, fill=cyan!15, xshift=10.8cm, yshift=-0.1cm] (Tp4) {};
 \node [block3,node distance = 1cm,text width=12em, draw=black!40, fill=cyan!20, xshift=10.6cm, yshift=-0.2cm] (Tp3) {};
 \node [block3,node distance = 1cm,text width=12em, draw=black!50, fill=cyan!25, xshift=10.4cm, yshift=-0.3cm] (Tp2) {};
 \node [block3,node distance = 1cm,text width=12em, draw=black!60, fill=cyan!35, xshift=10.2cm, yshift=-0.4cm] (Tp1) {};
 \node [block3,node distance = 1cm,text width=12em, xshift=10cm, yshift=-0.5cm] (Tp) {};
 \node [xshift=10.4cm, yshift=1.5cm]  {Observable Local Partial-Traces};
 \node [xshift=10cm,   yshift=-1.57cm]  {Local Partial-Trace $S_1(t)$};
% \node [xshift=11.45cm,yshift=-1.57cm, black!70]  { $.$};
% \node [xshift=11.5cm, yshift=-1.52cm, black!60]  { $.$};
% \node [xshift=11.55cm,yshift=-1.47cm, black!50]  { $.$};
% \node [xshift=11.6cm, yshift=-1.42cm, black!40]  { $.$};
 \node [xshift=10.4cm, yshift=-1.17cm, black!60]  {Local Partial-Trace $S_{\textit{\textbar}\setofschedulers\textit{\textbar}}(t)$};
 \node [block,text width=9em,node distance = 1cm,xshift=10cm, yshift=-0.7cm]  (Sp) {Local Events};
 \node [xshift=6cm, yshift=0.2cm]  {Transformation};
 \node [block2, text width=9em,xshift=10cm]  (BIPp) {Instrumented Distributed CBS};  
 \path [line] (Tp)  -- (Tw);
 \path [line,-,black!60] (Tp1)  -- (Tw);
 \path [line,-,black!50] (Tp2)  -- (Tw);
 \path [line,-,black!40] (Tp3)  -- (Tw);
 \path [line,-,black!30] (Tp4)  -- (Tw);
 
 \path [line,ultra thick] (BIPg)  -- (BIPp); 
 \node [circle1, text width=14em, xshift=9cm,yshift=-4.5cm,fill=purple!30]  (TL) {Set of paths of $\mathcal{L}$\\$\Pi$};
 \node [circle1, text width=7em, xshift=9cm,yshift=-7.8cm,fill=blue!30]  (TLP) {$\mathit{progression}(\Pi)$};
 \node [circle1, text width=3.5em, xshift=12cm,yshift=-7.8cm,fill=blue!30]  (TLPL) {$\eta^f.\Sigma$};
 \node [circle1, text width=7em, xshift=1.5cm,yshift=-7.8cm,fill=blue!30,dashed]  (TLPT) {$\PROG(\mathcal{P}(t))$};
 
 \node [xshift=15.5cm, yshift=-7.3cm,text width=12em, align=center]  {Online\\ Progression/Update on $\mathcal{L}$};
 \node [xshift=15.5cm, yshift=-8.2cm,text width=17em, align=center]  {\defref{def:progression}, \defref{def:update}};
 \node [xshift=15.3cm, yshift=-4.2cm,text width=17em, align=center]  { \defref{def:set-of-paths}};

 \node [xshift=9.23cm, yshift=-6.1cm,rotate=-90,text width=10em, align=center]  {Standard\\ Progression\\ \cite{bauer12}};
 \node [xshift=1.75cm, yshift=-6.1cm,rotate=-90,text width=10em, align=center]  {Standard\\ Progression\\ \cite{bauer12}};
 \node [circle1, text width=14em, xshift=1.5cm,yshift=-4.5cm,fill=purple!30,dashed]  (M) {Set of Compatible Global Traces\\ $\mathcal{P}(t)$};
 \node [xshift=9cm, yshift=-3.5cm,red]  {(Soundness)};
 \node [xshift=9cm, yshift=-3cm,red]  {\proporef{proposition:make_soundness}};
 \node [xshift=5.2cm, yshift=-3.5cm,red]  {(Completeness)};
 \node [xshift=5.2cm, yshift=-3.cm,red]  {\proporef{proposition:make_completeness}};

 \path [line, dashed] (Tg)  -- (M); 
 \path [line] (TL)  -- (TLP);
 \path [line] (Tw) -- (S);
 \path [line] (S)  |- (TL);
 \path [line] (S)  |- (TLPL);
 \path [line, dashed] (M)  -- (TLPT);
 
 \node at (5.2,-4.5)  {{\Huge $\equiv$}};
 \node at (10.85,-7.9) {{\huge $=$}};
 \node [xshift=10.85cm, yshift=-8.7cm]  {Theorem~\ref{theorem:soundness}};
 \node [xshift=10.85cm, yshift=-9.2cm]  {(Soundness)};
 \node at (5.2,-7.9)  {{\Huge $=$}};
 \node [xshift=5.2cm, yshift=-8.7cm]  {Theorem~\ref{theorem:completeness}};
 \node [xshift=5.2cm, yshift=-9.2cm]  {(Completeness)};
 
 \begin{scope}[on background layer]
 \filldraw[dashed, draw=black!30, top color=cyan!15,bottom color=purple!30](Tp4.south west)--(TL.south west)--(TL.south east)--(Tp4.south east)--cycle;
 \filldraw[dashed, draw=black!30, top color=cyan!20,bottom color=purple!30](Tp3.south west)--(TL.south west)--(TL.south east)--(Tp3.south east)--cycle; 
 \filldraw[dashed, draw=black!30, top color=cyan!25,bottom color=purple!30](Tp2.south west)--(TL.south west)--(TL.south east)--(Tp2.south east)--cycle;
 \filldraw[dashed, draw=black!30, top color=cyan!35,bottom color=purple!30](Tp1.south west)--(TL.south west)--(TL.south east)--(Tp1.south east)--cycle;
 \filldraw[dashed, draw=black!30, top color=cyan!40,bottom color=purple!30](Tp.south west)--(TL.south west)--(TL.south east)--(Tp.south east)--cycle;
 \end{scope}

\end{tikzpicture}}
\caption{Approach overview}
\label{fig:over-view-ltl}
\end{figure}

%% file: implementation.tex
%%%%%%%%%%%%%%%%%%%%%%%%%%%%%%%%%%%%%%%%%%
\section{Implementation} 
\label{sec:implem}
%%%%%%%%%%%%%%%%%%%%%%%%%%%%%%%%%%%%%%%%%%
%
We present an implementation of our monitoring approach in a tool called $\toolname$.
$\toolname$ is a prototype tool implementing algorithm $\make$ presented in \secref{sec:ltl}, written in the C++ programming language.
$\toolname$ takes as input a configuration file describing the architecture of the distributed system and a list of events.
The configuration file has the parameters of system such as the number of schedulers, the number of components, the initial state of the system, the LTL formula to be monitored, the mapping of each atomic propositions to the components.
The formula is monitored against the sequence of events by progression over the constructed computation lattice. 
$\toolname$ outputs the evaluation of the constructed lattice by reporting the number of observed events, the number of existing nodes of the constructed lattice, the number of nodes which have been removed from the lattice due to optimizing the size of the lattice, the vector clock of the frontier node, the number of paths from the initial node to the frontier node which have been monitored (the set of all compatible traces), the set of formulas associated to the frontier node.
Figure~\ref{fig:tool-over-view} depicts the work-flow of $\toolname$. 
\input{tooloverview}
%
%

%% file: tooloverview.tex
\pgfdeclarelayer{background}
\pgfdeclarelayer{foreground}
\pgfsetlayers{background,main,foreground}

\begin{figure}[t]
    \centering
%\resizebox{\textwidth}{!}{             
\tikzstyle{block} = [rectangle, draw, fill=gray!10, 
    text centered]   
\tikzstyle{block2} = [rectangle, draw,fill=yellow,  
    text width=5em, text centered, minimum height=2.5em]   
\tikzstyle{circle1} = [rectangle, draw,fill=yellow,  
    text width=5em, text centered, minimum height=2.5em]    
\tikzstyle{block3} = [rectangle, draw,fill=yellow!40,  
    text width=5em, text centered, minimum height=1em]     
\tikzstyle{line} = [draw, -latex']  
\def\myarm{1cm}
\def\myangle{0}
\tikzset{
  arm/.default=1cm,
  arm/.code={\def\myarm{#1}},
  angle/.default=0,
  angle/.code={\def\myangle{#1}} 
}   
\tikzset{
    myncbar/.style = {to path={
        let
            \p1=($(\tikztotarget)+(\myangle:\myarm)$)
        in
            -- ++(\myangle:\myarm) coordinate (tmp)
            -- ($(\tikztotarget)!(tmp)!(\p1)$)
            -- (\tikztotarget)\tikztonodes
    }}
} 
\begin{tikzpicture}[node distance = 1.5cm, auto]

 \node [block3,node distance = 1cm,text width=2em,   yshift=-0.5cm] (Tp4) {};
 \node [block3,node distance = 1cm,text width=2em,   yshift=-1cm] (Tp3) {};
 \node [block3,node distance = 1cm,text width=2em,   yshift=-1.5cm] (Tp2) {};
 \node [block3,node distance = 1cm,text width=2em,   yshift=-2cm] (Tp1) {};
 \node [block3,node distance = 1cm,text width=2em] (Tp) {};  
 
 \node [block3,node distance=1cm,text width=8em,minimum height=3em, fill=cyan!30 ,xshift=3.5cm, yshift=-1cm] (buffer) {Sequence of events};
 \node [block3,node distance=1cm,text width=6em, fill=black!20,xshift=7cm,yshift=-1cm](tool){$\toolname$};
 \node [block3,node distance=1cm,text width=8em, fill=green!30,xshift=10.5cm,yshift=-1cm](verdict){Verdicts};  
 \node [block3,node distance=1cm,text width=6em, fill=none,xshift=7cm,yshift=0cm](config){$\texttt{config.ini}$}; 
 \node [xshift=-0.8cm, yshift=-1cm,rotate=90]  {{\footnotesize Controllers of schedulers}};
 \node [xshift=1.3cm, yshift=-0.1cm,rotate=-16]  {local events};
 \path [line] (Tp)   -- (buffer);
 \path [line] (Tp1)  -- (buffer);
 \path [line] (Tp2)  -- (buffer);
 \path [line] (Tp3)  -- (buffer);
 \path [line] (Tp4)  -- (buffer);
 
 \path [line] (buffer)  -- (tool);
 \path [line] (config)  -- (tool);
 \path [line] (tool)  -- (verdict);

\end{tikzpicture}
%}
\caption{Overview of $\toolname$ work-flow}
\label{fig:tool-over-view}
\end{figure}

%% file: evaluation.tex
%
%%%%%%%%%%%%%%%%%%%%%%%%%%%%%%%%%%%%%%%%%%
\section{Evaluation}
\label{sec:evaluation}
%%%%%%%%%%%%%%%%%%%%%%%%%%%%%%%%%%%%%%%%%%
%
%We tested our monitoring algorithm on two systems modeled in BIP.
%%
%BIP (Behavior, Interaction, Priority) framework is a component-based framework with formal operational semantics. 
%%
%Coordination between components is achieved by using multiparty interactions and dynamic priorities for scheduling interactions.
%%
%BIP allows the generation of correct-by-construction distributed implementations from a high-level model.
%
We present the evaluation of our monitoring approach on two case studies carried out with $\toolname$.
\subsection{Case Studies}
We present a realistic example of a robot navigation and a model of two phase commit protocol (TPC).
\subsubsection{Deadlock Freedom of Robotic Application ROBLOCO}
The functional level of this navigating robot consists of a set of modules. 
ROBLOCO is in charge of the robot low-level controller. 
	It has a \textit{track} task associated to the activities \textit{TSStart/TSStop} (TrackSpeedStart, TrackSpeedStop). 
	\textit{TSStart} reads data from the dedicated \textit{speed} port and sends it to the motor \textit{controller}. 
	In parallel, the \textit{manager} module, which is associated to the \textit{odo} task  activities (OdoStart and OdoStop) reads the signals from the encoders on the wheels and produces a current position on the \textit{pos} port.
ROBLASER is in charge of the laser.
	It has a \textit{scan} task associated to the \textit{StartScan/StopScan} activities. 
	They produce the free space in the laser's range tagged with the position where the scan has been made.
ROBMAP aggregates the successive \textit{scan} data in the \textit{map} port. 
ROBMOTION has one task plan which, given a goal position, computes the appropriate speed to reach it and writes it on \textit{speed}, using the current position, and avoiding obstacles.
We deal with the most complex module, \ie ROBLOCO, involving three schedulers in charge of the execution of the dedicated actions.
ROBLOCO has 34 components and 117 multi-party interactions synchronizing the actions of components.
Since tasks are based on the specific sequence of the execution of interactions and tasks are not totally independent, there exist some share components which are involved in more than one task.
%
%\paragraph{Property:} 
To prevent deadlocks in the system, it is required that whenever the \textit{controller} is in \textit{free} state, at some point in future, the \textit{signal} module must reach the \textit{start} state before the \textit{manager} starts managing a new \textit{odo} task.
The deadlock freedom requirement can be defined as LTL formula $\varphi_1$.
\begin{itemize}
\item[$\varphi_1$:] $\textbf{G}\bigg( \texttt{ControlFree} \implies (\textbf{X}\neg\texttt{ManagerStartodo} \textbf{ U} \texttt{SignalStart})\bigg)$
\end{itemize}
%
%Property $\varphi$ states that whenever the \textit{controller} is in \textit{free} state, at some point in future, the \textit{signal} module must reach the \textit{start} state before the \textit{manager} start managing a new \textit{odo} task.
%
\subsubsection{Protocol Correctness of Two Phase Commit (TPC):}
We consider the distributed transaction commit~\cite{gray06} problem where a set of nodes called \textit{resource managers} $\{rm_1, rm_n, \ldots, rm_n\}$ have to reach agreement
on whether to \textit{commit} or \textit{abort} a transaction.
Resource managers are able to locally \textit{commit} or \textit{abort} a transaction based on a local decision. 
In a fault-free system, it is required the global system to commit as a whole if each resource manager has locally committed, and that it aborts as a whole if any of the resource managers has locally aborted. 
In case of global abort, locally-committed resource managers may perform roll-back steps to undo the effect of the last transaction~\cite{tanenbaum02}. 

Two phase commit protocol is a solution proposed by~\cite{gray78} to solve the transaction commit problem.
It uses a \textit{transaction manager} that coordinates between resource managers to ensure they all reach one global decision regarding a particular transaction. 
The global decision is made by the transaction manager based on the feedback it gets from resource managers after making their local decision ($\texttt{LocalCommit}/\texttt{LocalAbort}$).

The protocol, running on a transaction, uses a \textit{client}, a transaction manager and a non-empty set of resource managers which are the active participants of the transaction.
The protocol starts when client sends remote procedure to all the participating resource managers. 
Then each participating resource manager $rm_i$ makes its local decision based on its local criteria and reports its local decision to transaction manager.
$\texttt{LocalCommit}_i$ is true if resource manager $rm_i$ can locally-commit the transaction, and $\texttt{LocalAbort}_i$ is true if resource manager $rm_i$ cannot locally-commit the transaction.
Each participant resource managers stays in wait location until it hears back from transaction manager whether to perform a global commit or abort for the current transaction.
%
%The resource manager then resides in a state that represents the decision it made. 
%
After all local decisions have been made and reported to transaction manager, the latter makes a global decision ($\texttt{GlobalCommit}/\texttt{GlobalAbort}$) that all the system will agree upon. 
When $\texttt{GlobalCommit}$ is true, the system will globally-commit as a whole, and it will abort as a whole when $\texttt{GlobalAbort}$ is true.
We consider two specifications related to TPC protocol correctness:
\begin{itemize}
	\item[$\varphi_2$:] $\textbf{G}\bigg(\bigwedge_{i=1}^n (\texttt{LocalAbort}_i  \implies \textbf{X}(\neg\texttt{LocalAbort}_i \wedge  \neg\texttt{LocalCommit}_i ) \textbf{ U} \texttt{GlobalAbort})\bigg)$,
	\item[$\varphi_3$:] $ \textbf{G}\bigg( \bigwedge_{i=1}^n \texttt{LocalCommit}_i  \implies \textbf{X}\bigg( \bigwedge_{i=1}^n (\neg\texttt{LocalAbort}_i \wedge  \neg\texttt{LocalCommit}_i )\bigg) \textbf{ U} \texttt{GlobalCommit}\bigg)$.
\end{itemize}
Property $\varphi_2$ states that, sending locally abort in any resource managers for a current transaction implies the global abort ($\texttt{GlobalAbort}$) on that transaction before the resource manager locally aborts or commits the next transaction, that is, none of the resource managers commit.
Property $\varphi_3$ states that, if all the resource managers send locally commit for a current transaction, then all the resource managers commit the transaction ($\texttt{GlobalCommit}$) before the resource managers locally aborts or commits the next transaction.
For each system we applied the model transformation defined in Section~\ref{sec:instrumentation} and run them in a distributed setting. 
Each instrumented system produces a sequence of event which is generated and sent from the controllers of its schedulers.
The events are sent to the $\toolname$ where the associated configuration file is already given.
Upon the reception of each event, $\toolname$ applies the online monitoring algorithm introduced in Section~\ref{sec:ltl} and outputs the result consists of the information stored in the constructed computation lattice and evaluation of the desired LTL property so far.
%
%

%\subsection{Shared Component vs. Lattice Size}
%
In the following we investigate how the number of shared components effects on the size of the computation lattice over a very simple example of a distributed CBS with multiparty interaction.
%number of messages in a distributed system which is due to extra communication needed for vector clock exchange among the controllers of schedulers and shared components.
%
\input{plotsimple}
\begin{example}[Shared component, lattice size]
\label{exmp:simple} 
Let us consider a component-based system consists of four independent components $\textit{Comp}_1, \ldots, \textit{Comp}_4$. 
Each component has two actions $\textit{Action}_1, \textit{Action}_2$ which are designed to only be executed with the following order: $\textit{Action}_1 . \textit{Action}_2 . \textit{Action}_1$ and then the component terminates.
We distribute the execution of actions using four schedulers $\textit{Sched}_1, \ldots, \textit{Sched}_4$.
For the sake of simplicity, we consider each action of the components as a singleton interaction of the system such that $\Actions = \{\textit{Comp}_i.\textit{Action}_1, \textit{Comp}_i.\textit{Action}_2 \mid i \in [1,4]\}$.
Each scheduler manages a subset of $\Actions$.
We define various partitioning of the interactions to obtain the following settings:
\begin{enumerate}
	\item Each scheduler is dedicated to manage the actions of only one component, such that actions $\textit{Comp}_i.\textit{Action}_1$ and $ \textit{Comp}_i.\textit{Action}_2$ are managed by Scheduler $\textit{Sched}_i$ for $i \in [1,4]$.
	In this setting, no component has shared its actions to more that one scheduler. 
	\item Considering the previous setting with the only difference that action $\textit{Comp}_1.\textit{Action}_2$ by scheduler  $\textit{Sched}_2$. 
	In this setting, component $\textit{Comp}_1$ is a shared component.
	\item Considering the previous setting with the only difference that action $\textit{Comp}_2.\textit{Action}_2$ by scheduler  $\textit{Sched}_3$. 
	In this setting, components $\textit{Comp}_1$ and  $\textit{Comp}_2$ are  shared component.
	\item Considering the previous setting with the only difference that action $\textit{Comp}_3.\textit{Action}_2$ by scheduler  $\textit{Sched}_4$. 
	In this setting, components $\textit{Comp}_1$,  $\textit{Comp}_2$  and  $\textit{Comp}_3$ are shared component.
	\item Considering the previous setting with the only difference that action $\textit{Comp}_4.\textit{Action}_2$ by scheduler  $\textit{Sched}_1$. 
	In this setting, all the components are shared component.
\end{enumerate}
Since the components are designed to be involved only in three actions, the number of generated action/update events is equal in different settings (24 events in total), no matter which scheduler manages which action, and the only differences of events obtained through those setting are the vector clock of the action events and the sender of action/update events. 
Table~\ref{table:simple} and Figure~\ref{fig:simple} represent the results with respect to the above-mentioned settings.
Columns in Table~\ref{table:simple} have the following meaning:
\begin{itemize}
	\item Column \textit{shared component} indicates the number of shared components in each setting.
	%\item Column \textit{message overhead} indicates the overhead in the number of exchanged messages due to communication among the added controllers. 
	\item Column \textit{lattice nodes} shows the number of the nodes of the constructed lattice in each setting.
	\item Column \textit{removed nodes} indicates the number of removed nodes in the lattice using the optimization algorithm.
	\item Column \textit{path} indicates the number of paths of the constructed computation lattice.
\end{itemize}
Considering the first setting where there is no shared component in the system, results a set of independent action events where none of the two action events from two different schedulers are causally related, so that we construct a complete (maximal) computation lattice in order to cover all the compatible global traces.
The size of constructed lattice as well as the number of paths of the lattice is decreased by considering more shared components (see Figure~\ref{fig:nodes} and Figure~\ref{fig:paths}).
%
%Moreover, the number of extra messages increases proportionally with the number of shared components (see Figure~\ref{fig:msgs}).
%%
%More accurately, one can use function $\mathit{msg} : \IntActions \rightarrow \mathbb{N}$ which takes as input an interaction and outputs  an integer value corresponding to the number of messages exchanged among the controllers to execute that interaction, such that 
%$\mathit{msg}(a)=|I_a| + \sum\limits_{B_i \in I_a}|\setof{S_j \mid i \in \scope(j)}|$
%where $I_a=\setof{B_i \in B_{\rm s} \mid i \in \involved(a)}$.
%
\end{example}
\input{tablesimple}
\subsection{Results and Conclusion}
Table~\ref{table:experiments} and Figure~\ref{fig:lattice optimization} present the results checking specifications \textit{deadlock freedom} on ROBLOCO and \textit{protocol correctness} on TPC.
The columns of the table have the following meanings:
\begin{itemize}
	\item Column $|\varphi|$ shows the size of the monitored LTL formula.
	Note, the size of formulas are measured in terms of the operators entailment inside it, \eg $\textbf{G} (a \wedge b ) \vee \textbf{X}c$ is of size 2.
	\item Column \textit{observed event} indicates the number of action/update events sent by the controllers of the schedulers.
	\item Column \textit{lattice size} reports the size of constructed lattice using optimization algorithm is used vs. the size of constructed lattice when non-optimized algorithm is used. 
	\item Column \textit{frontier node VC} indicated the vector clock associated to the frontier node of the constructed lattice.
\end{itemize}
Figures~\ref{fig:robloco-opt},~\ref{fig:tcp-opt} show how the size of constructed lattice varies in two systems as they evolve.
Having shared components in system is not the only reason to have a small lattice size, what is more important is how  often the shared components are used as a part of executed interactions. The more execution with shared components results the more dependencies in the generated events and thus the smaller lattice size.

In ROBLOCO system, after receiving 3463 events, the size of the obtained computation lattice is 17, whereas the size of non-optimized lattice is 10602 which is a quite large in terms of storage space and iteration process. 
It shows how efficient our optimization algorithm minimize and optimize the monitoring process.
Figures~\ref{fig:robloco-opt} shows how the size of constructed lattice varies by the time the ROBLOCO  systems evolves.
Although what we need in the constructed computation lattice as the verdicts of the monitor output is only stored in the frontier node, but the rest of the nodes are necessary to be kept at runtime in order to extend the lattice in case of reception of new events.

In TPC system we also obtained a very small size of the lattice after the reception of 4709 events.
As it is shown in Table~\ref{table:experiments} the size and complexity of the LTL property does not change the structure of the constructed lattice, it only effects on the progression process. 
The frontier-node vector clock shows that how many interactions have been executed by each scheduler at the end of the system run.

Our monitoring algorithm implemented in $\toolname$ provide a lightweight tool to runtime monitor the behavior of a distributed CBS. 
$\toolname$ keeps the size of the lattice as small as possible even for a long run.
\begin{figure}[t]
  \center
  \begin{subfigure}{1\textwidth}
    \begin{tikzpicture}
  \begin{axis}[
  		width=\linewidth,height=7cm,
  		grid=major, 
  		grid style={dashed,gray!30},
  		xmin=0, xmax=1567,
  		ymin=0, ymax=170,
  		ylabel=Number of nodes,
  		xlabel=Number of action events
  		]
    \addplot[mark=none,blue] table [x expr=\coordindex,y index=0] {LatticeSizeROB.text};
  \end{axis} 
  \end{tikzpicture}
  \caption{ROBLOCO}
  \label{fig:robloco-opt}
  \end{subfigure}
  \begin{subfigure}{1\textwidth}
  \begin{tikzpicture}
    \begin{axis}[
  		width=\linewidth,height=7cm,
  		grid=major, 
  		grid style={dashed,gray!30},
  		xmin=0, xmax=1807,
  		ymin=0, ymax=55,
  		ylabel=Number of nodes,
  		xlabel=Number of action events
  		]
    \addplot[mark=none,blue] table [x expr=\coordindex,y index=0] {LatticeSizeTPC.text};
  \end{axis} 
\end{tikzpicture}
\caption{TPC}
\label{fig:tcp-opt}
\end{subfigure}
    \caption{Optimization algorithm effect on the size of the constructed lattice}
    \label{fig:lattice optimization}
\end{figure}
\begin{table}[t]
\centering
\caption{Results of monitoring ROBLOCO and TPC with $\toolname$}
\label{table:experiments}
\begin{tabular}{|c|c|c|c|c|c|c|}
\hline
\multirow{2}{*}{System} & \multirow{2}{*}{property} & \multirow{2}{*}{$|\varphi|$} & \multirow{2}{*}{\# observed events} & \multicolumn{2}{c|}{\# lattice size} & \multirow{2}{*}{frontier node VC} \\ \cline{5-6}
 &  &  &  & optimized & not-optimized &  \\ \hline
ROBLOCO & $\varphi_1$ & 3 & 3463 & 17 & 10602 & (730,352,485) \\ \hline
\multirow{2}{*}{TPC} & $\varphi_2$ & 10 & \multirow{2}{*}{4709} & \multirow{2}{*}{11} & \multirow{2}{*}{2731} & \multirow{2}{*}{(402,402,402,601)} \\ \cline{2-3}
 & \multicolumn{1}{c|}{$\varphi_3$} & 26 &  &  &  &  \\ \hline
\end{tabular}
\end{table}

%% file: plotsimple.tex
  \begin{figure}[t]
  \center
  \begin{subfigure}{0.49\textwidth}
  \begin{tikzpicture}
    \begin{axis}[
  		width=\linewidth,height=6cm,
  		grid=major, 
  		grid style={dashed,gray!30},
  		ytick={18,34,56,88,175},
  		ylabel=Number of nodes,
  		xlabel=Number of shared components
  		]
    \addplot[mark=square,blue] table [x index=0,y index=2] {msg.text};
  \end{axis} 
\end{tikzpicture}
\caption{Number of lattice nodes}
\label{fig:nodes}
\end{subfigure}
  \begin{subfigure}{0.49\textwidth}
  \begin{tikzpicture}
    \begin{axis}[
  		width=\linewidth,height=6cm,
  		grid=major, 
        xtick={0,1,2,3,4},
  		grid style={dashed,gray!30},
  		ylabel=Number of compatible paths,
  		xlabel=Number of shared components
  		]
    \addplot[mark=square,blue] table [x index=0,y index=3] {msg.text};
  \end{axis} 
\end{tikzpicture}
\caption{Number of lattice paths}
\label{fig:paths}
\end{subfigure}
\caption{Lattice construction vs. number of shared components}
\label{fig:simple}
\end{figure}

%% file: tablesimple.tex
\begin{table}[t]
\centering
\caption{Results of lattice construction w.r.t different settings of Example~\ref{exmp:simple}}
\label{table:simple}
\begin{tabular}{|c|c|c|c|}
\hline
\begin{tabular}[c]{@{}c@{}}shared\\ component\end{tabular} & \begin{tabular}[c]{@{}c@{}}lattice\\ nodes\end{tabular} & \begin{tabular}[c]{@{}c@{}}removed\\ nodes\end{tabular} & paths \\ \hline
0 & 175 & 81 & 10,681,263 \\ \hline
1 & 88 & 72 & 1,616,719 \\ \hline
2 & 56 & 60 & 572,847 \\ \hline
3 & 34 & 52 & 316,035 \\ \hline
4 & 18 & 47 & 251.177 \\ \hline
\end{tabular}
\end{table}

%% file: rw.tex
%\newpage
\section{Related Work}
\label{sec:rw}
A close work to the approach presented in this paper has been exposed in~\cite{bauer12}. 
In this setting, multiple components in a system each observe a subset of some global event trace. 
Given an LTL property $\varphi$, their goal is to create sound formula derived from $\varphi$ that can be monitored on each local trace, while minimizing inter-component communication.
Similar to our approach, the monitor synthesis is based on the internal structure of the monitored system and the projection of the global trace upon each component is well-defined and known in advance. 
Moreover, all components consume events from the trace synchronously.
Compare to our setting, we target a distributed component system with asynchronous executions. 
Hence, instead of having a global trace at runtime, we are dealing with a set of possible global traces which possibly could happen during the run of the system. 

%%
%%%%%%%%%%%%%%%%%%%%%%%%
%\subsection{Consistent detection of global predicates.}
%%%%%%%%%%%%%%%%%%%%%%%%
%%
In~\cite{CooperM91}, Cooper and Marzullo present three algorithms for detecting global predicates based on the construction of the lattice associated with a distributed execution. The first algorithm determined that the predicate was \textit{possibly} true at some point in the past; the second algorithm determines that the predicate was \textit{definitely} true in the past; while the third algorithm establishes that the predicate is \textit{currently} true, but to do so it may delay the execution of certain processes.

%%
%%%%%%%%%%%%%%%%%%%%%%%%
%\subsection{Reachability analysis on distributed executions.}
%%%%%%%%%%%%%%%%%%%%%%%%
%%
In~\cite{DiehlJR93}, Diehl, Jard and Rampon present basic algorithm for trace checking of distributed programs by building the lattice of all reachable states of the distributed system under test, based on the on-the-fly observation of the partial order of message causality.
Compare to our approach, in our distributed setting schedulers don't communicate directly by sending-receiving messages. 
Moreover, no monitor has been proposed in~\cite{DiehlJR93} for the purpose of verification whereas in our algorithm we synthesize a runtime monitor which evaluate on-the-fly the behavior of the system based on the reconstructed computation lattice of partial-states.

%%%%%%%%%%%%%%%%%%%%%%%
%\subsection{Efficient decentralized monitoring of safety in distributed systems.}
%%%%%%%%%%%%%%%%%%%%%%%
%
In~\cite{SenVAR04}, Sen and Vardhan design a method for monitoring safety properties in distributed systems using the past-time linear temporal logic (PLTL).
The distributed monitors gain knowledge about the state of the system by piggybacking on the existing communication among processes.
That is, if processes rarely communicate, then monitors exchange very little information and, hence, some violations of properties may remain undetected. 
In that paper, a tool called DIANA (distributed analysis) introduce in order to implement the proposed monitoring method.
The main noteworthy difference between~\cite{SenVAR04} and our work is that we evaluate the behavior of the distributed system based on all of the possible global traces of the distributed system. 
%
 
%%%%%%%%%%%%%%%%%%%%%%%
%\subsection{Efficient online monitoring of LTL properties for asynchronous distributed systems.}
%%%%%%%%%%%%%%%%%%%%%%%
%
In~\cite{MassartM06}, Massart and Meuter define an online monitoring method which collect the trace and checks on the fly that is satisfies a requirement, given by any LTL property on finite sequence. 
Their method explores the possible configurations symbolically, as it handles sets of configurations.  
Our approach mainly differs from~\cite{MassartM06} in that we target distributed CBSs with multi-party interactions where the execution traces are defined over the set of the partial states of the system.

%%%%%%%%%%%%%%%%%%%%%%%
%\subsection{Three-valued asynchronous distributed runtime verification.}
%%%%%%%%%%%%%%%%%%%%%%%
%
In~\cite{ScheffelS14}, Scheffel and Schmitz studied runtime verification of distributed asynchronous systems against Disributed Temporal Logic (DTL) properties. 
DTL combines the three-valued Linear Temporal Logic (LTL$_3$) with past-time Distributed Temporal Logic (ptDTL). 
In that paper, a distributed system is modeled as $n$ agents and each agent has a local monitor. 
These monitors work together to check a property, but they only communicate by adding some data to the messages already sent by the agents. 
They can not force their agent to send a message or even communicate on their own.
%

%%%%%%%%%%%%%%%%%%%%%%%
%\subsection{Decentralized Runtime Verification of LTL Specifications in Distributed Systems.}
%%%%%%%%%%%%%%%%%%%%%%%
%
In~\cite{MostafaB15}, a decentralized algorithm for runtime verification of distributed programs is proposed. 
Proposed algorithm conducts runtime verification for the 3-valued semantics of the linear temporal logic (LTL$_3$). 
In that paper, they adapt the distributed computation slicing algorithm for distributed online detection of conjunctive predicates, and also the lattice-theoretic technique is adapted for detecting global-state predicates at run time.
%

%%%%%%%%%%%%%%%%%%%%%%%
%\subsection{Detecting temporal logic predicates on the happened-before model.}
%%%%%%%%%%%%%%%%%%%%%%%
%
In~\cite{SenG01}, Sen and Garg use a temporal logic, CTL, for specifying properties of distributed computation and interpret it on a finite lattice of global states and check that a predicate is satisfied for an observed single execution trace of the program.
Compare to our approach, we deal with a set of events at runtime generated by the schedulers which results in  a infinite lattice of partial-states. Although the computation lattice in our method is made based of the observed partial states, we could check the satisfaction of temporal predicates defined over the global states of the system, which mean that we could monitor the system even if the global state of the is not defined.

%%%%%%%%%%%%%%%%%%%%%%%
%\subsection{Detecting temporal logic predicates in distributed programs using computation slicing.}
%%%%%%%%%%%%%%%%%%%%%%%
%
In~\cite{SenG04}, Sen and Garg used computation slicing for offline predicate detection in the subset of CTL with the following three properties; i) temporal operators, ii) atomic propositions are regular predicates and iii) negation operator has been pushed onto atomic propositions. 
They called this logic \textit{Regular CTL plus} (RCTL+), where plus denotes that the disjunction and negation operators are included in the logic. 
In that paper, authors gave the formal definition of RCTL+ which uses regular predicates as atomic propositions and implemented their predicate detection algorithms, which use computation slicing, in a prototype tool called Partial Order Trace Analyzer (POTA).
Morover, the authors developed this work in~\cite{SenG07}, by presenting the central online algorithm with respect to properties expressed in RCTL+.

%% file: conclusion.tex
%
%%%%%%%%%%%%%%%%%%%%%%%%%%%%%%%%%%%%%
\section{Conclusions and Future Work}
\label{sec:conc}
%%%%%%%%%%%%%%%%%%%%%%%%%%%%%%%%%%%%%
%
We draw conclusions and outline avenues for future work.
%%%%%%%%%%%%%%%%%%%%%%%%%%%%%%
\subsection{Conclusions}
%%%%%%%%%%%%%%%%%%%%%%%%%%%%%%
%
In this paper, we tackled the problem of runtime verification of distributed component-based system with multi-party interactions.
The goal is to verify the satisfaction or violation of properties referring to the global states of the system on-the-fly.
To this end, we introduced an abstract semantic model of CBSs consisting of a set of components, each of which is endowed with a set of actions, and a set of schedulers.
Each scheduler is in charge of executing the dedicated subset of multi-party interactions (joint actions).
The execution of each interaction triggers the set of actions of corresponding components involved in the interaction.
In the distributed setting, schedulers execute interactions by knowing the partial states of the system.
Therefore, the observable trace of the system is a sequence of partial states which is not suitable for verifying global-state properties.
Moreover, the total order of the executions of the interactions is not observable.

Our technique consists of two steps, (i) model instrumentation of the CBSs to extract the events of the system, and (ii) synthesizes a centralized monitor which collects the events, reconstructs the corresponding global trace(s), and verifies the desired properties on-the-fly.

In this context, the instrumented system outputs a sequence of partially-ordered events.
Since the total ordering of the events is not observable, we deal with a set of compatible partial traces of the system.
We showed that each compatible partial trace could have occurred as the actual run of the system.
Moreover, for each compatible partial trace, there exists a corresponding-compatible global trace.
The set of compatible global traces is represented in the form of a lattice.
In this setting, our approach (i) integrates a centralized observer which collects the local events of all schedulers (ii) constructs the computation lattice, and (iii) verifies on-the-fly any LTL property over the constructed lattice.
We introduced a novel online LTL monitoring technique on the computation lattice, so that each nodes carries a set of formulas evaluating the set of paths start from the initial node and end up with the node.
The set of formulas attached to the frontier node of the constructed lattice represents the evaluation of all the compatible global traces with respect to the given LTL formula.

We implemented our monitoring approach in a prototype tool called {\tooldist}.
{\tooldist} executes in parallel with the distributed system and takes as input the events generated from each scheduler and outputs the evaluated computation lattice.
The experimental results show that, thanks to the optimization applied in the online monitoring algorithm, the size of the constructed computation lattice is insensitive to the the number of received events, and the lattice size is kept reasonable.
%
%
%
%
%%
%In our setting, global joint actions are partitioned among a set of distributed scheduler.
%%
%Each scheduler is in charge of execution of the dedicated subset of global joint actions.
%%
%execution of each global actions, triggers the set of actions of corresponding component involved in the joint action.  
%%
%Our proposed technique consists in (i) transformation of the given distributed system to generate locally observed events by each distributed scheduler, (ii) synthesizing a centralized observer which collects the local events of all schedulers (iii) introducing an algorithm to reconstruct on-the-fly the set of possible ordering among the received events which forms a computation lattice, (iv) augmentation of the reconstructed computation lattice with a verification method in the sense that the observer plays the role of a runtime monitor while it is building the lattice.
%%
%We showed that the set of paths of the constructed lattice  represents the set of compatible traces, such that each of them could have occurred as the actual run of the system.
%%
%The experimental results show that even for a long run of a system, that is having many generated events, using the optimization algorithm keeps the size of the lattice minimal.
%%
%Moreover, the set of formulas attached to the frontier node of the constructed lattice represents the evaluation of all the compatible traces with respect to the given LTL formula.
%%
%%%%%%%%%%%%%%%%%%%%%%%%%%%%%%
\subsection{Future Work}
%%%%%%%%%%%%%%%%%%%%%%%%%%%%%%
%
%
Several research perspectives can be considered.

A first direction for monitoring distributed CBSs is to decentralizing the runtime monitor, such that the satisfaction or violation of specifications can be detected by local monitors alone.
%
%For distributed CBSs with large number of schedulers, a centralized algorithm requires a single process to perform high number of computations, and to store very large data.
%%
%By distributing the monitors we indeed decrease the load of monitoring process on a single process. 
%%
%Indeed, the centralized monitoring execution steps must be decomposed into multiple steps, so that these steps are executed by each local monitor independently.
%

Another direction is to use static analysis to detect a set of global states that can never occur at runtime, so that some nodes of the lattice can be ignored and the paths consisting of these nodes are never explored.
%
%Moreover, using static analysis leads to design an adaptive model instrumentation, in the sense that the runtime monitor only receives events that potentially affect the truth value of the property.
% 
Thus, the computation lattice size and monitoring load is decreased.

Runtime verification might provide a sufficient assurance to check whether or not the desired property is satisfied. 
However, for some classes of systems \eg safety-critical systems, a misbehavior might be not acceptable. 
To prevent this, a possible solution is to enforce the desired property so that the monitor does not only observe the current program execution, but it also controls it in order to ensure that the expected property is fulfilled.
Runtime enforcement was initiated by the work of Schneider~\cite{schneider00} on security automata.
Runtime enforcement consists in using a monitor to watch the current execution sequence and after it whenever it deviates from the property by for instance halting the system.
This line of work has been also studied for program monitoring in~\cite{ligatti05,falcone08,ligatti09,falcone11,pinisetty14,FalconeJMP16} and for monitoring sequential component-based systems in~\cite{falcone16}.

Another possible direction is to extend the proposed framework to runtime verify and enforce timed specifications on timed components.
Recently, the real-time multi-threaded and distributed CBSs~\cite{triki13,triki15} have been proposed, where each interaction has a timing constraint.
%
%In this setting, schedulers take into account the timing constraints before execute any interactions, \eg before triggering an interaction, the scheduler has to ensure that no interaction with an earlier deadline is enabled.
%
%
%In~\cite{demirbas13}, the authors presented a technique for augmenting vector clocks with real time to enable better ordering of events.

%% file: appendix.tex
%
%%%%%%%%%%%%%%%%%%%%%%%%%%%%%%%%%%%%%%%%%%%%%%
\section{Correctness Proof of the Approach}
\label{sec:proofs}
%%%%%%%%%%%%%%%%%%%%%%%%%%%%%%%%%%%%%%%%%%%%%%
%
In the following, we consider a distributed system $\model$ consisting a set of schedulers and components $\{S_1,\ldots,S_{|\setofschedulers|},B_1,$ $\ldots,B_{|\setofcomponents|}\}$ with the global behavior $( Q , \GActions , \trans{}{}) $ as per Definition~\pgref{def:global_behavior} and the transformed version of $\model$ due to monitoring purposes,  that is $\model^c$ consisting instrumented schedulers and shared components $\{S_1 \otimes \mathcal{C}^s_1,\ldots,S_{|\setofschedulers|} \otimes \mathcal{C}^s_{|\setofschedulers|},B'_1,$ $\ldots,B'_{|\setofcomponents|}\}$ with behavior $( Q_c , \GActions_c , \trans{c}{}) $  as per Definition~\pgref{def:instrumented_system} where $\mathcal{C}^s_j$ for $ j \in [1 \upto {|\setofschedulers|}]$ is the controller of scheduler $S_j$ as per Definition~\pgref{def:controller_scheduler} such that  $\forall i \in [1\upto {|\setofcomponents|}] \quant B'_i = B_i \otimes \mathcal{C}^b_i$  such that $\mathcal{C}^b_i$ is the controller of the share component $B_i$ if $B_i \in \setofsharedcomponents$  as per Definition~\pgref{def:controller_shared} and $B'_i = B_i$  otherwise.

We consider the global state of system $\model$ as $q=(q_{s_1},\ldots,q_{s_{|\setofschedulers|}},q_{b_1},\ldots,q_{b_{|\setofcomponents|}}) \in Q$, where $q_{s_j}$ is the state of scheduler $S_j$ for $j\in[1 \upto |\setofschedulers|]$, and  $q_{b_i}$ is the state of component $B_i$ for $i\in[1\upto |\setofcomponents|]$.
Moreover, the global state of system $\model^c$ is  $q'=(q'_{s_1},q_{sc_1},\ldots,q'_{s_{|\setofschedulers|}},q_{sc_{|\setofschedulers|}},q'_{b_1},q_{bc_1},\ldots,q'_{b_{|\setofcomponents|}},q_{bc_{|\setofcomponents|}}) \in Q_c$, where $q_{sc_j}$ is the state of the controller of scheduler $S_j$ for $j\in[1 \upto |\setofschedulers|]$, $q_{bc_i}$  is the state of the controller of shared component $B_i$ for $i\in[1\upto |\setofcomponents|]$ if $B_i \in B_{\rm s}$  and empty otherwise.
The first result concerns the correctness of the transformed model $\model^c$ through the instrumentation defined in~\ppsecref{sec:instrumentation}.

The instrumented model $\model^c$ generates events and sends them the observer in order to reconstruct the set of compatible global traces, that is, the computation lattice $\mathcal{L}$.
The second results concerns the correctness (soundness and completeness) of computation lattice construction presented in~\ppsecref{sec:lattice} with respect to the obtained events from the instrumented model.

The third result states the correctness (soundness and completeness) of the monitoring algorithm applied on the constructed lattice presented in \ppsecref{sec:ltl}, that is, our algorithm verifies the set of compatible global traces of the system. 
\paragraph{Proof outline.}
The following proofs are organized as follows.
Some intermediate definitions and lemmas are introduced in Appendix~\pgref{sec:intermidiate1} in order to prove Proposition~\ppproporef{proposition:sim} in Appendix~\pgref{proof:sim}.
The proof of Property~\pgref{property:ext} is in Appendix~\pgref{proof:unique_extenting_node}.
The proof of Property~\pgref{property:meet} is in Appendix~\pgref{proof:meet_existance}.
The proof of Proposition~\pgref{proposition:make} is in Appendix~\pgref{proof:make}.
The proof of \ppproporef{proposition:make_soundness} is in Appendix~\pgref{proof:make_soundness}.
The proof of~\ppproporef{proposition:make_completeness} is in Appendix~\pgref{proof:make_completeness}.
The proof of~\ppproporef{proposition:prog} is in Appendix~\pgref{proof:prog}.
The proof of Theorem~\pgref{theorem:soundness} is in Appendix~\pgref{proof:th_soundness}.
The proof of Theorem~\pgref{theorem:completeness} is in Appendix~\pgref{proof:th_completeness}.
%
%
%
%
%

%
%
%
%
%
%
%%%%%%%%%%%%%%%%%%%%%%%%%%%%%%%%%%%%%%%%%%%%%% 
\subsection{Intermediate Definition and Lemma}
\label{sec:intermidiate1}
%%%%%%%%%%%%%%%%%%%%%%%%%%%%%%%%%%%%%%%%%%%%%% 
%
%
We give some intermediate definition and lemma that are needed to prove Proposition~\pgref{proposition:sim} to prove the bi-simulation of $\model$ and $\model^c$.
First we define a relation between the states of two systems.
\begin{definition}
\label{def:equality}
Relation $\equ \subseteq Q \times Q_c$ is the smallest set that satisfies the following rule.
\[
(q,q') \in \equ \implies  q_{s_j}=q'_{s_j} \wedge  q_{b_i}=q'_{b_i}, \forall j\in[1\upto |\setofschedulers|] , \forall i\in[1\upto |\setofcomponents|]
\]
\end{definition}
Two states $q \in Q$ and $q' \in Q_c$ are in relation $\equ$ where the states of scheduler $S_j$ for  $j\in[1 \upto |\setofschedulers|]$ and the state of component $B_i$ for $i\in[1\upto |\setofcomponents|]$ in global state $q$ are the same as they are in global state $q'$.

The following lemma is a direct consequence of Definition~\ref{def:equality}.
\begin{lemma}
A global action $\alpha\in \GActions$ is enabled in the initial system at global state $q$, and in the transformed system at global state $q'$ if $(q,q') \in \equ$. 
\end{lemma}
Since controllers do not induce any restriction in the system in the sense that they do not hold the execution of the system, any enabled action in the initial system at state $q$ is also enabled in the augmented system at state $q'$ if $(q,q') \in \equ$.
%
%%%%%%%%%%%%%%%%%%%%%%%%%%%%%%%%%%%%%%%%%%%%%%%%%%%%%%%%%%
\subsection{Proof of \proporef{proposition:sim}  (p.~\pageref{proposition:sim})}
\label{proof:sim}
%%%%%%%%%%%%%%%%%%%%%%%%%%%%%%%%%%%%%%%%%%%%%%%%%%%%%%%%%%
%
We shall prove the existence of a bi-simulation between initial and transformed model, that is relation \\ $R = \setof{(q,q')\in Q\times Q_c \mid (q,q') \in \equ}$ satisfies the following property:
\[
\left( (q,q')\in R \wedge \exists \alpha \in \GActions, \exists z \in Q \quant q \longtrans{}{\alpha}  z \right) \implies \exists \alpha \in \GActions_c, \exists z'\in Q_c \quant \left(q' \longtrans{c}{\alpha} z' \wedge (z,z')\in R \right)
\]
%
%\item[(ii)]
%$\left( (q,q')\in R \wedge \exists \alpha \in \GActions, \exists z \in Q :q \longtrans{}{\alpha}  z \right) \implies \exists z'\in Q', q' \longtrans{}{\alpha} z' \wedge (z,z')\in R$.
%
\begin{proof}
After executing global action $\alpha$, states of the schedulers managed the interactions of the global action and components involved in the interactions are changed following the semantic rules defined in Definition~\pgref{def:global_behavior}.
Moreover, because of the equality of $q$ and $q'$, global action $\alpha$ is enabled at state $q'$ in the transformed system.
Execution of global action $\alpha$ in the transformed system following the semantic rules defined in Definition~\pgref{def:semantics-controller-scheduler} results the new states of the schedulers managed the action $\alpha$ which are the same as the states of the schedulers in the initial model after the execution of global action $\alpha$.

If any shared component is involved in global action $\alpha$, according to the semantic rules defined in Definition~\pgref{def:semantics-controller-shared-component}, state of the shared component in the transformed system after the execution of global action $\alpha$ is the same as it is in the initial model after the execution of $\alpha$.

Therefore, we can conclude that $(z,z')\in R$.
\end{proof}
%
%
%%%%%%%%%%%%%%%%%%%%%%%%%%%%%%%%%%%%%%%%%%%%%% 
\subsection{Proof of Property~\ref{property:ext} (p.~\pageref{property:ext})}
\label{proof:unique_extenting_node}
%
%%%%%%%%%%%%%%%%%%%%%%%%%%%%%%%%%%%%%%%%%%%%%% 
%
We shall prove that a computation lattice can be extended by an action event $e\in E_a$ from just one node. 
\begin{proof}
The proof is done by contradiction.
Let us assume that there exists two nodes $\eta,\eta' \in \mathcal{L}.\mathit{nodes}$ in such that $\upda(\eta, e)$ and $\upda(\eta', e)$ both are defined. 
According to Definition~\pgref{def:jrelation}, the vector clocks of the nodes  $\eta$ and $\eta'$ are in relation $\mathcal{J}_{\mathcal{L}}$, that is $(\eta.\mathit{clock},\eta'.\mathit{clock}) \in \mathcal{J}_{\mathcal{L}}$.
Moreover, its concluded that nodes  $\eta$ and $\eta'$ are associated to two concurrent action events.
%
%$\eta = ((q_1,\ldots,q_n),(\vc(1),\ldots,\vc(m)))$ and $\eta' = ((q'_1,\ldots,q'_n),(\vc'(1),\ldots,\vc'(m)))$
%
Based on the definition~\pgref{def:jointnode}, joint node of  $\eta$ and $\eta'$ has the same vector clock as $e.\mathit{clock}$, which means that we received an action event whose vector clock is already dedicated to another node in the lattice.
Reception of an action event with a vector clock similar to the vector clock of the joint node of $\eta$ and $\eta'$ defeats the concurrency between $\eta$ and $\eta'$ and contradict the assumption.
Therefore there exist at most one node in the lattice for with function $\upda$ is defined.
%
% are concurrent, therefore the joint of them, that is $\vartriangle(\eta,\eta')$, already exists in the lattice.
%
%We never receive an event $e= (a, \vc)\in E_a$ when in the computation lattice we have a node with the same vector clock $\vc$.
\end{proof}
%
%
%%%%%%%%%%%%%%%%%%%%%%%%%%%%%%%%%%%%%%%%%%%%%% 
\subsection{Proof of Property~\ref{property:meet} (p.~\pageref{property:meet})}
\label{proof:meet_existance}
%
%%%%%%%%%%%%%%%%%%%%%%%%%%%%%%%%%%%%%%%%%%%%%% 
%
We shall prove that the meet of two nodes $\eta,\eta' \in \mathcal{L}.\mathit{nodes}$ in relation $\mathcal{J}_{\mathcal{L}}$ exists in $\mathcal{L}.\mathit{nodes}$.
\begin{proof}
According to \ppdefref{def:node-extend}, for a node $\eta$ in the lattice where $\eta.\mathit{clock}= (c_1, \ldots,c_{|\setofschedulers|})$ we have $\exists \eta'\in \mathcal{L}.\mathit{nodes}\quant $ $\eta'.\mathit{clock}= (c'_1, \ldots,c'_{|\setofschedulers|})$ such that $\exists ! j\in [1 \upto |\setofschedulers|] \quant c'_j= c_j -1$.
Node $\eta'$ is the node that lattice $\mathcal{L}$ has been extended from, to make node $\eta$. 
If two nodes of the lattice satisfy relation $\mathcal{J}_{\mathcal{L}}$, According to \ppdefref{def:jrelation}, they have extended the lattice from a unique node which exists in the lattice according to the above argument.
\end{proof}
\subsection{Proof of Proposition~\ref{proposition:make} (p.~\pageref{proposition:make})}
\label{proof:make}
%%%%%%%%%%%%%%%%%%%%%%%%%%%%%%%%%%%%%%%%%%%%%% 
%
%$\forall \zeta ,\zeta' \in E^*: \zeta = e_1 \cdot e_2 \cdot e_3 \cdots e_z ,  \zeta' = e'_1 \cdot e'_2 \cdot e'_3 \cdots e'_r$
%
%$(z=r) \wedge \setof{e_1 , e_2 , e_3, \ldots ,e_z}=\setof{e'_1 , e'_2 , e'_3 ,\ldots ,e'_r} \implies \make(\zeta)=\make(\zeta')$
%
We shall prove that reception of certain events results computing a unique lattice and unique queue of stored events.
\[
\zeta ,\zeta' \in E^* \quant \left( \forall S_j \in \setofschedulers \quant \zeta \downarrow_{S_j} = \zeta'\downarrow_{S_j} \right) \implies  \make(\zeta) = \make(\zeta')
\]
%
%, no matter in which order they have been received. 
%
%$\forall \zeta ,\zeta' \in E^* \quant codom (\zeta) = codom (\zeta') \implies  \make(\zeta)=\make(\zeta')$.
%
\begin{proof}
The proof is done by induction over the length of the sequence of received events.
\begin{itemize}
	\item Base case. for the sequences of events with the length of one the proposition holds.
	\item Let us suppose that the proposition holds for two sequences of events $\zeta$ and $\zeta'$ such that $\make(\zeta)=\make(\zeta')$. By Extending $\zeta$ and $\zeta'$ via two events $e$ and $e'$ in different order such that $\zeta \cdot e_1 \cdot e_2$ and $\zeta' \cdot e_2 \cdot e_1$ are the new sequences, we have following possible cases:
	\begin{itemize}
		\item If either $e_1,e_2 \in E_a$ such that $e_1 \not\!\rightarrowtail e_2 \vee e_2\not\!\rightarrowtail e_1$ or  $e_1,e_2 \in E_\beta$, these events are said to be independent events in the sense that using one of them in order to extend/update the lattice or storing it in the queue does not depend to the reception of the other event.
		\item If  $e_1,e_2 \in E_a$ and there exist happened-before relation between them such that for instance $e_1 \!\rightarrowtail e_2$, and $e_2$ is received before $e_1$.
		After the reception of event $e_2$, one can't find node $\eta$ in the lattice in which $\upda(\eta,e_2)$ is defined, thus event $e_2$ is added to the queue.
		After the reception of event $e_1$ and extending the lattice with the associated node, the algorithm recalls event $e_2$ in the queue, as if event $e_1$ has been received earlier than $e_2$.
		 In other words, the algorithm reorders the received events by using the queue $\kappa$.
		\item If $e_1 \in E_a, e_2 \in E_\beta$, where $e_2$ is an update event contains the state of component $B_i$ for $i\in[1 \upto |\setofcomponents|]$.
		 Updating the lattice with event $e_2$ or storing event $e_2$  in the queue depends on the other events in the queue such that if there exists an action event associated to execution of an action concerning the component $B_i$, then $e_2$ must be stored in the queue.
		However, based on our assumption, it never be the case that from  a specific scheduler, the observer receives an update event associated to component component $B_i$ earlier than receiving the action event associated to the execution of an action concerning component $B_i$. 
		Therefore, updating the lattice with event $e_2$ or storing event $e_2$ in the queue does not depend on event $e_1$ which is going to be received later.
		
		Moreover, extending the lattice with the action event $e_2$ or storing  the action event $e_2$ in the queue does not depend on any update event.
	\end{itemize}
\end{itemize} 
\end{proof}
%
%
%%%%%%%%%%%%%%%%%%%%%%%%%%%%%%%%%%%%%%%%%%%%%% 
\subsection{Proof of~\proporef{proposition:make_soundness} (p.~\pageref{proposition:make_soundness})}
\label{proof:make_soundness}
%%%%%%%%%%%%%%%%%%%%%%%%%%%%%%%%%%%%%%%%%%%%%% 
%
We shall prove that for any possible set of events $\zeta$ of a given global trace $t$, the projection of the paths in the constructed lattice on scheduler $S_j$ with $j\in [1 \upto |\setofschedulers| ]$ results  the refined local trace of the scheduler.
$\forall \zeta \in \Theta(t) , \forall \pi \in \Pi\big(\make\left(\zeta\right).\mathit{lattice}\big) , \forall j\in[1 \upto |\setofschedulers| ]  \quant  \pi\downarrow_{S_j} = \mathcal{R}_\beta(s_j(t))$.
\begin{proof}
The proof is done by induction over the length of the global trace $t$.
\begin{itemize}
	\item Base case. The proposition holds initially where $t=\init$.
	\item Let us suppose that the proposition holds for a global trace $t$.
	
	We have two cases for the next action of the system, which leads to the extension of the global trace: 
	\item Any extension of the global trace by execution of an action $a \in \IntActions$ (\ie $t \cdot a \cdot q$) generates the associated action event $e=(a,\vc)$, and if there exist a node in the lattice $\eta \in \mathit{lattice}$ such that $\upda(\eta, e)$ is defined, then event $e$ extends the lattice from $\eta$ toward the direction of the scheduler manages interaction $a$. 
We consider two cases whether or not the lattice is extended from the frontier node:
	\begin{itemize}
		\item if $\eta$ is the frontier node (\ie $\eta=\eta^f$)
		
		$\Pi\big(\make\left(\zeta.e\right).\mathit{lattice}\big)= 
	 \setof{\pi \cdot a \cdot \eta' \lmid \pi \in \Pi\big(\make\left(\zeta\right).\mathit{lattice} \wedge \eta' = \upda(\eta, e) \big)} $. 
	 We have two cases $\forall j\in[1 \upto |\setofschedulers|]$:
	 \begin{itemize}
	 	\item if $ S_j \neq \managed(a)$ we have $s_j(t \cdot a \cdot q)=s_j(t)$, and\\ $\Pi\big(\make\left(\zeta.e\right).\mathit{lattice}\big) \downarrow_{S_j} = \Pi\big(\make\left(\zeta\right).\mathit{lattice}\big) \downarrow_{S_j}$,
	 	\item if $S_j = \managed(a)$ we have $s_j(t \cdot a \cdot q)=s_j(t) \cdot a \cdot q $, and $\Pi\big(\make\left(\zeta.e\right).\mathit{lattice}\big) \downarrow_{S_j} = \Pi\big(\make\left(\zeta\right).\mathit{lattice}\big) \downarrow_{S_j} \cdot a \cdot (\eta'.\mathit{state})$ where $\eta'$ is the new node.
	 \end{itemize}
		 
	 	\item if $\eta$ is not the frontier node
	 	
%	 	 $\Pi\big(\make\left(\zeta.e\right).\mathit{lattice}\big)= \Pi\big(\make\left(\zeta\right).\mathit{lattice}\big) \cup \\

	 $\{\pi(0 \ldots k) \cdot a \cdot \eta' \mid \pi \in \Pi\big(\make\left(\zeta\right).\mathit{lattice}\big) \wedge k \in[0 \upto \length(\pi)] \wedge \pi(k)=\eta \wedge \eta' = \upda(\eta, e) \big)\} $.
	 A set of new paths starting from the initial node and ending with node $\eta'$ is added to the set of paths.

	 First, Algorithm $\make$ extends the lattice by generating node $\eta' = \upda(\eta, e)$.
	  
	 Since $\eta$ is not the frontier node, there exists node $\eta'' \in \mathcal{L}.\mathit{nodes}$ such that $(\eta'.\mathit{clock},\eta''.\mathit{clock}) \in \mathcal{J}_{\mathcal{L}}$, meaning that execution of interaction $a$ is concurrent with the execution associated to node $\eta''$.\\
	 Extending the lattice with the possible joints leads to an extended lattice up to a new frontier $\eta^f_{\mathit{new}}= \upda(\eta^f, (a,\max(\vc,\eta^f.\mathit{clock})))$, where $\eta^f_{\mathit{new}}.\mathit{state}=q$.
	 
	 For each path $\pi$ in the initial lattice where $\pi = \pi_1 \cdot a' \cdot \pi_2$ such that $a'$ is concurrent to action $a$, and the vector clock of the first node of $\pi_2$ is in relation $\mathcal{J}_{\mathcal{L}}$ with the vector clock of node $\eta'$, we have a set of new paths $\pi'=\setof{ (\pi_1 \cdot a \cdot q_1 \cdot a' \cdot \pi'_2),(\pi_1 \cdot a' \cdot q_2 \cdot a \cdot \pi'_2), (\pi_1 \cdot (a\cup a') \cdot \pi'_2)}$ where $\pi'_2$ has the same sequence of actions with $\pi_2$ with the difference that $\pi'_2$ begins from the joint of $\eta'$ and the first node of $\pi_2$ and $\pi'_2$ ends up with node $\eta^f_{\mathit{new}}$.
	 Projection of each new path on schedulers results the following $\forall j\in[1 \upto |\setofschedulers|]$:
	 \begin{itemize}
	 	\item if $S_j \neq \managed(a)$ then $\pi' \downarrow_{S_j} = \pi \downarrow_{S_j}$
	 	\item if $S_j = \managed(a)$ then $\pi' \downarrow_{S_j} = \pi \downarrow_{S_j} \cdot a \cdot \pi'_2 \downarrow_{S_j}$
	\end{itemize}

	\end{itemize}
	\item Any extension of the global trace by execution of a busy action  (\ie $t \cdot \beta_i \cdot q$) generates an update event $e=(\beta_i,q_i)$.
	According to Algorithm $\make$, procedure \textsc{UpdateEvent} uses the state information of event $e$ and update the busy state of the nodes.
	Similarly, refine function updates the busy states associated to the component $B_i$ using $\upd$ function.  
	Therefore, for each updated path of the lattice $\pi$ we have $\forall j\in[1 \upto |\setofschedulers| ]  \quant  \pi\downarrow_{S_j} = \mathcal{R}_\beta(s_j(t\cdot \beta_i \cdot q))$.
\end{itemize}	
In either case, the proposition holds. 
%
%$\upda (e,\eta)= \eta' \Rightarrow \eta.\mathit{state} \longtrans{}{a} \eta'.\mathit{state}$
\end{proof}
%
%%%%%%%%%%%%%%%%%%%%%%%%%%%%%%%%%%%%%%%%%%%%%% 
\subsection{Proof of~\proporef{proposition:make_completeness} (p.~\pageref{proposition:make_completeness})}
\label{proof:make_completeness}
%%%%%%%%%%%%%%%%%%%%%%%%%%%%%%%%%%%%%%%%%%%%%% 
%
We shall prove that for any possible set of events $\zeta$ of a given global trace $t$, there exists a unique path in the constructed lattice associated to each compatible trace, such that $\forall \zeta \in \Theta(t) , \forall t' \in\mathcal{P}(t) , \exists ! \pi \in \Pi\bigg(\make\left(\zeta\right).\mathit{lattice}\bigg) \quant \pi = \mathcal{R}_\beta(t')$
\begin{proof}
The proof is done by induction over the length of the global trace $t$.
\begin{itemize}
	\item Base case. The proposition holds initially where $t=\init$.
	\item Let us assume that the proposition holds for a global trace $t$.
	
	\item We have two cases for the next action of the system, which leads to the extension of the global trace: 
	\begin{itemize}

	\item Any extension of the global trace by execution of a global action $a \in 2^{\IntActions}$ (\ie $t \cdot \alpha \cdot q$) results the new set of compatible traces $\mathcal{P}(t \cdot \alpha \cdot q)$ in which for each trace $t' \in \mathcal{P}(t)$	we have a set of extended traces considering all the possible ordering of the actions in $\alpha$ that is  
	$\mathcal{P}(t \cdot \alpha \cdot q) = \setof{t'' \cdot t'''  \mid t'' \in \mathcal{P}(t),  t''' \in \mathcal{P}(\last(t'') \cdot \alpha \cdot q)}$.
	According to~\pproporef{proposition:make_soundness} execution of an action causes the lattice extension considering all the possible orderings with the rest of the execution of the system.
	 Therefore for each trace $t'$ in set $\mathcal{P}(t \cdot \alpha \cdot q)$ there exists a corresponding path $\pi$ in the lattice such that  $\pi = \mathcal{R}_\beta(t')$.
	\item Any extension of the global trace by execution of a global action $\beta \subseteq \bigcup_{i \in [1\upto |\setofcomponents|]} \setof{ \beta_i }$ (\ie $t \cdot \beta \cdot q$) does not extend the lattice but the state of the nodes of the lattice. 
	Moreover, according to~\ppdefref{def:node_update}, updating the nodes of the path corresponding to a compatible trace $t$ is done in such a way that the refine function updates the partial states of $t$ as per~\ppdefref{def:refine}.
	Therefore, based on our assumption updated path $\pi$ is still equal to the corresponding refined compatible trace trace $t'$ after update by the internal action $\beta$.
	\end{itemize}
\end{itemize}
For both cases, the proposition holds. 
\end{proof}

%
%%%%%%%%%%%%%%%%%%%%%%%%%%%%%%%%%%%%%%%%%%%%%% 
\subsection{Proof of~\proporef{proposition:prog} (p.~\pageref{proposition:prog})}
\label{proof:prog}
%%%%%%%%%%%%%%%%%%%%%%%%%%%%%%%%%%%%%%%%%%%%%% 
%
We shall prove that for a given an LTL formula $\varphi$ and a global trace $t$, there exists a global trace $t'$ such that $\PROG(\varphi, t') =\mathit{progression} (\varphi, \mathcal{R}_\beta(t'))$ with:

\[
t' = \begin{dcases}
    t  &\text{ if } \last(t)[i] \in Q_i^{\rm r} \text{ for all } i \in [1\upto |\setofcomponents|], \\
   t \cdot \beta \cdot q  &\text{ otherwise.}
   \end{dcases}
\]
Where $ \beta \subseteq \bigcup_{i \in [1\upto |\setofcomponents|]} \setof{ \beta_i }$, $\forall i \in [1\upto |\setofcomponents|], q[i] \in Q_i^{\rm r}$ and $\mathit{progression}$ is the standard progression function.
\begin{proof}
The proof is done by induction over the length of the global trace $t$.
\begin{itemize}
	\item Base case. The proposition holds initially where\\ $\PROG(\varphi, \init) = \mathit{progression} (\varphi, \mathcal{R}_\beta(\init)) = \mathit{progression} (\varphi, \init)$.
	\item Let us assume that the proposition holds for a global trace $t$.
	\item We shall prove that the proposition holds for any possible extension of $t$.
	We consider two cases based on the last element of $t$: 
	\begin{itemize}
		\item if $\last(t)[i] \in Q_i^{\rm r}$ for all $i \in [1\upto |\setofcomponents|]$, the next action only could be in set $\alpha \subseteq 2^{\IntActions}$ because all the components are ready and no $\beta$ action is possible.
		Then, the new global trace is $t \cdot \alpha \cdot q$ and $q$ is a global state in which the state of components involved in $\alpha$ are busy state.
		According to~\ppdefref{def:bigprog}, function $\PROG(\varphi, t \cdot \alpha \cdot q)$ uses sub-function $\prog$ which postpones the formula evaluation by using $\textbf{X}_\beta$ until the busy components execute their busy actions.
		As soon as the busy components are done with their computations and the corresponding schedulers generate the update events, function $\PROG$ evaluates all postponed propositions.
		Therefore, there exists a stabilized global trace $t'=t \cdot \alpha \cdot q \cdot \beta \cdot q'$ with ready states for all components in $q'$.
		Moreover, according to~\ppdefref{def:refine}, the refined trace of $t'$ consists of sequence of global ready state so that the construction of ready states is postponed until the occurrence of busy actions  of busy components.
		Hence, $\PROG(\varphi, t \cdot \alpha \cdot q \cdot \beta \cdot q')=  \mathit{progression} (\varphi, \mathcal{R}_\beta(t \cdot \alpha \cdot q \cdot \beta \cdot q'))$. 
		\item if $\last(t)[i] \not\in Q_i^{\rm r}$ for all $i \in [1\upto |\setofcomponents|]$,
		% based on the assumption after stabilizing $t$ by $\beta$ actions of the busy components the proposition holds. 
		%
		we consider two cases based on the next action:
		\begin{itemize}
			\item If the next action is a busy action $\beta \subseteq \bigcup_{i \in [1\upto |\setofcomponents|]} \setof{ \beta_i } $, it follows our base assumption, therefore the proposition holds.
			\item If the next action contains actions in set $\alpha \subseteq 2^{\IntActions}$, the components involved in $\alpha$ are added to the set of busy components and their states evaluations are postponed until their next ready state (following the first case argument).  
		\end{itemize}
	\end{itemize}
In both cases, the proposition holds.
\end{itemize}
\end{proof}
%
%%%%%%%%%%%%%%%%%%%%%%%%%%%%%%%%%%%%%%%%%%%%%% 
\subsection{Proof of Theorem~\ref{theorem:soundness} (p.~\pageref{theorem:soundness})}
\label{proof:th_soundness}
%%%%%%%%%%%%%%%%%%%%%%%%%%%%%%%%%%%%%%%%%%%%%% 
%
We shall prove that for a global trace $t$ and LTL formula $\varphi$, each formula attached to the frontier node of the reconstructed computation lattice $\mathcal{L}$ corresponds to the evaluation of a compatible trace, that is $ \forall \varphi' \in \eta^f.\Sigma , \exists t' \in \mathcal{P}(t) \quant \PROG(\varphi, t')= \varphi'$.
\begin{proof}
The proof is done by induction over the length of the global trace $t$.
\begin{itemize}
	\item According to \ppdefref{def:lattice-agumentation}, initially lattice has only one node $\init^{\varphi}_{\mathcal{L}}= (\init,(0,\ldots,0),\{\varphi\})$. 
$\mathcal{P}(t)= \setof{\init}$ and $\PROG(\varphi, \init)= \varphi$ therefor the theorem holds for the initial state.
	\item We assume that for a global trace $t$ progression of all the compatible traces of $t$ exists in the set of formula of the frontier node.
	\item any extension creates the corresponding nodes. if the extension take take place from the frontier node $\eta^f$, then the set of formulas of the new frontier is $\Sigma = \{ \prog(\LTL',q) \mid   \LTL' \in \eta^f.\Sigma \}$.
	and for all compatible trace $t' \in \mathcal{P}(t)$ the corresponding extended compatible trace is $(t \cdot a \cdot q) \in \mathcal{P}(t \cdot a \cdot q)$.
	According to~\ppdefref{def:bigprog}, evaluation of each formula   $\PROG(\varphi, t \cdot a \cdot q)= \prog (\varphi, q)= \prog (\prog(\varphi, \eta^f.\mathit{state}), q)$. Base on our assumption $\prog(\varphi, \eta^f.\mathit{state})$ is the associated progressed formula of the compatible trace $t'$ which exists in $\eta^f.\Sigma$.
	\item Extension of the lattice from a non-frontier node also leads to a new frontier node such that the number of compatible traces are increased as well as the number of path of the lattice (see~\pproporef{proposition:make_completeness}).
	For each newly generated path in the lattice there exists a corresponding compatible trace. 
	The set of formulas associated to the new frontier node $\eta^f.\Sigma= \{ \prog(\LTL,\eta^f.\mathit{state}) \mid \LTL \in \eta.\Sigma \wedge ( \eta \xrightarrowtail[] \eta^f \vee \exists N \subseteq \mathcal{L}^\varphi.\mathit{nodes} \quant \eta=\meet(N,\mathcal{L}) \wedge \eta^f =\joint(N,\mathcal{L}) )\}$.
	Each formula associated to the new frontier is evaluation of one path of the lattice. 
	Similarly, the progression of global trace by function $\PROG$ is done using function $\prog$, so that the evaluation of each compatible global trace results a formula which exists in the set of formula of new frontier node associated to the compatible global trace.
	\item Any extension of the global trace by busy actions $\beta$ on one hand updates the formulas of lattice nodes using function $\updfct_\varphi$ in order to ascertain the truth of falsity of associated $\textbf{X}_\beta$ modalities, and on the other hand function $\PROG$ updates the evaluation of the compatible traces, \ie $\PROG(\varphi, t \cdot \beta \cdot q)= \UPD \left( \varphi, Q^r \right)$ where $Q^r$ is the set of update states after busy actions.
	According to~\ppdefref{def:bigprog}, function $\UPD$ uses function $\updfct_\varphi$ so the the progressed formula of each compatible global trace is updated in the similar way as the formulas in the frontier node are updated.
\end{itemize}
\end{proof}
%
%%%%%%%%%%%%%%%%%%%%%%%%%%%%%%%%%%%%%%%%%%%%%% 
\subsection{Proof of Theorem~\ref{theorem:completeness} (p.~\pageref{theorem:completeness})}
\label{proof:th_completeness}
%%%%%%%%%%%%%%%%%%%%%%%%%%%%%%%%%%%%%%%%%%%%%% 
%
We shall prove that the set of formula in the frontier node of the constructed lattice consists in the evaluation of the compatible global traces of the system, that is $\eta^f.\Sigma = \setof{\PROG(\varphi, t') \lmid t' \in \mathcal{P}(t) }$.
\begin{proof}
The proof is straightforward since the lattice construction is complete (\pproporef{proposition:make_completeness}), in the sense that for each compatible global trace there exists a path in the constructed lattice and according to Theorem~\pgref{theorem:soundness} each formula in the frontier node corresponds to the evaluation of a compatible global trace. 
\end{proof}